\documentclass{article}
\usepackage{graphicx}
\usepackage{float}
\usepackage{subfigure}
\usepackage[super]{nth}
\usepackage{amsmath}
\usepackage{bm}
\usepackage{hyperref}
\usepackage[dvipsnames]{xcolor}  
\usepackage{lineno} 
\usepackage{placeins}
\usepackage{array}
\usepackage{tabu}
\usepackage{amssymb}
\usepackage{mathtools}
\usepackage{lineno}
\usepackage[ruled,vlined]{algorithm2e}
\usepackage{algorithmic}
\usepackage{wrapfig}
\usepackage{booktabs}       
\usepackage{float}
\usepackage{authblk}
\usepackage{enumerate}
\usepackage{xcolor}

\usepackage{amsthm}
\providecommand{\keywords}[1]{\textbf{\textit{Keywords: }} #1}

\newcommand{\D}{\mathcal{D}}

\newcommand{\argmax}{\operatornamewithlimits{argmax}}

\title{Solving Inverse Stochastic Problems from Discrete Particle Observations Using the Fokker-Planck Equation and Physics-informed Neural Networks\thanks{
This work was funded by the DOE PhILMS grant DE-SC0019453, by the OSD/ARO/MURI  W911NF-15-1-0562, and by the NIH grant U01 HL142518. The work of the first author was supported by the China Scholarship Council under 201806160038. The first two authors contributed equally to this paper.
Corresponding author: George Em Karniadakis (george\_karniadakis@brown.edu) }}

\author[1,2]{Xiaoli Chen}
\author[2]{Liu Yang}
\author[3]{Jinqiao Duan}
\author[2,4]{George Em Karniadakis}
\affil[1]{Center for Mathematical Sciences, Huazhong
University of Science and Technology, Wuhan 430074, China}
\affil[2]{Division of Applied Mathematics, Brown University, Providence, RI 02912, USA}
\affil[3]{Department of Applied Mathematics, College of Computing, Illinois Institute of Technology, Chicago, IL 60616, USA}
\affil[4]{Pacific Northwest National Laboratory, Richland, WA 99354, USA}

\date{\vspace{-5ex}}

\begin{document}

\maketitle
\begin{abstract}
The Fokker-Planck (FP) equation governing the evolution of the probability density function (PDF) is applicable to many disciplines but it requires specification of the coefficients for each case,
which can be functions of space-time and not just constants, hence requiring the development of a data-driven modeling approach. When the data available is directly on the PDF, then there exist methods for inverse problems that can be employed to infer the coefficients and thus determine the FP equation and subsequently obtain its solution. Herein, we address a more realistic scenario, where only sparse data are given on the particles' positions at a few time instants, which are not sufficient to accurately construct directly the PDF even at those times from existing methods, e.g., kernel estimation algorithms. To this end, we develop a general framework based on physics-informed neural networks (PINNs) that introduces a new loss function using the Kullback-Leibler divergence to connect the stochastic samples with the FP equation, to simultaneously learn the equation {\em and} infer the multi-dimensional PDF at all times. In particular, we consider two types of inverse problems, {\em type I} where the FP equation is known but the initial PDF is unknown, and {\em type II} in which, in addition to unknown initial PDF, the drift and diffusion terms are also unknown. In both cases, we investigate problems with either Brownian or  L\'{e}vy noise or a combination of both.  We demonstrate the new PINN framework in detail in the one-dimensional case (1D) but we also provide results for up to 5D demonstrating that we can infer both the FP equation {\em and} dynamics {\em simultaneously at all times} with high accuracy using only very few discrete observations of the particles.
\end{abstract}

\keywords{physics-informed learning, small data,  deep neural networks, L\'evy noise, Kullback-Leibler divergence, non-locality. }


\section{Introduction}

Stochastic dynamical systems, defined as deterministic systems
of differential equations perturbed by random disturbances that are not necessarily small, have been used effectively across disciplines, including gene regulatory networks, physics, chemistry, population dynamics and medicine. The general form of equations may be similar, however the specific noise type and the constant or variable coefficients in these equations are case-dependent. Hence, it is important to use data to infer these coefficients and infer the stochastic dynamical system and forecast its dynamics. 

There exist a few methods to  determine stochastic differential equations from observations. Ruttor et al \cite{ruttor2013approximate} used a Gaussian process prior over the drift as a function of the state vector, and developed an approximate expectation maximization algorithm to deal with the unobserved latent dynamics between observations.  
Archambeau et al \cite{archambeau2008variational} presented a variational approach for the approximate inference of stochastic differential equations (SDEs) from a finite set of noisy observations.
Subsequently, Mapper \cite{opper2019variational} used the variational inference approach to approximate the distribution over the unknown path of the SDE conditioned on the observations and  provided approximations for the intractable likelihood of the drift. 
Also, in \cite{boninsegna2018sparse}, the authors used the sparse learning method to learn the stochastic dynamical systems.
In \cite{YangLiu-GAN-SODE}, the authors used small samples from just a few snapshots of unpaired data to infer the drift and diffusion terms in SDEs with Brownian and L\'{e}vy noise in high-dimensions.
Here, we follow the recent work of \cite{YangLiu-GAN-SODE} but we replace the SDE with the Fokker-Planck (FP)  equation, which governs the dynamics of the probability density function (PDF).  

For the forward problem, there are several methods to solve the FP directly \cite{deng2009finite,gao2016fokker,zheng2015novel,chen2019most} assuming that the initial condition is known. Recently,
Xu et al \cite{xu2020solving} used PINNs to solve the one-dimensional stationary FP equation (for Brownian motion)  with homogeneous Dirichlet boundary conditions, where the PINN loss was augmented with an additional loss for the constraint of numerical integration. 
More broadly, there are several papers by now using deep learning to solve general partial differential equations \cite{raissi2019physics,pang2019fpinns,han2018solving,pang2020npinns,chen2019learning}. In these works, to solve the forward problem, the initial condition and the boundary condition is needed, while for the inverse problem, observation of the solution in some points in space-time is required. 

\begin{figure}[H]
\centering
\includegraphics[width = 0.8\textwidth]{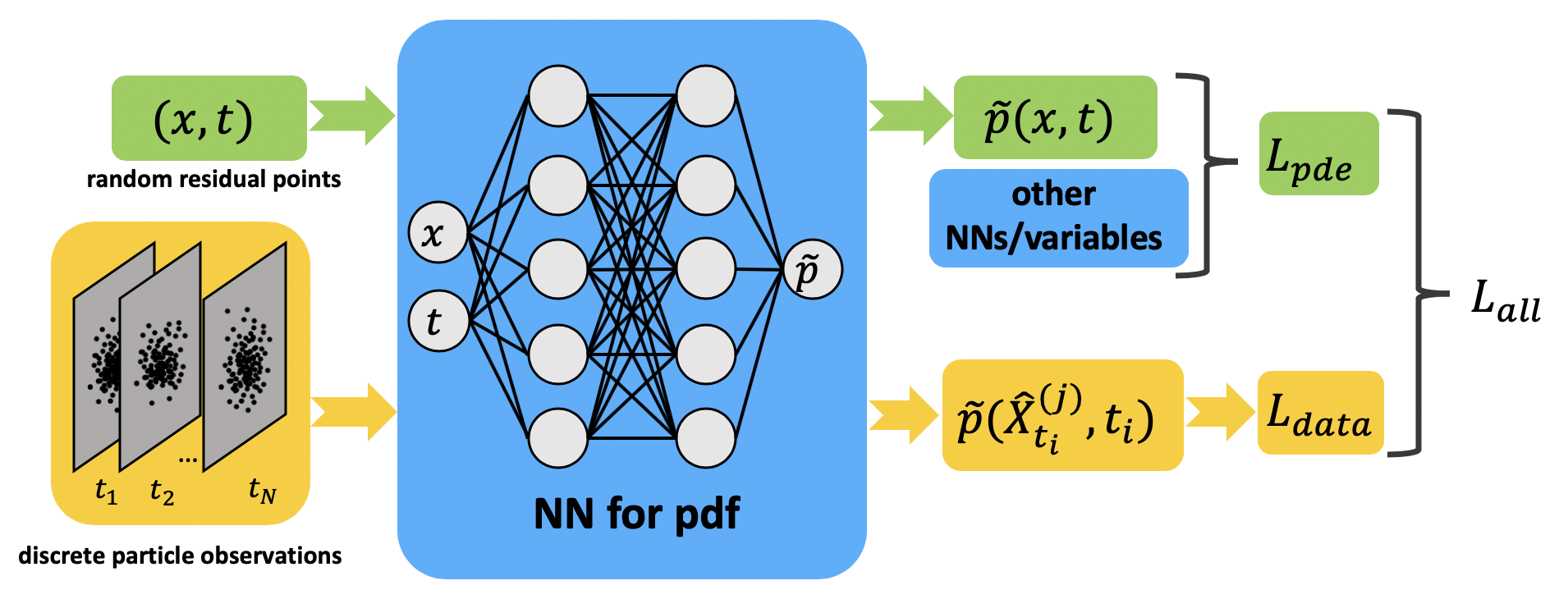}
\caption{\textbf{} Schematic of the PINNs for solving inverse stochastic problems from discrete particle observations. The total loss consists of a loss from residual of the Fokker-Planck equation and another loss from particle observations.}
\label{NN}
\end{figure}

In the current paper, we assume that we only have the observation data of the sample paths at just a few discrete times but we do not know the initial PDF. We consider problems with both Brownian and L\'{e}vy noise as well as a combination of the two.
Our aim is to combine the observation data and the corresponding parametrized FP equation with unknown coefficients to infer its precise form and hence obtain the PDF at all times, including the initial condition. In particular,   we employ the variational form of the Kullback-Leibler divergence \cite{nguyen2010estimating} to connect and combine the observations of particle positions and the corresponding FP equation.
To this end, we employ a physics-informed neural network (PINN) as shown in Figure \ref{NN} that enables a seamless integration of the data and the FP equation using automatic differentiation, hence avoiding the need for grid generation, which is computationally prohibitive in high-dimensions.

The paper is organized as follows. In Section 2, we introduce the methodology to solve the FP equation and the connection between the trajectories and the PDF.  In Section 3, we present the results of PINNs for solving the two types of inverse problems of the FP equation for up to five dimensions (5D). Finally, we conclude with a short summary. We have included more detailed results in the Appendices (A-F), studying both Brownian noise and L\'{e}vy noise cases.

\section{\label{Methodology}Methodology}

\subsection{Problem setup}
Consider the following stochastic differential equation:
\begin{align}  \label{stomodel}
  dX_t &= a(X_t)dt + \sigma dB_t+ \varepsilon dL_{t}^{\alpha}, X_t \in \mathbb{R}^n,
 \end{align}
where $a(\cdot)$ is the vector drift term, $\sigma$ and $\varepsilon$ are $n\times n$ matrices,  $B_t$ is a Brownian motion in $\mathbb{R}^n$, and $L_{t}^{\alpha}$ is a symmetric $\alpha$-stable L\'{e}vy process in $\mathbb{R}^n$. The generating triplet of the  L\'{e}vy process is $(0,0,\nu_\alpha)$. The processes $B_t$ and $L_t^\alpha$ are taken to be independent.

The generator of the SDE \ref{stomodel} is \cite{applebaum2009levy,duan2015introduction}:
\begin{align}  \label{generator}
  Au=&\sum_{i}a_i\frac{\partial u}{\partial x_i}+\frac{1}{2}\sum_{i,j}(\sigma \sigma^T)_{i,j}\frac{\partial^2f}{\partial x_i \partial x_j}+\varepsilon^\alpha \int_{\mathbb{R}^n\setminus \{0\}}[u(x+y)-u(x)]\nu_{\alpha}(dy) \nonumber
 \end{align}
where $\nu_{\alpha}(dy)$ is the $\alpha$-stable L\'{e}vy measure and $\nu_{\alpha}(dy)=C_{n,\alpha}||y||^{-n-\alpha}dy$, $C_{n,\alpha}=\frac{\alpha \Gamma((n+\alpha)/2) }{2^{1-\alpha} \pi^{n/2}\Gamma(1-\alpha/2)}.$

The FP equation describing the time evolving PDF of the stochastic process $X_t$, starting from $p_0$ at $T_0$, is written as follows:
\begin{equation}
    \begin{aligned}
&\partial_t p=A^{*}p(x,t), (x, t) \in \mathbb{R}^n \times [T_0, T_1],   \\
&p(x,T_0) = p_0(x),
 \end{aligned}
 \end{equation}
where $A^{*}$ is the adjoint operator of the generator $A$ and has the following form:
\begin{equation}\label{ajoint}
    \begin{aligned} 
  A^{*}p=&-\sum_{i=1}^{n}\frac{\partial }{\partial x_i}(a_i p)+\frac{1}{2}\sum_{i,j=1}^{n}\frac{\partial}{\partial x_i \partial x_j}((\sigma \sigma^T)_{i,j} p) \\
  & +\varepsilon^\alpha \int_{\mathbb{R}^n\setminus \{0\}}[p(x+y)-p(x)]\nu_{\alpha}(dy).
 \end{aligned}
 \end{equation}
For simplicity, we only consider the case where the $\alpha$-stable L\'{e}vy noises are independent across dimensions, which means the L\'{e}vy measure has the formula \cite{chen2019most}:
\begin{equation}
    \begin{aligned} 
\nu_{\alpha}(dx)=&\nu_{\alpha}(dx_1,dx_2,...,dx_n)\nonumber\\
=&\nu_{\alpha}(dx_1)\delta_0(x_2)\nu_{\alpha}(dx_3)...\nu_{\alpha}(dx_n)+\nu_{\alpha}(dx_1)\nu_{\alpha}(dx_2)\delta_0(x_3)...\nu_{\alpha}(dx_n)+...\nonumber\\
&+\nu_{\alpha}(dx_1)\nu_{\alpha}(dx_2)\nu_{\alpha}(dx_3)...\delta_0(x_n),\
 \end{aligned}
 \end{equation}
where $\delta_0$ is the Dirac delta measure concentrated at 0.

Consequently, 
\begin{equation}\label{ajoint1}
    \begin{aligned}
    A^{*}p=&-\sum_{i}\frac{\partial }{\partial x_i}(a_i p)+\frac{1}{2}\sum_{i,j}\frac{\partial}{\partial x_i \partial x_j}((\sigma \sigma^T)_{i,j} p)\\
  &  +\sum_i \varepsilon_i^\alpha \int_{\mathbb{R}\setminus \{0\}}[p(x_1,x_2,...,x_{i-1},x_i+y_i,x_{i+1},...,x_n,t)\\
  & -p(x_1,x_2,...,x_{i-1},x_i,x_{i+1},...,x_n,t)]\nu_{\alpha}(dy_i).
    \end{aligned}
\end{equation}

We consider the scenario where the data available are the discrete observations of the particles at different time instants $\{t_i\}_{i=1}^N$ with $T_0 \le t_1 < t_2 ... <t_N \le T_1$. More formally, the available data are $\{\mathcal{D}_i\}_{i=1}^N$ with $\D_i = \{\hat{X}_{t_i}^{(j)}\}_{j=1}^{n_i}$, where $\hat{X}_{t_i}^{(j)}$ are i.i.d. random samples of $X_{t_i}$ drawn from $p(\cdot,t_i)$. We refer to $\D_i$ as a snapshot for each $i$. Our objective is to infer the possibly unknown terms in the FP equation as well as the PDF at all times by utilizing the equation and the observation data in a unified learning system. More specifically, we consider the following two types of problems.
\begin{enumerate}
    \item {\em Problem I}: We have full knowledge of the operator $A^*$ but we do not know the initial condition $p_0$. Instead, we have a few snapshots consisting of limited samples, and wish to learn the PDF as a function of time, including the initial condition.
    \item {\em Problem II}: While we know the general form of $A^*$, we do not have full knowledge of some terms in $A^*$, e.g., the drift term $a$, or the entries in the coefficient matrices $\sigma$ and $\varepsilon$. We wish to use the snapshots to infer the unknown terms in the FP  equation.
\end{enumerate}

\subsection{PDF estimation using neural networks}

Given two distributions $P$ and $Q$ on $\mathbb{R}^n$ with PDF $p(x)$ and $q(x)$, respectively, with $p,q>0$ for simplicity, the Kullback-Leibler (KL) divergence from $Q$ to $P$ is formulated as:
\begin{equation}\label{eqn:KL divergence}
    \begin{aligned}
    D_{KL}(P||Q) =\int p(x)\log(\frac{p(x)}{q(x)}) dx.
    \end{aligned}
\end{equation}

From~\cite{nguyen2010estimating} we have the following variational form of the KL divergence:
\begin{equation}\label{eqn:KL_variational}
    \begin{aligned}
     D_{KL}(P||Q) = \sup_{g>0} \left( \mathbb{E}_P [\log(g(x))] - \mathbb{E}_Q [g(x)]  + 1\right),
    \end{aligned}
\end{equation}
where the supremum is attained at 
\begin{equation}\label{eqn:attain_sup}
    \begin{aligned}
    g^*(x) = \frac{p(x)}{q(x)}.
    \end{aligned}
\end{equation}
In other words,
\begin{equation}\label{eqn:pdf_sup}
    \begin{aligned}
    \frac{p(x)}{q(x)} = \argmax_g\left( \mathbb{E}_P [\log(g(x))] - \mathbb{E}_Q [g(x)]  + 1\right).
    \end{aligned}
\end{equation}

Note that Eqn. \ref{eqn:pdf_sup} is a key to our formulation and provides an effective means of  estimating the PDF from particle observations. We denote the estimator of $p(x)$ as $\hat{p}(x)$. Suppose that we are given a finite number of i.i.d. samples from the unknown distribution $P$ on $\mathbb{R}^n$ with PDF $p(x)$, for which the empirical distribution is denoted as $\hat{P}$. We define $Q$ as a known distribution with PDF $q(x) > 0$. Then
\begin{equation}\label{eqn:pdf_variational}
    \begin{aligned}
    p(x) &= q(x)\argmax_{g>0}\left( \mathbb{E}_P[\log(g(x))] - \mathbb{E}_Q [g(x)]  + 1\right) \\
    &= q(x)\argmax_{g>0}\left( \mathbb{E}_P[\log(g(x))] - \int q(x)g(x)dx  + 1\right).
    \end{aligned}
\end{equation}

While other distributions, e.g., the Gaussian distribution also applies in principle, for simplicity, we set $Q$ as an uniform distribution on $\D$, i.e., $q(x) = 1/\mu(\D)$ for $x \in \D$, where $\mu$ denotes the Lebesgue measure and $\D$ is a hypercube that contains the support of $\hat{P}$. We also set the estimator $\hat{p}(x) = 0$ for $x$ outside $\D$. Then, from \ref{eqn:pdf_variational} we have 
\begin{equation}\label{eqn:estimator_variational}
    \begin{aligned}
    \hat{p}(x) = \begin{cases}
    \frac{1}{\mu(\D)}\argmax_{g\in \mathcal{G}}\left( \mathbb{E}_{\hat{P}}[\log(g(x))] - \frac{1}{\mu(\D)}\int_{\D} g(x)dx  + 1\right) & x\in D\\
    0 & \text{otherwise}\\
    \end{cases},
    \end{aligned}
\end{equation}
where the integral over $\D$ can be calculated via Monte Carlo sampling or quadrature. $\mathcal{G}$ is a specific class of positive real value functions. Intuitively, we wish $\mathcal{G}$ to be rich enough so that it can contain $p(x)/q(x)$, while also exclude ``morbid'' functions to avoid overfitting. A natural idea is to use a family of neural networks as $\mathcal{G}$ due to the huge capacity of neural networks as well as our empirical knowledge that the post-training neural networks tend to be ``regularized''.

\subsection{Solving the FP equation using discrete particle observations}
We use a deep neural network 
 $\tilde{p}(x,t)$ with the concatenation of $x$ and $t$ as input and a real number as output to approximate $p(x,t)$, the unknown solution of the FP equation. Different from vanilla neural networks, we add an additional \textit{softplus} activation function 
\begin{equation}
     \textit{softplus}(x) = \ln(1+\exp(x))
\end{equation}
before the output layer to make $\tilde{p}(x,t)$ strictly positive. Based on the FP equation and observations of discrete particle observations, we have two requirements for the post-training $\tilde{p}(x,t)$, as follows:
\begin{enumerate}
    \item $\tilde{p}(x,t)$ should approximately satisfy the FP equation. 
    \item At each time frame $t_i$, $\tilde{p}(x,t_i)$ should match the samples $\{\hat{X}_{t_i}^{(j)}\}_{j=1}^{n_i}$.
\end{enumerate}
For the first requirement, inspired by physics informed neural networks (PINNs), we can encode the FP equation into the training of neural networks by introducing the PDE loss:
\begin{equation}\label{eqn:loss_pde}
    \begin{aligned}
    L_{pde} &= \frac{1}{N_r}\sum_{i=1}^{N_r} (\partial_t\tilde{p}(x_i,t_i) - A^*\tilde{p}(x_i,t_i))^2, \\
    \end{aligned}
\end{equation}
where $\{(x_i,t_i)\}_{i=1}^{N_r}$ are $N_r$ randomly sampled so-called ``residual points'' on the spatio-temporal domain $\D\times[T_0, T_1]$.
 The differentiation applied to the neural network $\tilde{p}(x,t)$ could be performed via automatic differentiation, while the integral operator in $A^*$ can be estimated by a quadrature as in the fractional PINNs~\cite{pang2019fpinns}. 
 
 More specifically, in 1D problem, where 
 \begin{equation}\label{eqn:FP-1D}
    \begin{aligned}
A^* p(x,t) =- (a(x)p(x,t))_x&+\frac{\sigma^2}{2}\frac{\partial^2 p(x,t)}{\partial x^2}+\varepsilon^\alpha \int_{\mathbb{R}\setminus \{0\}}[p(x+y,t)-p(x,t)]\nu_{\alpha}(dy),\nonumber\\
    \end{aligned}
\end{equation}
we truncate the solution on the domain $(-L,L)$, and use the following discretization scheme \cite{gao2016fokker} to approximate the non-local part at point $(x_i,t_j)$:
 \begin{equation}
    \begin{aligned}
 &\int_{\mathbb{R}\setminus \{0\}}[p(x_i+y,t_j)-p(x_i,t_j)]\nu_{\alpha}(dy)\nonumber\\
 =&\int_{-\infty}^{-L-x_i}[p(x_i+y,t_j)-p(x_i,t_j)]\nu_{\alpha}(dy)+\int_{-L-x_i}^{L-x_i}[p(x_i+y,t_j)-p(x_i,t_j)]\nu_{\alpha}(dy)\\
 &+\int_{L-x_i}^{\infty}[p(x_i+y,t_j)-p(x_i,t_j)]\nu_{\alpha}(dy)\\
=&\frac{C_\alpha}{\alpha}\!\big[\frac{1}{(L\!\!+\!\!x_i)^\alpha}\!+\!\frac{1}{(L\!\!-\!\!x_i)^\alpha}\big]\!p(x_i,t_j)\!+\! C_\alpha \int_{\!-L\!-\!x_i}^{L-x_i} \frac{p(x_i\!+\!y,  t_j)\!-\!p(x_i, t_j)}{|y|^{1\!+\!\alpha}} dy\\
\approx &\frac{C_\alpha}{\alpha}\!\big[\frac{1}{(L\!\!+\!\!x_i)^\alpha}\!+\!\frac{1}{(L\!\!-\!\!x_i)^\alpha}\big]\!p(x_i,t_j)+C_{h}p_{xx}(x_i,t_j)   +C_\alpha \!h\!\!\!\!\! \sum^{J\!-\!i}_{k\!\!=\!\!-\!J\!-\!i,k\!\neq\! 0}\!\!\!\!\!\!\!\!\!{''} \;
    {\frac{p_{i\!+\!k,j} \!-\! p_{i,j}}{|x_{k}|^{1\!+\!\alpha}} }\\
\end{aligned}
\end{equation}

where $C_{h}\!\!=\!\!- C_\alpha\zeta(\alpha\!-\!1)h^{2\!-\!\alpha}$ and $\sum^{''}$ means that the quantities corresponding to the two end summation indices are multiplied by $\frac{1}{2}$.

In problems with dimension larger than one, the scheme of non-local part in \ref{ajoint1} in each dimension is the same as in the 1D case.

For the second requirement, recall that $\tilde{p}(x,t_i)$ as an estimator of $p(x,t_i)$ should satisfy
\begin{equation}\label{eqn:pde_estimator_variational}
    \begin{aligned}
   \tilde{p}(x,t_i) \mu(\D) = \begin{cases}
    \argmax_{g\in \mathcal{G}}\left( \mathbb{E}_{\hat{P_i}}[\log(g(x))] - \frac{1}{\mu(\D)}\int_{\D} g(x)dx  + 1\right) & x\in \D\\
    0 & \text{otherwise}\\
    \end{cases},
    \end{aligned}
\end{equation}
where $\hat{P_i}$ is the empirical distribution induced by samples $\{\hat{X}_{t_i}^{(j)}\}_{j=1}^{n_i}$. Inspired by Eqn. \ref{eqn:pde_estimator_variational}, we then introduce the data loss:
\begin{equation}\label{eqn:loss_data}
    \begin{aligned}
    L_{data,i} &= -\mathbb{E}_{\hat{P_i}}[\log(\tilde{p}(x,t_i)\mu(\D))] + \frac{1}{\mu(\D)}\int_{\D} \tilde{p}(x,t_i)\mu(\D)dx  - 1, \\
     &\simeq -\mathbb{E}_{\hat{P_i}}[\log(\tilde{p}(x,t_i))] + \int_{\D} \tilde{p}(x,t_i)dx \\
     &= -\frac{1}{n_i}\sum_{j=1}^{n_i}\log(\tilde{p}( \hat{X}_{t_i}^{(j)} ,t_i)) + \int_{\D} \tilde{p}(x,t_i)dx, 
    \end{aligned}
\end{equation}
for $i = 1,2...N$, where $\simeq$ represents equality up to a constant. 
While the first term in the right-hand side is averaged over all the samples in $\D_i = \{\hat{X}_{t_i}^{(j)}\}_{j=1}^{n_i}$, in practice we can take the average over a minibatch of size $b$ smaller than $n_i$. The integral over $\D$ in the second term can be estimated by quadrature or Monte Carlo sampling.

The total loss is then a combination of PDE loss and data loss:
\begin{equation}\label{eqn:loss_all}
    \begin{aligned}
    L_{all} &= \tau L_{pde} + \sum_i^N L_{data,i},
    \end{aligned}
\end{equation}
where $\tau$ is the weight to scale and balance the losses from the FP equation and data.

We note that when we have observations of $X_t$, we can also use the kernel density estimation to obtain the PDF, denote as as $p_{KD}$. Correspondingly, we can replace the data loss in \ref{eqn:loss_all} with another loss from the estimated density, as follows:
\begin{align}\label{loss_KD}
L_{all}=\tau L_{pde} + \sum_i^N L_{KD,i},
\end{align}
where 
\begin{align}\label{loss_KD_1}
L_{KD,i}&=\frac{1}{N_p}\sum_{j=1}^{N_p}[\tilde{p}(x_{j},t_i)-p_{KD}(x_{j},t_i)]^2,
\end{align}
and $\{(x_{j},t_i)\}_{j=1}^{N_p}$ are $N_p$ collocation points in the spatial domain $\D$ at $t = t_i$.

\section{Computational Results}
All the neural networks in our tests have 4 hidden layers and 20 neurons per layer, with $\tanh$ activation function. We use the Adam optimizer with a learning rate of $10^{-4}$. We have two methods to compute the inverse problem II. In the first one, we use a polynomial to approximate the drift term, while in the second one, we use directly one or more neural networks to approximate the drift term, which could be any general function. The training takes $100,000$ iterations for problem I, while $200,000$ iterations for problem II, except 3D, 4D and 5D cases where the training iterations are shown in the corresponding figures.

\subsection{One-dimensional Case}

Consider the following 1D SDE:
\begin{align}  \label{stomodel-1D}
  dX_t &= a(X_t)dt + \sigma dB_t+ \varepsilon dL_{t}^{\alpha},
 \end{align}
 the corresponding FP equation is:
\begin{equation}\label{eqn:FPE-1D}
    \begin{aligned}
\partial_t p(x,t) =- (a(x)p(x,t))_x&+\frac{\sigma^2}{2}\frac{\partial^2 p(x,t)}{\partial x^2}+\varepsilon^\alpha \int_{\mathbb{R}\setminus \{0\}}[p(x+y,t)-p(x,t)]\nu_{\alpha}(dy),\nonumber\\
    \end{aligned}
\end{equation}

\subsubsection{Problem I: Brownian noise}
In this section, we use the deep neural network to compute problem I for the FP equation driven by Brownian noise. We consider the drift term $a(x)=x-x^3$, $\sigma=1$ and $\varepsilon=0$. We choose the observation data of $X_t$ at several times, and use the the Monte-Carlo method with 5,000,000 samples to obtain the reference PDF for estimating the accuracy of our method.

For the inverse problem I, we assume that the observation data of samples of $X_t$ are available only at one snapshot.
To assess the accuracy of our predictions, we consider $100$, $1000$ and $10000$ samples at $t=0.3$, i.e., $N=1$. The minibatch size is set as the training data size. When we compute \eqref{eqn:loss_data}, we use the compound trapezoid formula with $301$ points to approximate the integral part. The number of residual points for evaluating the FP residual is $N_r = 1000$.

\begin{figure}[H]
\begin{minipage}[]{0.2 \textwidth}
 \leftline{~~~~~~~\tiny\textbf{(a1)}}
\centerline{\includegraphics[width=3.5cm,height=3.2cm]{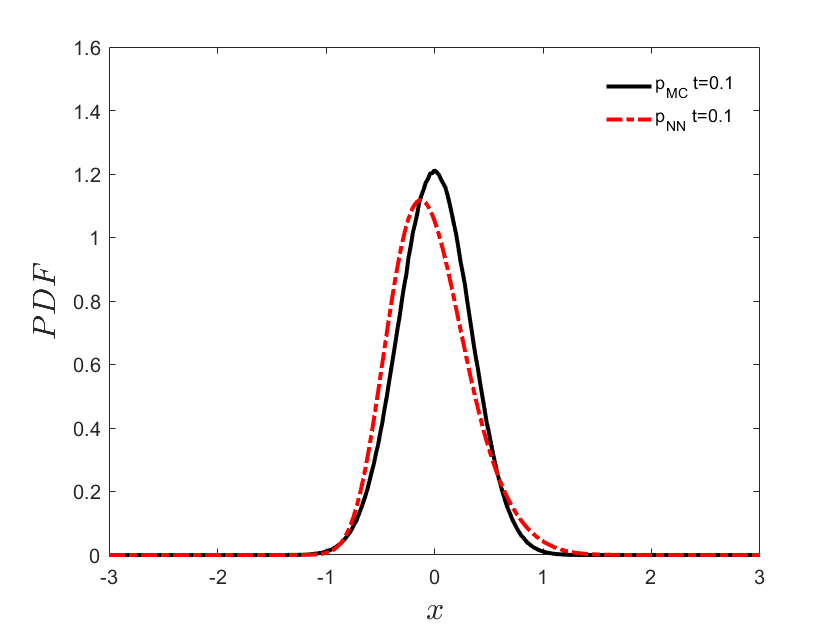}}
\end{minipage}
\hfill
\begin{minipage}[]{0.2 \textwidth}
 \leftline{~~~~~~~\tiny\textbf{(a2)}}
\centerline{\includegraphics[width=3.5cm,height=3.2cm]{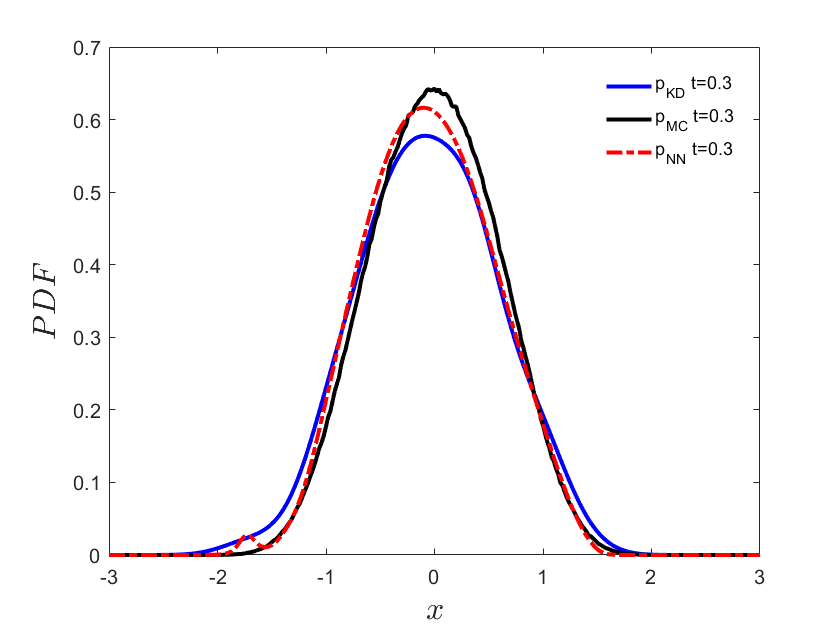}}
\end{minipage}
\hfill
\begin{minipage}[]{0.2 \textwidth}
 \leftline{~~~~~~~\tiny\textbf{(a3)}}
\centerline{\includegraphics[width=3.5cm,height=3.2cm]{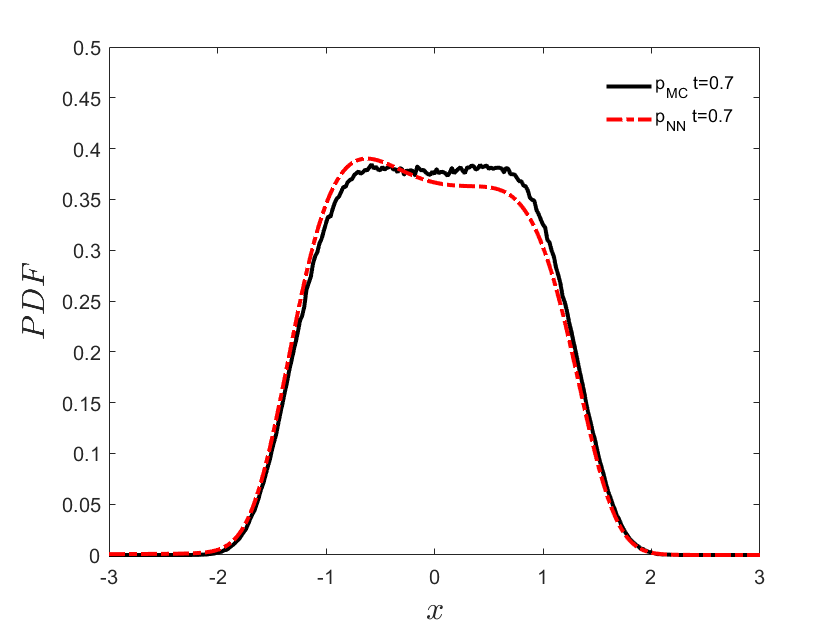}}
\end{minipage}
\hfill
\begin{minipage}[]{0.2 \textwidth}
 \leftline{~~~~~~~\tiny\textbf{(a4)}}
\centerline{\includegraphics[width=3.5cm,height=3.2cm]{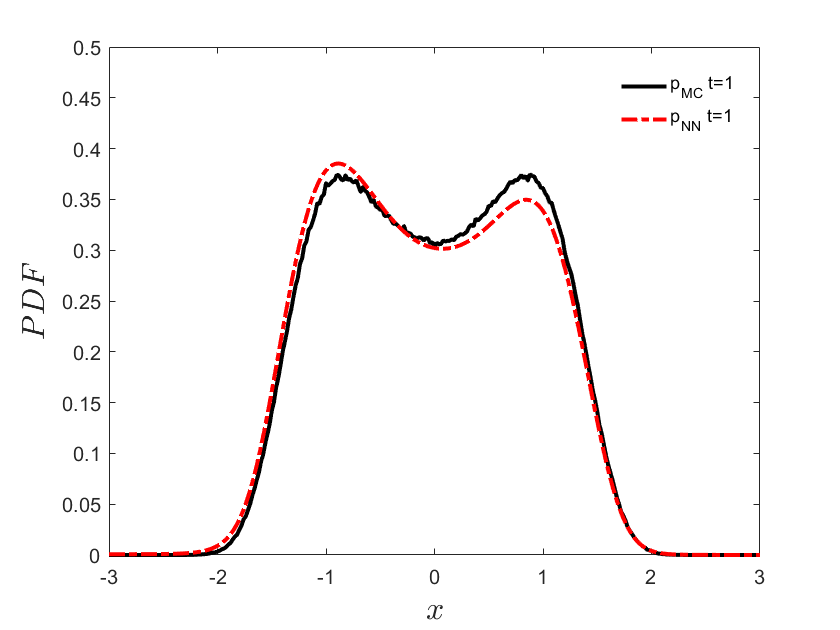}}
\end{minipage}
\hfill
\begin{minipage}[]{0.2 \textwidth}
 \leftline{~~~~~~~\tiny\textbf{(b1)}}
\centerline{\includegraphics[width=3.5cm,height=3.2cm]{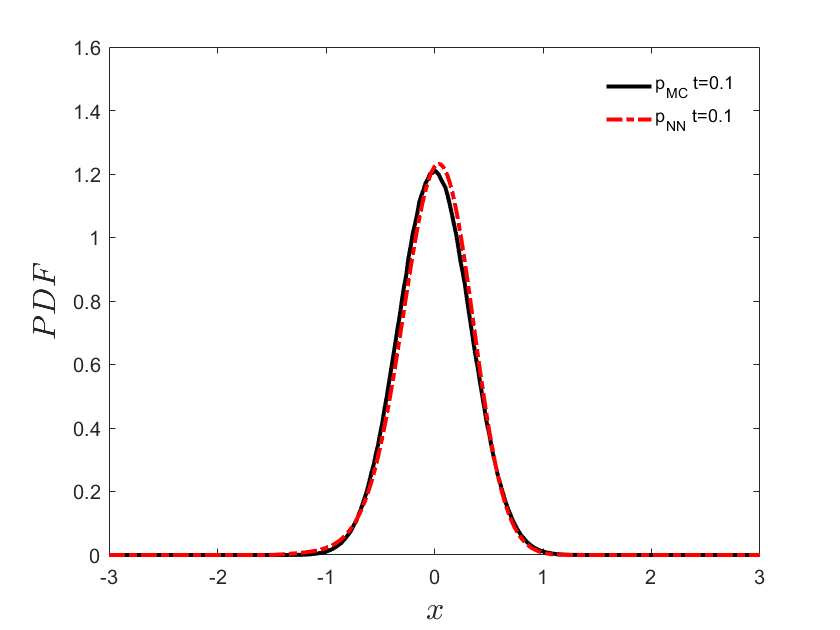}}
\end{minipage}
\hfill
\begin{minipage}[]{0.2 \textwidth}
 \leftline{~~~~~~~\tiny\textbf{(b2)}}
\centerline{\includegraphics[width=3.5cm,height=3.2cm]{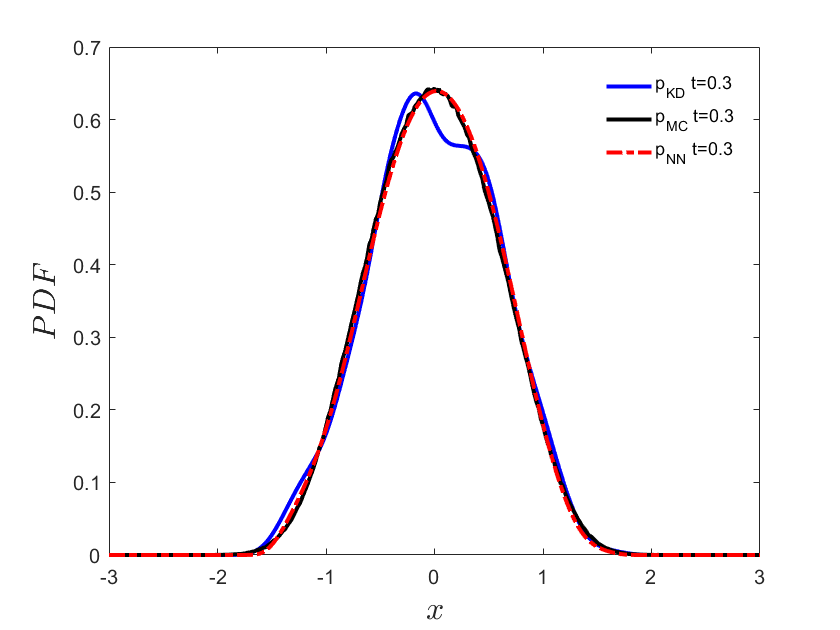}}
\end{minipage}
\hfill
\begin{minipage}[]{0.2 \textwidth}
 \leftline{~~~~~~~\tiny\textbf{(b3)}}
\centerline{\includegraphics[width=3.5cm,height=3.2cm]{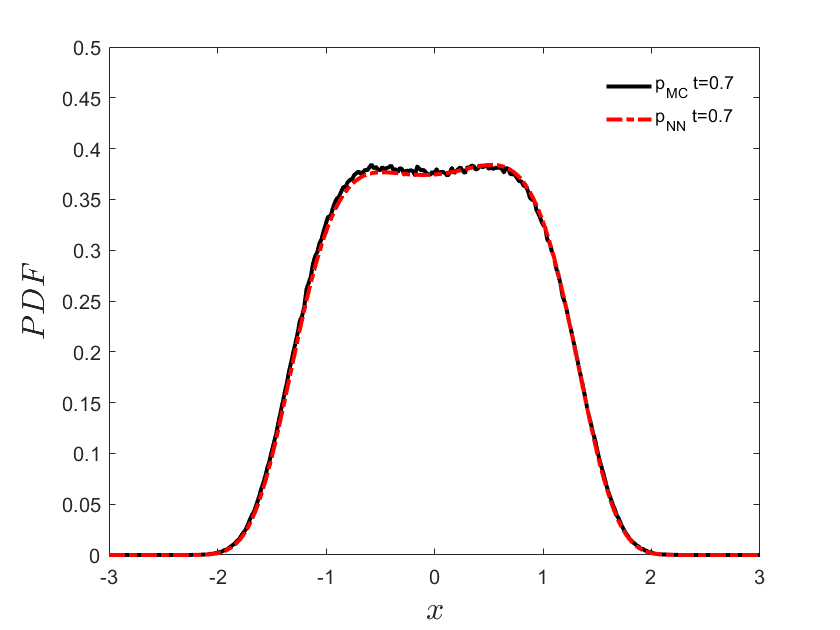}}
\end{minipage}
\hfill
\begin{minipage}[]{0.2 \textwidth}
 \leftline{~~~~~~~\tiny\textbf{(b4)}}
\centerline{\includegraphics[width=3.5cm,height=3.2cm]{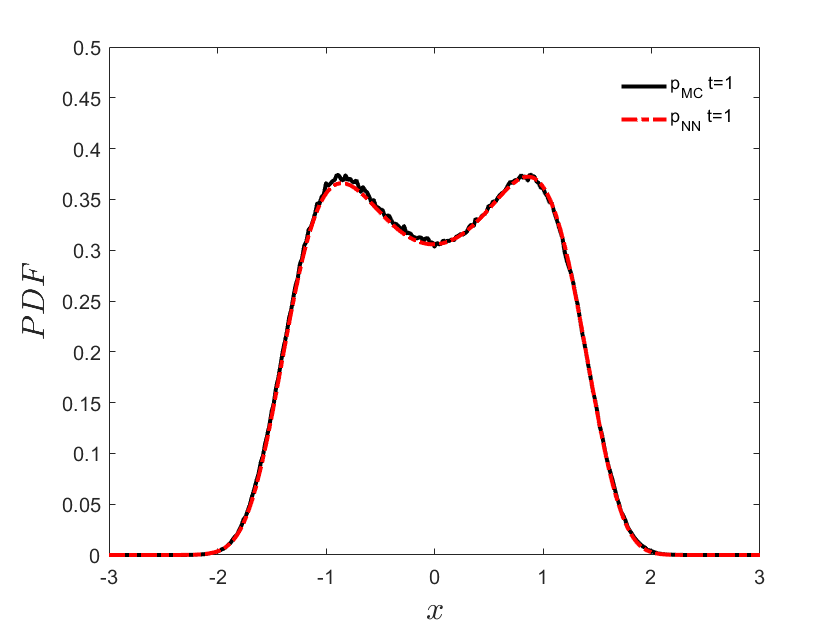}}
\end{minipage}
\begin{minipage}[]{0.2 \textwidth}
 \leftline{~~~~~~~\tiny\textbf{(c1)}}
\centerline{\includegraphics[width=3.5cm,height=3.2cm]{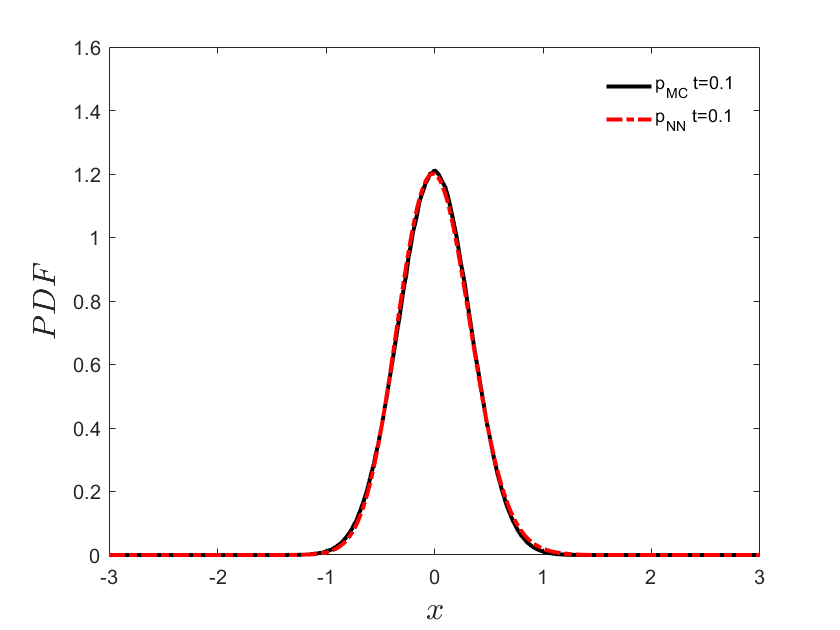}}
\end{minipage}
\hfill
\begin{minipage}[]{0.2 \textwidth}
 \leftline{~~~~~~~\tiny\textbf{(c2)}}
\centerline{\includegraphics[width=3.5cm,height=3.2cm]{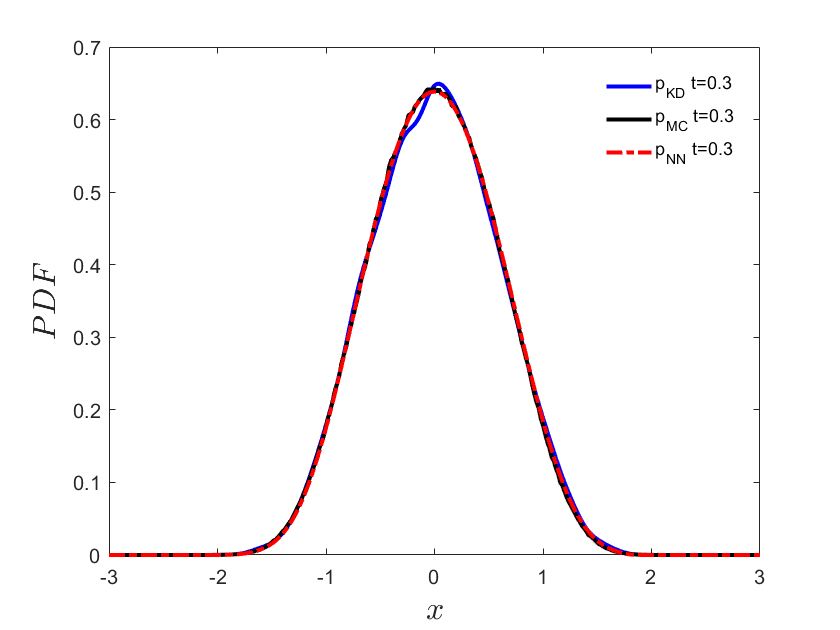}}
\end{minipage}
\hfill
\begin{minipage}[]{0.2 \textwidth}
 \leftline{~~~~~~~\tiny\textbf{(c3)}}
\centerline{\includegraphics[width=3.5cm,height=3.2cm]{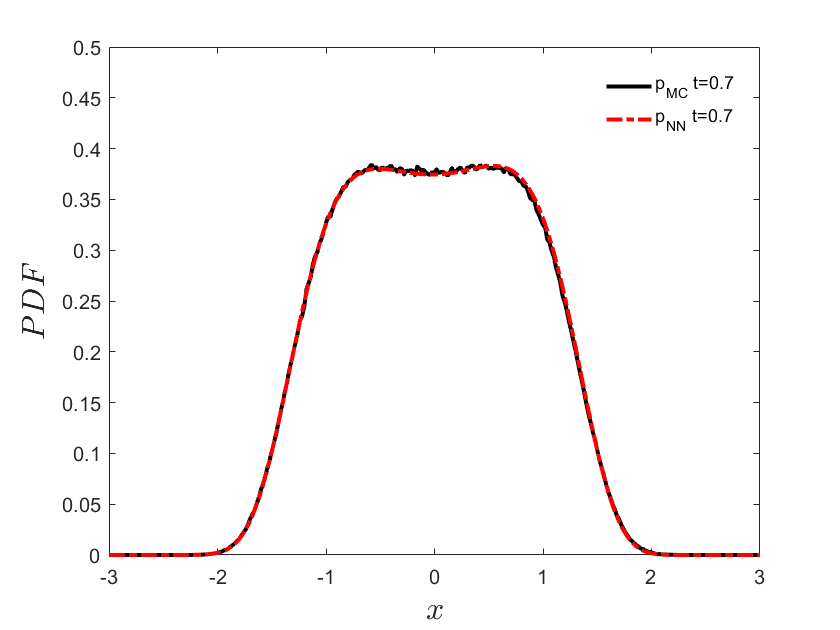}}
\end{minipage}
\hfill
\begin{minipage}[]{0.2 \textwidth}
 \leftline{~~~~~~~\tiny\textbf{(c4)}}
\centerline{\includegraphics[width=3.5cm,height=3.2cm]{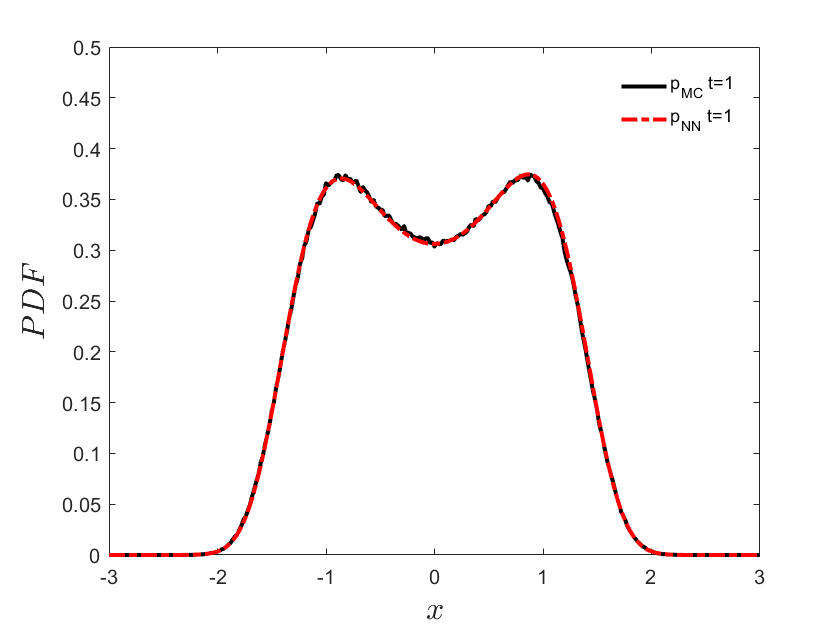}}
\end{minipage}
\caption{\textbf{Inverse problem I - Brownian noise:} Evolution of PDF for $a(x)=x-x^3$ and $\sigma=1$. Observation data of samples of $X_t$ are available at $t=0.3$. (a) 100 samples of $X_t$; (b) 1000 samples of $X_t$; (c) 10000 samples of $X_t$. The red curves are the PINN predictions and the black curves are the reference solutions obtained by MC. Also, shown with blue color at time $t=0.3$ are the PDFs obtained based on the kernel density estimation method.}
\label{f1-forward-1d-BM}
\end{figure}

The predicted results are shown in Figure \ref{f1-forward-1d-BM} at times $t=0.1, 0.3, 0.7, 1$.
We can see that the predicted PDFs are better if we have more observation data, as expected; however, even with only 100 samples, the two peaks of the PDF can be captured accurately. The results here shows that physics-informed learning can capture the dynamics accurately at all times given data of one snapshot only.

\begin{figure}[H]
\begin{minipage}[]{0.3 \textwidth}
\centerline{\includegraphics[width=4cm]{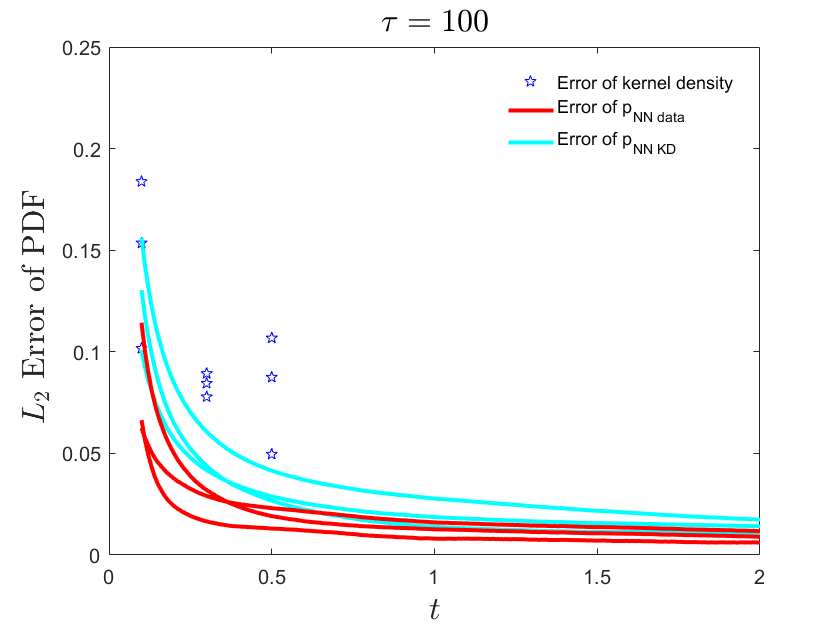}}
\end{minipage}
\hfill
\begin{minipage}[]{0.2 \textwidth}
\centerline{\includegraphics[width=4cm]{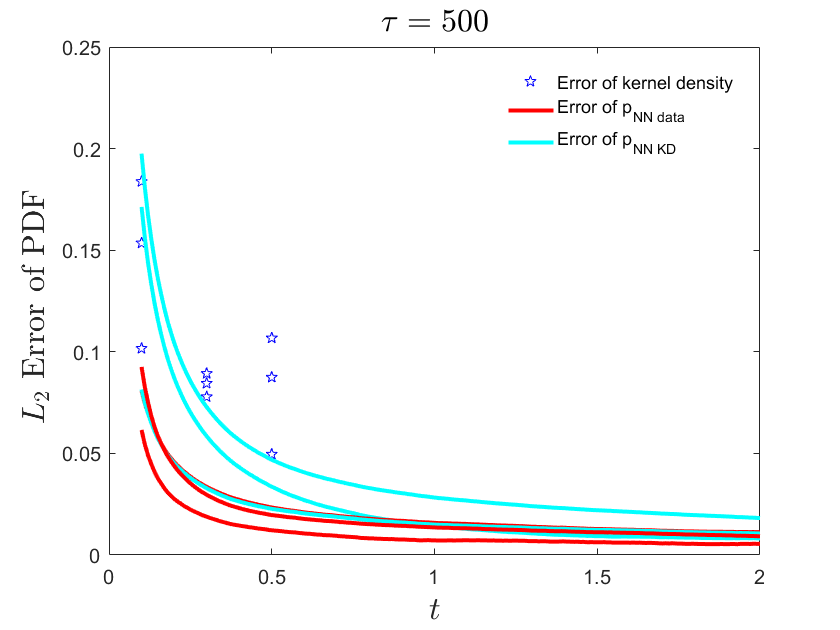}}
\end{minipage}
\hfill
\begin{minipage}[]{0.3 \textwidth}
\centerline{\includegraphics[width=4cm]{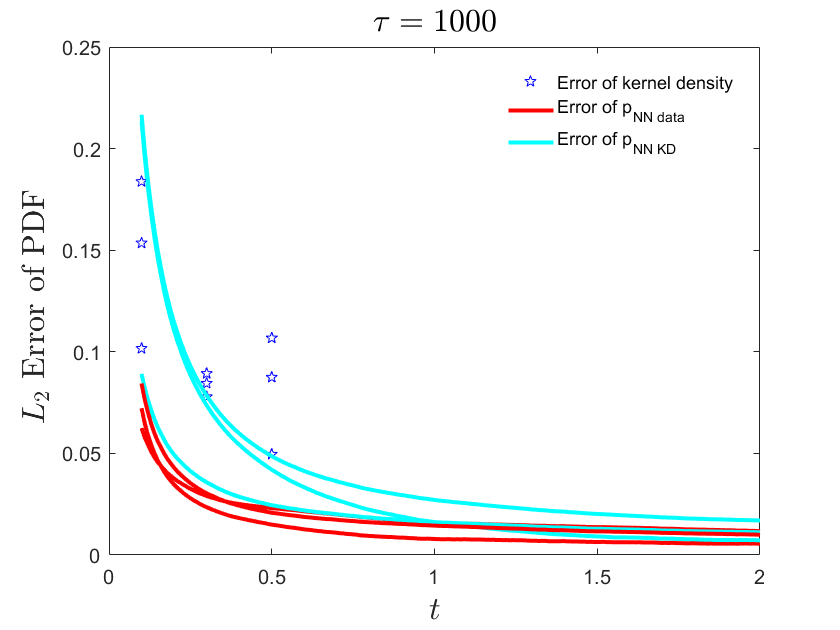}}
\end{minipage}
\begin{minipage}[]{0.3 \textwidth}
\centerline{\includegraphics[width=4cm]{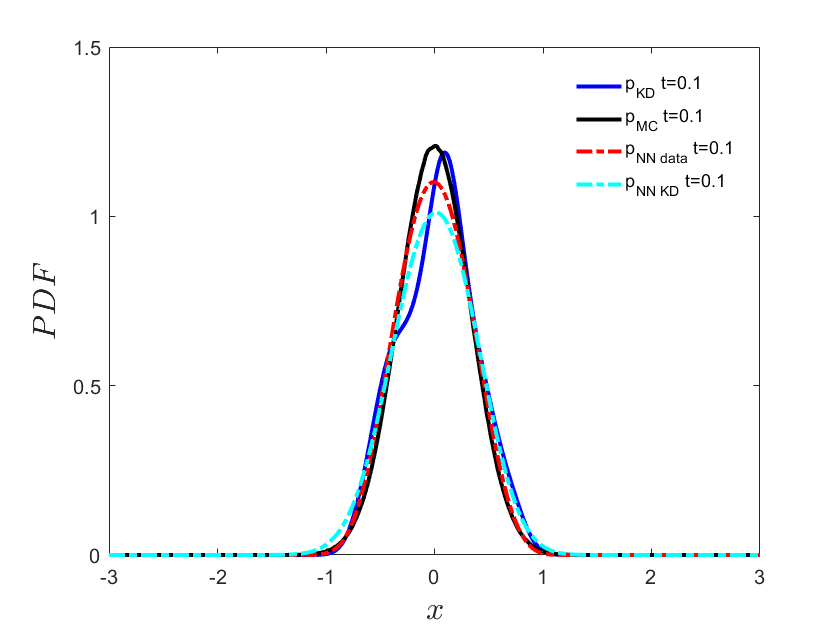}}
\end{minipage}
\hfill
\begin{minipage}[]{0.2 \textwidth}
\centerline{\includegraphics[width=4cm]{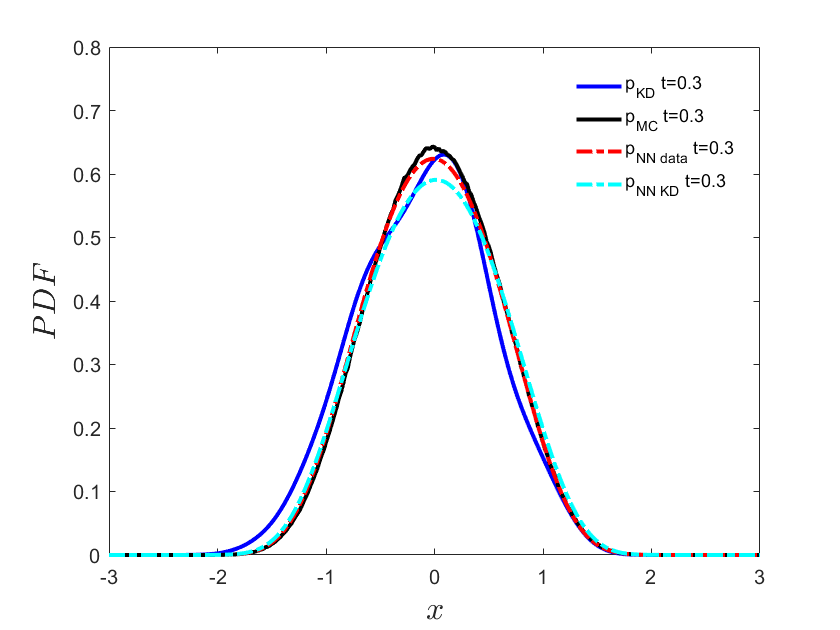}}
\end{minipage}
\hfill
\begin{minipage}[]{0.3 \textwidth}
\centerline{\includegraphics[width=4cm]{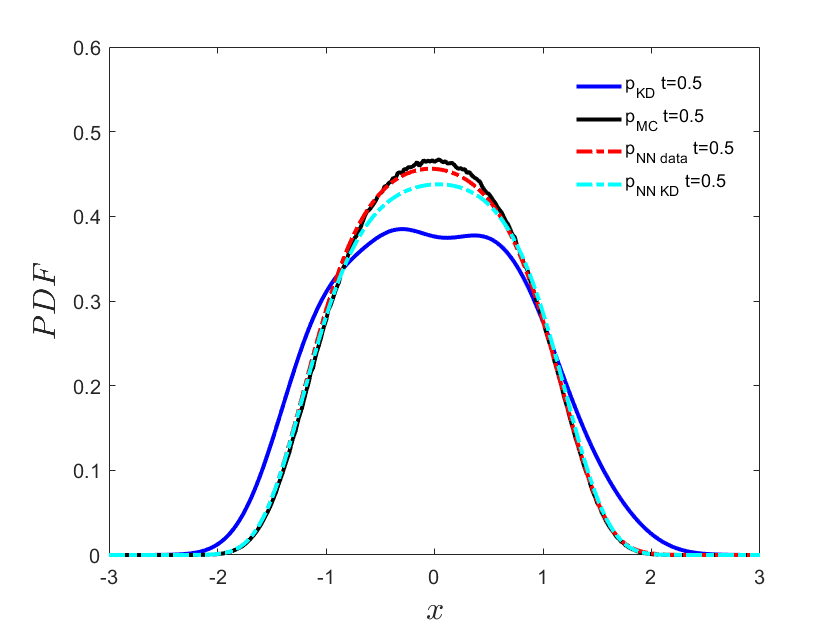}}
\end{minipage}
\begin{minipage}[]{0.3 \textwidth}
\centerline{\includegraphics[width=4cm]{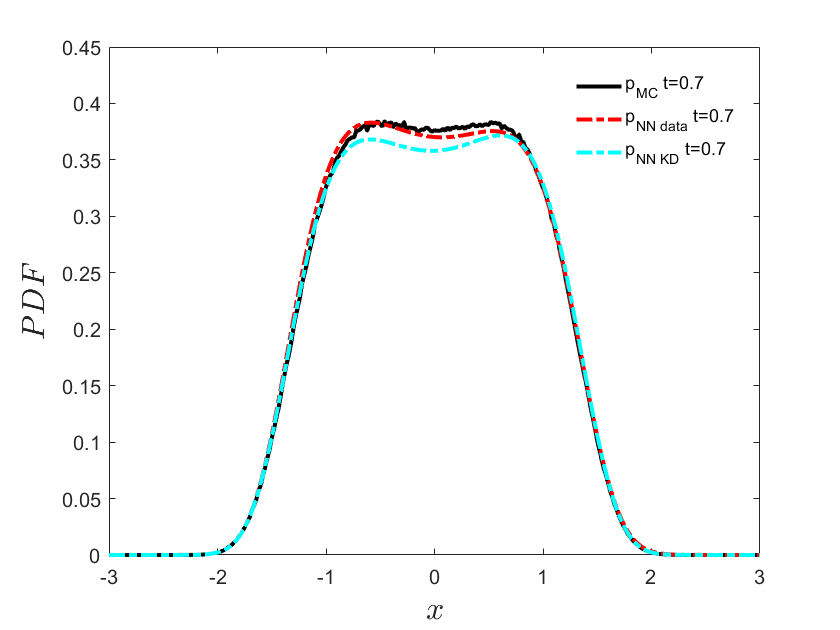}}
\end{minipage}
\hfill
\begin{minipage}[]{0.2 \textwidth}
\centerline{\includegraphics[width=4cm]{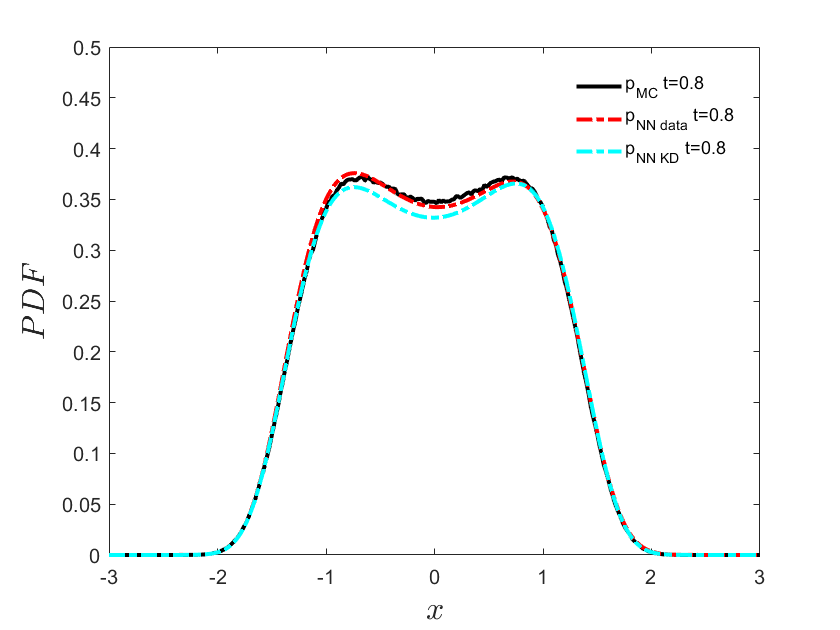}}
\end{minipage}
\hfill
\begin{minipage}[]{0.3 \textwidth}
\centerline{\includegraphics[width=4cm]{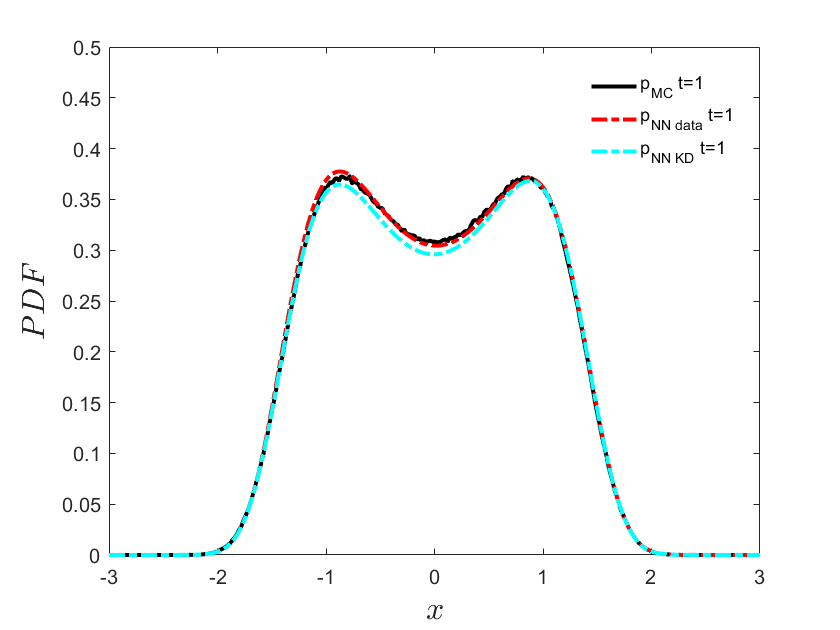}}
\end{minipage}
\caption{\textbf{Inverse problem I - Brownian noise:} Error as a function of time 
for different values of the weight $\tau$ (first row) and evolution of PDF for $a(x)=x-x^3$ and $\sigma=1$ at times $t=0.1, 0.3, 0.5, 0.7, 0.8, 1$ (second and third rows). 
First row: the same color curves correspond to different runs. The blue stars represent the error of $p_{KD}$.
Second and third rows: Observation data of 100 samples of $X_t$ are available only at $t=0.1,0.3,0.5$. The red curves are the PINN predictions while the black curves are the reference solutions obtained by MC. Also, shown with blue color at time $t=0.1, 0.3, 0.5$ are the PDFs obtained based on the kernel density estimation method, and with cyan color the PINN predictions using the estimated PDF as data in the loss function.}
\label{BM-KD-NN}
\end{figure}

 Next, we also compute the inverse problem I of Brownian noise with more observation data at $t=0.1,0.3,0.5$. For each time, we have $100$ independent samples. We use the kernel density estimation to approximate the density using the observation data ($p_{KD}$). When we obtain the $p_{KD}$ at $t=0.1,0.3,0.5$, we use this approximate density as observation data for the PDF and use the neural network to train the model to obtain the density at other times; we call $p_{NN~KD}$ the output of the PINN, whose loss function is defined in \eqref{loss_KD}. We use $N_p = 301$ equidistant points as the collocation points in \eqref{loss_KD}. 
We use the compound trapezoid formula with $301$ point to approximate the integral part of \eqref{eqn:loss_data} and $1000$ residual points for the FP residual in \eqref{eqn:loss_pde}.
We compare these results with the PINN results obtained directly from the observation data in \eqref{eqn:loss_all}. For each case we test different values of weights $\tau$ in \eqref{eqn:loss_all}, and for each $\tau$ we run the code for 3 times with different samples, initialization and random seeds. The errors are shown in the top line of the Figure \ref{BM-KD-NN}. For the second and third row, we choose one of the three cases with $\tau=100$. 
The comparison is shown in Figure \ref{BM-KD-NN} in terms of the $L_2$ error of the two methods. We can see that the results obtained with the PINN using directly the particle observations are better than the PINN results that uses the estimated PDF as input data. Also, our results are superior to $p_{KD}$ from the kernel density estimation for individual time instants, showing that physics-informed learning is very important in capturing the dynamics and integrating the data at multiple time instants.

\subsubsection{Problem I: L\'{e}vy noise}
Similar to the results above, in Appendix A we present a systematic study for problem I assuming that the SDE is driven by L\'{e}vy noise.

\subsubsection{Problem II: Brownian  and L\'{e}vy noise}
~ We now consider problem II, where we want to infer in addition to the initial PDF, the drift and diffusion terms from snapshots. In Appendix B, we present a systematic study for the case of Brownian noise, and in Appendix C we present results for the case of the L\'{e}vy noise. Instead, here in this section, we  consider the SDE driven by both Brownian noise and L\'{e}vy noise. Specifically, we consider the drift term $a(x)=x-x^3$, $\alpha=1.5$, $\sigma=1$ and $\varepsilon=1$.
We use the compound trapezoid formula with $201$ points to approximate the integral part of \eqref{eqn:loss_data} and $800$ residual points to compute the FP residual in \eqref{eqn:loss_pde}.

\begin{figure}[H]
\begin{minipage}[]{0.3 \textwidth}
 \leftline{~~~~~~~\tiny\textbf{(a)}}
\centerline{\includegraphics[width=4cm]{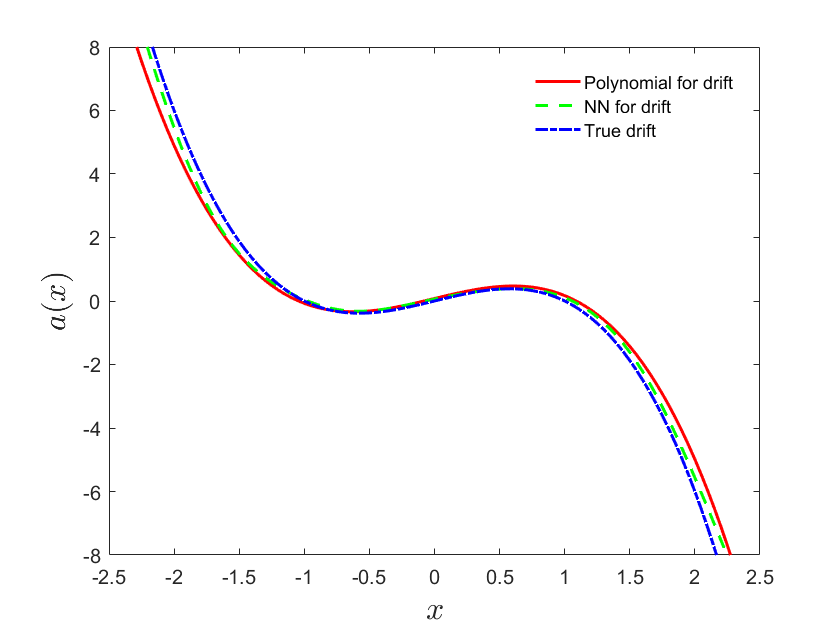}}
\end{minipage}
\hfill
\begin{minipage}[]{0.3 \textwidth}
 \leftline{~~~~~~~\tiny\textbf{(a1)}}
\centerline{\includegraphics[width=4cm]{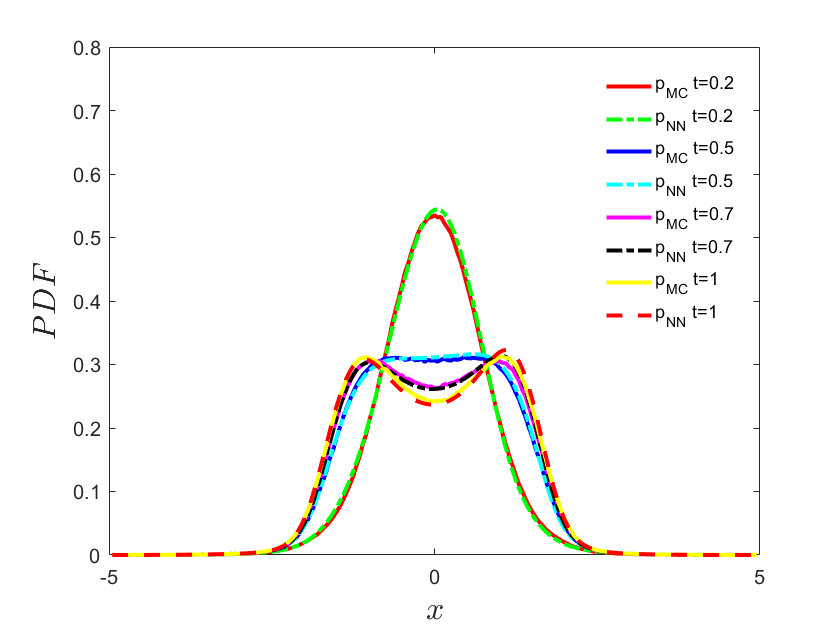}}
\end{minipage}
\hfill
\begin{minipage}[]{0.3 \textwidth}
 \leftline{~~~~~~~\tiny\textbf{(a2)}}
\centerline{\includegraphics[width=4cm]{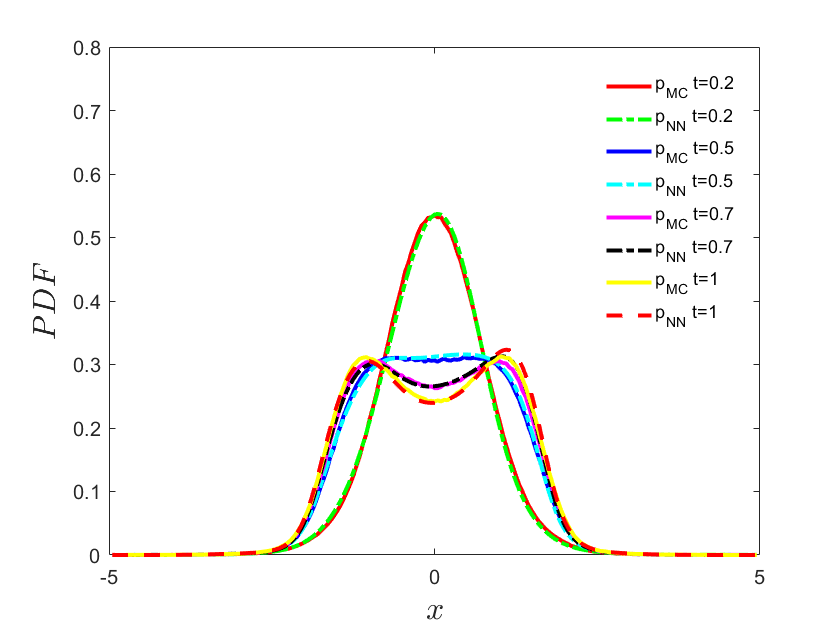}}
\end{minipage}
\hfill
\begin{minipage}[]{0.3 \textwidth}
 \leftline{~~~~~~~\tiny\textbf{(b)}}
\centerline{\includegraphics[width=4cm]{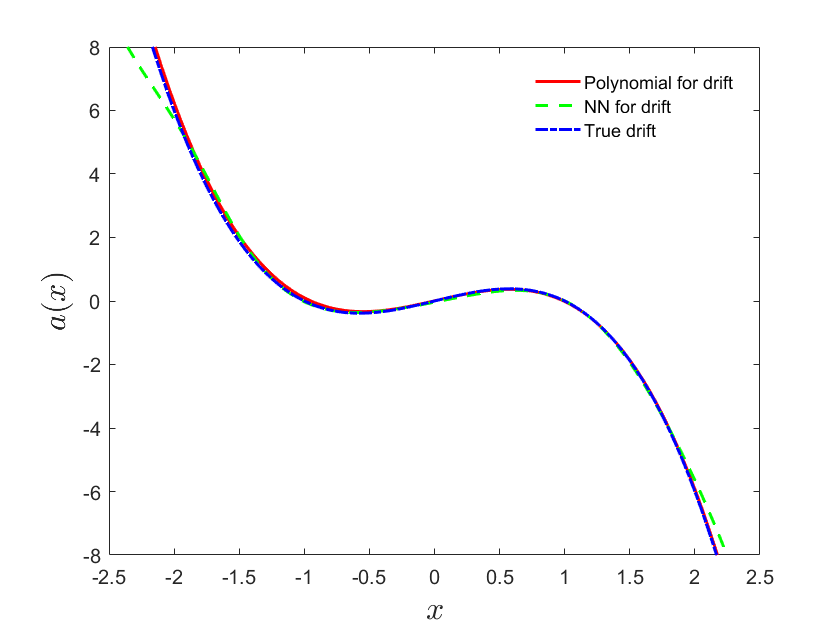}}
\end{minipage}
\hfill
\begin{minipage}[]{0.3 \textwidth}
 \leftline{~~~~~~~\tiny\textbf{(b1)}}
\centerline{\includegraphics[width=4cm]{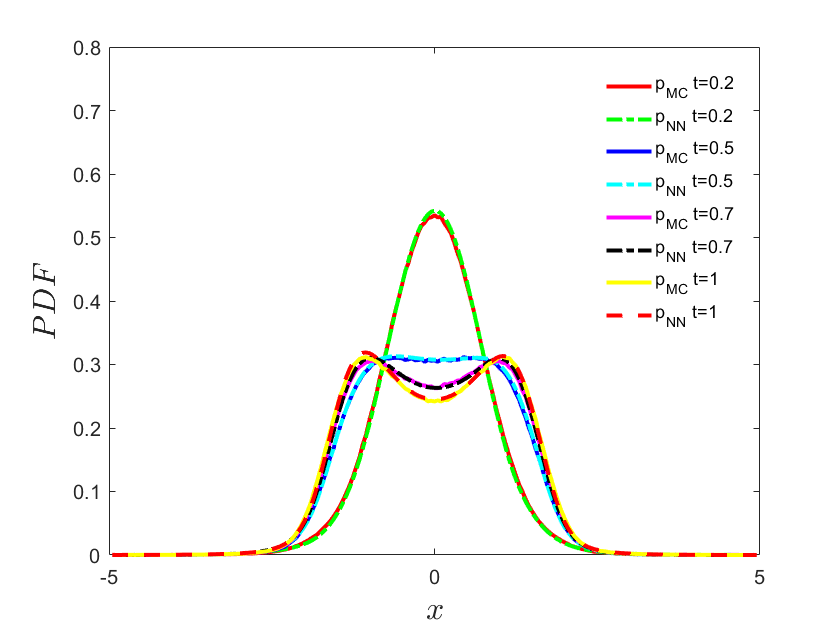}}
\end{minipage}
\hfill
\begin{minipage}[]{0.3 \textwidth}
 \leftline{~~~~~~~\tiny\textbf{(b2)}}
\centerline{\includegraphics[width=4cm]{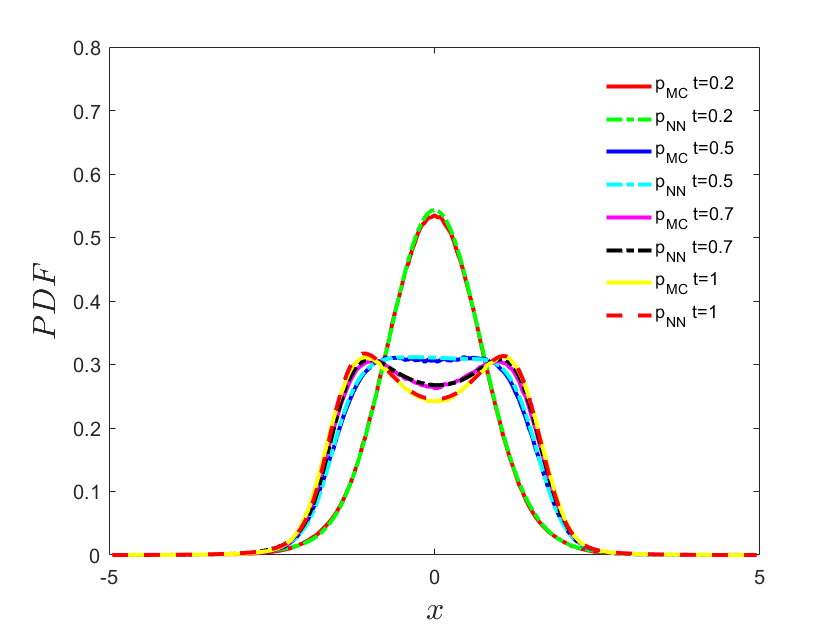}}
\end{minipage}
\caption{\textbf{Inverse problem II - Brownian and L\'{e}vy noise:} Drift term and PDF inference for $a(x)=x-x^3$, $\sigma=1$ and $\varepsilon=1$ with 10,000 samples of $X_t$ at different times: (a)  Observation data at $t=0.2,0.5,1$; (a1) Observation data at $t=0.2,0.5,1$ and use of polynomial fit to learn the drift term; (a2) Observation data at $t=0.2,0.5,1$ and use of neural network to learn the drift term; (b) Observation data at $t=0.1,0.4,0.7,1$; (b1) Observation data at $t=0.1,0.4,0.7,1$ and use of polynomial fit to learn the drift term; (b2) Observation data at $t=0.1,0.4,0.7,1$ and use of neural network to learn the drift term.}
\label{f1-inverse-1d-BM-Levy}
\end{figure}

We compute the following cases assuming different snapshots of available data and different methods of representing the drift term, which is an unknown function. For case (a1), the observation data is available at $t=0.2,~0.5,~1$ and we use a polynomial to approximate the drift term. For case (a2), the observation data is available at $t=0.2,~0.5,~1$ and we use a neural network to approximate the drift term. For case (b1), the observation data is available at $t=0.1,~0.4,~0.7,~1$ and we use a polynomial to approximate the drift term. For case (b2), the observation data is available at $t=0.1,~0.4,~0.7,~1$ and we use a neural network to approximate the drift term. For each case, we assume that we have 10,000 samples at each time instant. We set the minibatch size $b = 1000$. The results for the approximation of the drift term and the evolution of the inferred PDF for all four cases are shown in Figure \ref{f1-inverse-1d-BM-Levy}. The corresponding results for the parameter estimation for the polynomial approximation as well as the magnitudes of the diffusion coefficients of the Brownian term
and the L\'{e}vy term are shown in Table \ref{tab:1d-bm-levy}. It seems that the drift term is better represented using the polynomial approximation but the diffusion coefficients are predicted more accurately using the extra neural network to represent the drift function. Overall, the accuracy  
is higher when we consider separately the Brownian noise and the L\'{e}vy noise as shown in Appendices B and C, respectively, compared to the combined case shown here.

\begin{table*}[ht]
\scriptsize
\begin{center}
\caption{Problem II -- Brownian and L\'{e}vy noise: Parameter estimation for 1D cases.}
\begin{tabular}{ c cc cc cc cc cc c cc cc cc cc ccc c}
\hline
& Parameter           & $\lambda_0$  & $\lambda_1$  &$ \lambda_2$ & $\lambda_3$ &$ \sigma$&$ \varepsilon$\\[1ex]
& True parameter &$0$  &$1$  &$0$  &$-1$  &$1$   &$1$         \\[1ex]
& Case (a1)     &$0.0790$  &$0.9860$  &$-0.0314$  &$-0.8644$  &$0.6966$   &$1.0318$      \\[1ex]
& Case (a2)     &$*$  &$*$  &$*$  &$*$  &$0.7336$ &$1.0516$        \\[1ex]
& Case (b1)  &$0.0042$  &$0.9391$  &$-0.0380$  &$-0.9960$  &$0.7076$   &$1.2092$      \\[1ex]
& Case (b2)  &$*$  &$*$  &$*$  &$*$  &$0.9634$ &$1.0375$        \\[1ex]
\hline
\end{tabular}\label{tab:1d-bm-levy}
\end{center}
\end{table*}

\subsection{Two-dimensional (2D) case}
Consider the following 2D SDE:
\begin{align}  \label{stomodel-2D}
  dX_t &= a_1(X_t,Y_t)dt + \sigma_1 dB_{1,t}+\sigma_2 dB_{2,t}+\varepsilon_x dL_{1,t}^{\alpha},\nonumber\\
  dY_t &= a_2(X_t,Y_t)dt +  \sigma_3 dB_{1,t}+\sigma_4 dB_{2,t}+\varepsilon_y dL_{2,t}^{\alpha},
 \end{align}
where $B_{1,t}$, $B_{2,t}$, $L_{1,t}^{\alpha}$ and $L_{2,t}^{\alpha}$ are independent stochastic processes.

The corresponding FP equation is:
\begin{align}
p_{t}=&-  (a_1 p)_x+\frac{\sigma_1^2+\sigma_2^2}{2} p_{xx}-(a_2 p)_y+\frac{\sigma_3^2+\sigma_4^2}{2} p_{yy}+(\sigma_1 \sigma_3+\sigma_2 \sigma_4)p_{xy}\nonumber\\
&+\varepsilon_x^\alpha \int_{\mathbb{R}\setminus \{0\}}[p(x+x',y)-p(x,y)]\nu_{\alpha}(dx')\nonumber  \\
&+\varepsilon_y^\alpha \int_{\mathbb{R}\setminus \{0\}}[p(x,y+y')-p(x,y)]\nu_{\alpha}(dy').\nonumber\
\end{align}

In the next sections we present results for the 2D Brownian and L\'{e}vy noise. In Appendices D and E we also show results of easier problems, where $X_t$ and $Y_t$ are independent of each other.

\subsubsection{2D Brownian noise}

~~~~We consider the following specific case:
\begin{equation}
d\left( \begin{array}{ccc}
X_t\\
Y_t
\end{array}
\right )=
\left( \begin{array}{c}
\partial_{X_t} \Phi(X_t, Y_t)\\
\partial_{Y_t} \Phi(X_t, Y_t)
\end{array}
\right )dt+\left[ \begin{array}{cc}
\sigma_1 & \sigma_2 \\
0& \sigma_3\\
\end{array}
\right ]  d \left( \begin{array}{c}
B_{1,t}\\
B_{2,t}
\end{array}
\right )    
\end{equation}
where the potential $\Phi(x,y)=-(x+\lambda_0)^2(y+\lambda_1)^2 -(x+\lambda_2)^2(y+\lambda_3)^2$ with $\lambda_0 = \lambda_1 = 1$, $\lambda_2 = \lambda_3 = -0.5$, and $\sigma_1=\sigma_2=\sigma_3=1$. We set the initial distribution as $\mathcal{N}(0,1)$, and we use the compound trapezoid formula with $301\times 301$ points to approximate the integral part of \eqref{eqn:loss_data} and $5000$ points to compute the FP residual.

First, we compute the case with observation data given at $N=7$ snapshots, namely at $t=0, 0.1,0.2,0.3,0.5,0.7,1$. We also present partial results in terms of parameter estimation only for $N=5$, namely at $t= 0,0.3,0.5,0.7,1$. For each time snapshot, we give $100,000$ samples. We set the minibatch size $b=10,000$ to train the first part in \eqref{eqn:loss_data}.

\begin{figure}[H]
\begin{minipage}[]{0.4 \textwidth}
\centerline{\includegraphics[width=6cm,height=5cm]{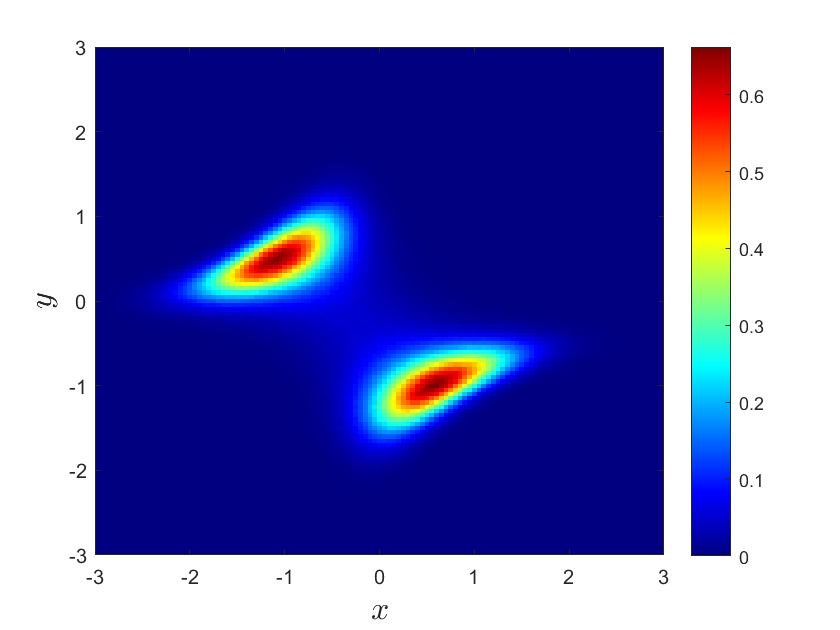}}
\end{minipage}
\hfill
\begin{minipage}[]{0.4 \textwidth}
 \leftline{~~~~~~~\tiny\textbf{Case a}}
\centerline{\includegraphics[width=6cm,height=5cm]{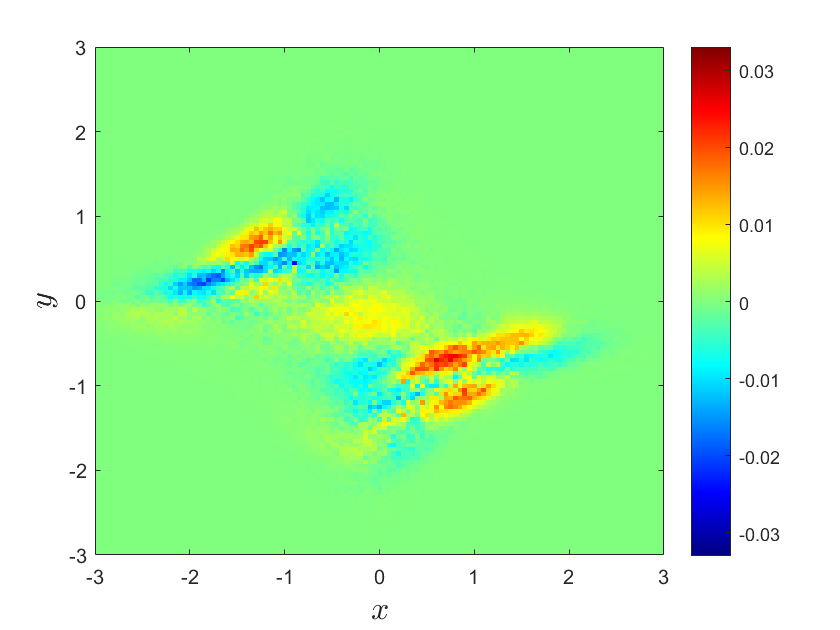}}
\end{minipage}
\hfill
\begin{minipage}[]{0.4 \textwidth}
 \leftline{~~~~~~~\tiny\textbf{Case b}}
\centerline{\includegraphics[width=6cm,height=5cm]{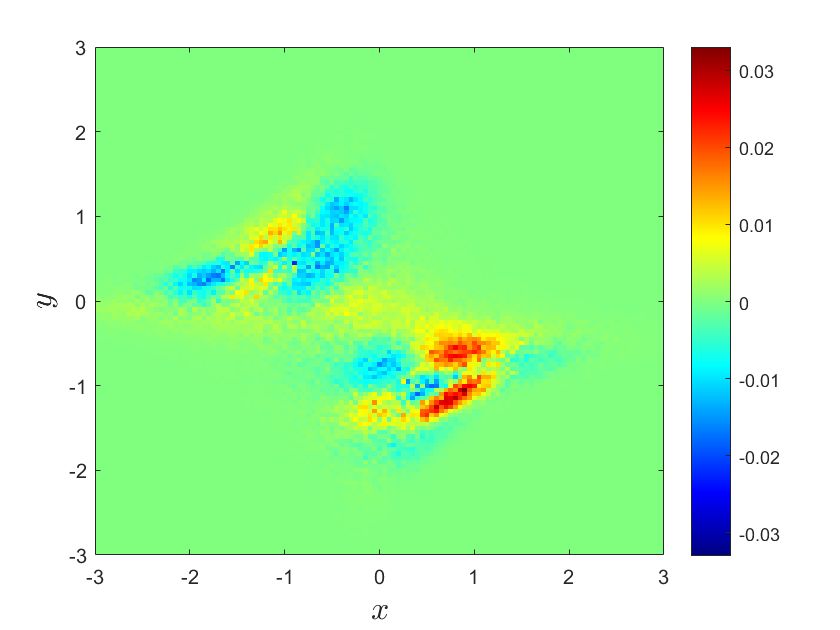}}
\end{minipage}
\hfill
\begin{minipage}[]{0.4 \textwidth}
 \leftline{~~~~~~~\tiny\textbf{Case c}}
\centerline{\includegraphics[width=6cm,height=5cm]{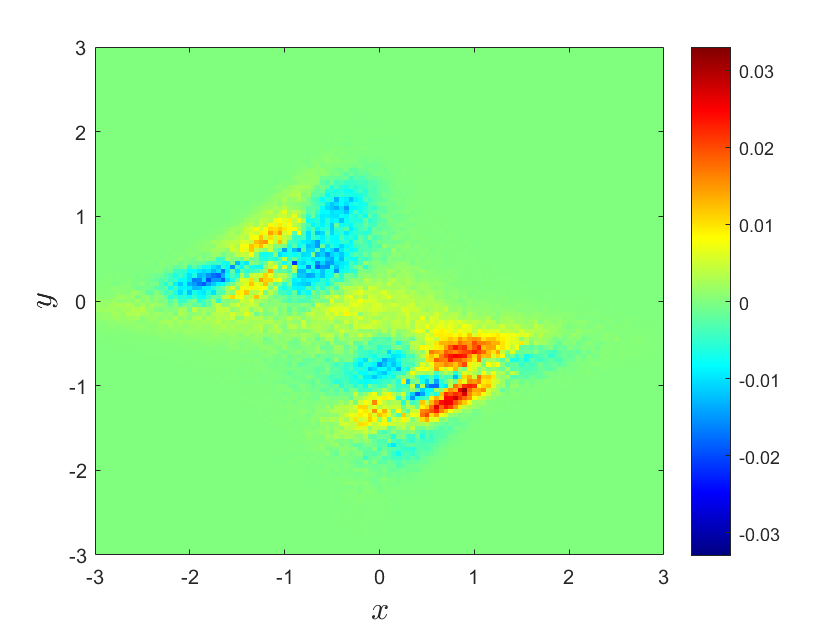}}
\end{minipage}
\caption{\textbf{Inverse problem II - 2D  Brownian noise:} Top left: Reference PDF  when $t=1$ using the Monte-Carlo method. The rest of the plots show the PDF errors for cases a-c at $t=1$. Here $N=7$ snapshots are used as described in the text.}
\label{f2-inverse-2d-BM-case2-error}
\end{figure}

\begin{figure}[H]
\begin{minipage}[]{0.3 \textwidth}
 \leftline{~~~~~~~\tiny\textbf{(a)}}
\centerline{\includegraphics[width=4cm,height=4cm]{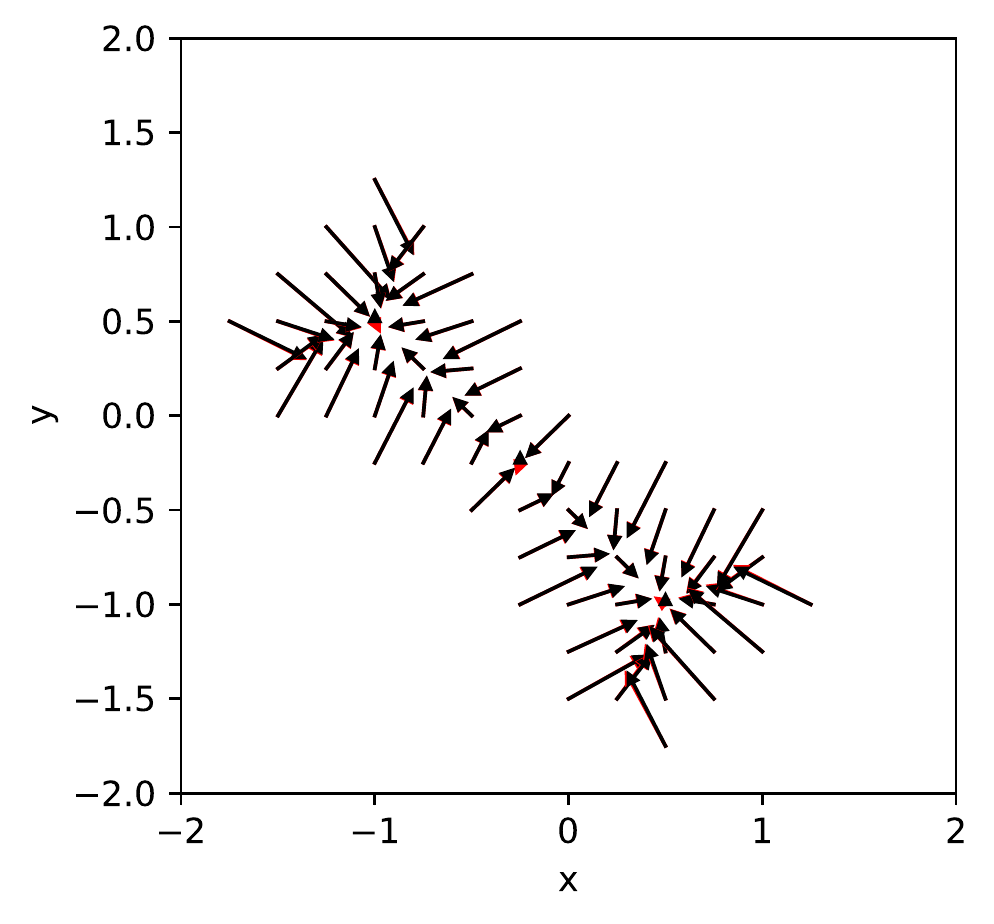}}
\end{minipage}
\hfill
\begin{minipage}[]{0.2 \textwidth}
 \leftline{~~~~~~~\tiny\textbf{(b)}}
\centerline{\includegraphics[width=4cm,height=4cm]{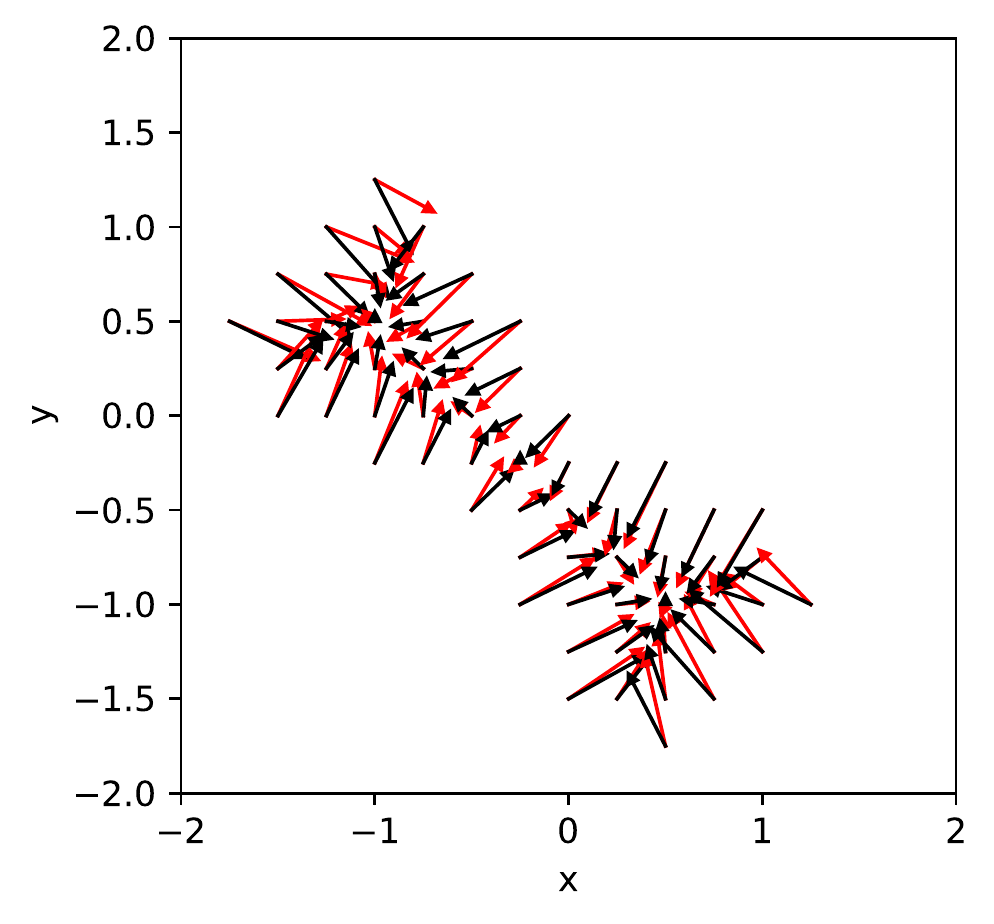}}
\end{minipage}
\hfill
\begin{minipage}[]{0.3 \textwidth}
 \leftline{~~~~~~~\tiny\textbf{(c)}}
\centerline{\includegraphics[width=4cm,height=4cm]{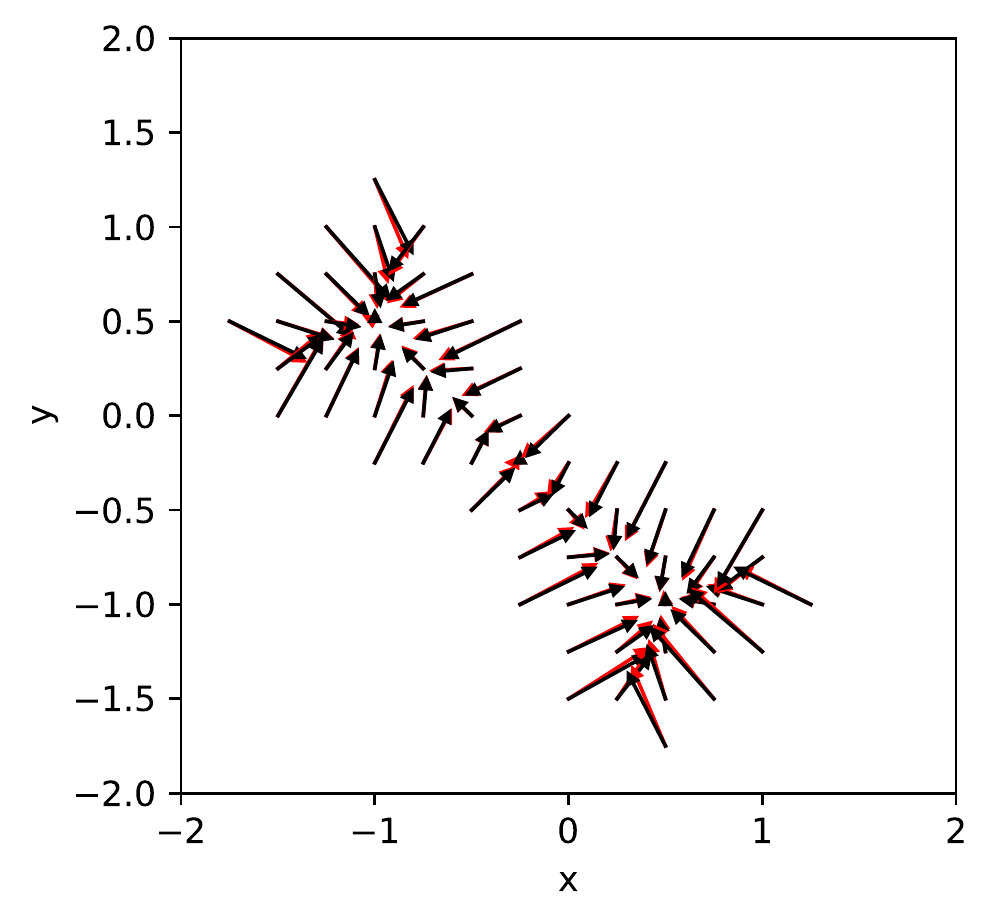}}
\end{minipage}
\caption{\textbf{Inverse problem II - 2D Brownian noise:} Learning the drift term (a) using parameters, (b) using two NNs, and (c) using one NN. The reference solution is denoted by black arrows while the inferred solution by red arrows. For each arrow, the length represents the norm of the drift at the starting point of the arrow, scaled by $0.1$. Here $N=7$ snapshots are used as described in the text.}
\label{f2-inverse-2d-BM-case2-drift}
\end{figure}

In sub-case (a), we construct the NN for the PDF and learn the parameters $\lambda_0$, $\lambda_1$, $\lambda_2$, $\lambda_3$, $\sigma_1$, $\sigma_2$ and $\sigma_3$. The results are shown in Figure \ref{f2-inverse-2d-BM-case2-error}(a).
In sub-case (b), we construct the NN for the PDF and two different NNs to approximate the two drift terms, and learn the parameters $\sigma_1$, $\sigma_2$ and $\sigma_3$. The results are shown in Figure \ref{f2-inverse-2d-BM-case2-error}(b). 
In sub-case (c), we construct the NN for the PDF and use one NN to approximate the potential of the drift terms and take the gradient to obtain the drift term, and learn the parameters $\sigma_1$, $\sigma_2$ and $\sigma_3$. The results are shown in Figure \ref{f2-inverse-2d-BM-case2-error}(c).  The corresponding results of parameter estimation are shown in Table \ref{tab:2d-bm-case2}. We can see that if we use the PINN to approximate the PDF and learn the parameter in the drift term and noise intensity, the results will be better even if we only have five snapshots of the observation data. However, if we also use a neural network to approximate the drift term, the errors in estimating the parameters are larger than $10 \%$.  
We also give the phase diagram of the drift term in Figure \ref{f2-inverse-2d-BM-case2-drift}. For the drift term results, we can see that the best results are obtained by learning the parameters and fitting the drift term explicitly.

\begin{table*}[ht]
\scriptsize
\begin{center}
\caption{Inverse problem II -- 2D Brownian noise: Parameter estimation for $N=5$ and $N=7$ snapshots of available data.}
\begin{tabular}{ c cc cc cc cc cc c cc cc cc cc ccc c}
\hline
& & Parameter           & $\lambda_{0}$  & $\lambda_{1}$  &$ \lambda_{2}$ & $\lambda_{3}$ &$ \sigma_1 $ & $\sigma_2 $ &$ \sigma_3 $\\[1ex]
& & True parameter &$1$  &$1$  &$-0.5$  &$-0.5$  &$1$   &$1$  &$1$        \\[1ex]
\hline
&& Case (a)     &$1.0022$  &$0.9939$  &$-0.4955$  &$-0.4908$  &$1.0036$   &$0.9690$  &$1.0088$    \\[1ex]
&$N=5$& Case (b)     &$*$  &$*$  &$*$  &$*$  &$0.9357$ &$0.9618$  &$0.8272$    \\[1ex]
&& Case (c)    
&$*$  &$*$  &$*$  &$*$  &$ 1.0090$ &$0.9929$  &$0.8792$  
\\[1ex]
\hline
&& Case (a)     &$0.9981$  &$0.9950$  &$-0.4928$  &$-0.4959$  &$1.0108$   &$0.9721$  &$1.0063$    \\[1ex]
&$N=7$& Case (b)     &$*$  &$*$  &$*$  &$*$  &$1.0023$ &$1.1198$  &$1.0433$      \\[1ex]
&& Case (c)     &$*$  &$*$  &$*$  &$*$  &$1.0143$ &$1.0700$  &$1.0179$ 
\\[1ex]
\hline
\end{tabular}\label{tab:2d-bm-case2}
\end{center}
\end{table*}

\subsubsection{2D L\'{e}vy noise}

We consider the following case:

\begin{equation}
d\left( \begin{array}{ccc}
X_t\\
Y_t
\end{array}
\right )=
\left( \begin{array}{c}
\partial_{X_t} \Phi(X_t, Y_t)\\
\partial_{Y_t} \Phi(X_t, Y_t)
\end{array}
\right )dt+\left[ \begin{array}{cc}
\varepsilon_x & 0 \\
0& \varepsilon_y\\
\end{array}
\right ]  d \left( \begin{array}{c}
L_{1,t}^{\alpha}\\
L_{2,t}^{\alpha}
\end{array}
\right ),
\end{equation}
where the potential $\Phi(x,y)=-(x+\lambda_0)^2(y+\lambda_1)^2 -(x+\lambda_2)^2(y+\lambda_3)^2$ with $\lambda_0 = \lambda_1 = 1$, $\lambda_2 = \lambda_3 = -0.5$, and $\varepsilon_x=\varepsilon_y=1$. We set the initial distribution as $\mathcal{N}(0,1)$. We use the compound trapezoid formula with $201\times 201$ points to approximate the integral part of \eqref{eqn:loss_data} and $160,000$ residual points for the FP equation.

First, we compute the case with observation data given at $N=7$ snapshots, namely at $t=0, 0.1,0.2,0.3,0.5,0.7,1$. We also present partial results in terms of parameter estimation only for $N=5$, namely at $t= 0,0.3,0.5,0.7,1$. For each time snapshot, we give $100,000$ samples. We set the minibatch size as $5,000$ for the first part in \eqref{eqn:loss_data}. 

\begin{figure}[ht]
\begin{minipage}[]{0.4 \textwidth}
\centerline{\includegraphics[width=6cm,height=5cm]{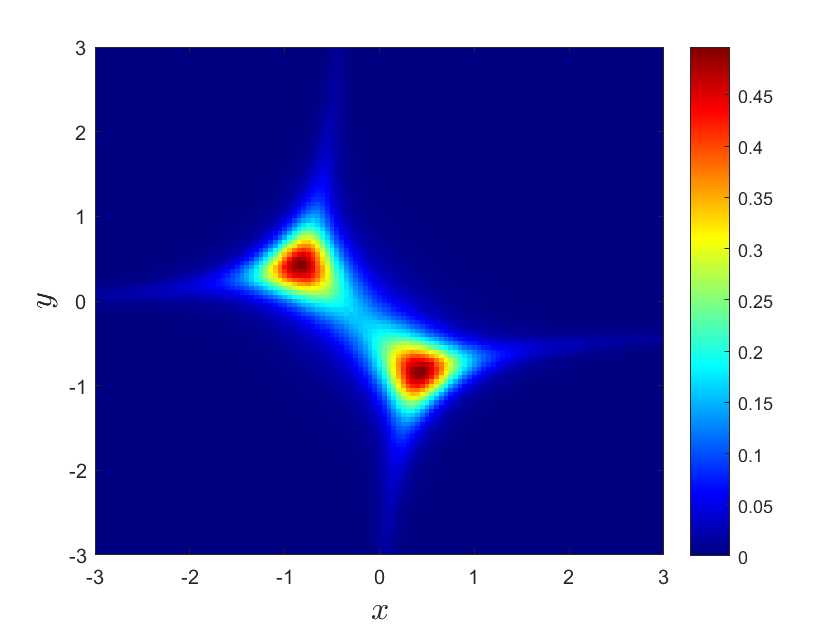}}
\end{minipage}
\hfill
\begin{minipage}[]{0.4 \textwidth}
 \leftline{~~~~~~~\tiny\textbf{Case a}}
\centerline{\includegraphics[width=6cm,height=5cm]{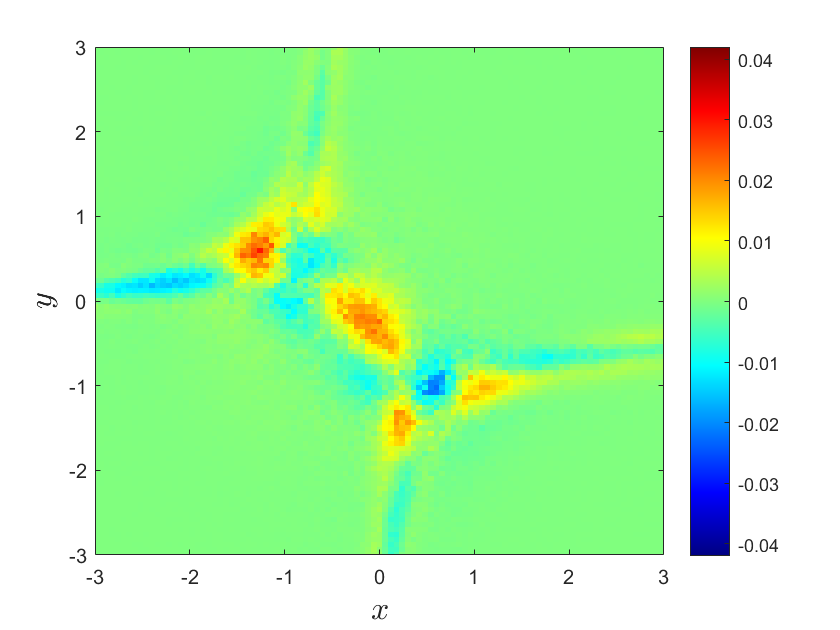}}
\end{minipage}
\hfill
\begin{minipage}[]{0.4 \textwidth}
 \leftline{~~~~~~~\tiny\textbf{Case b}}
\centerline{\includegraphics[width=6cm,height=5cm]{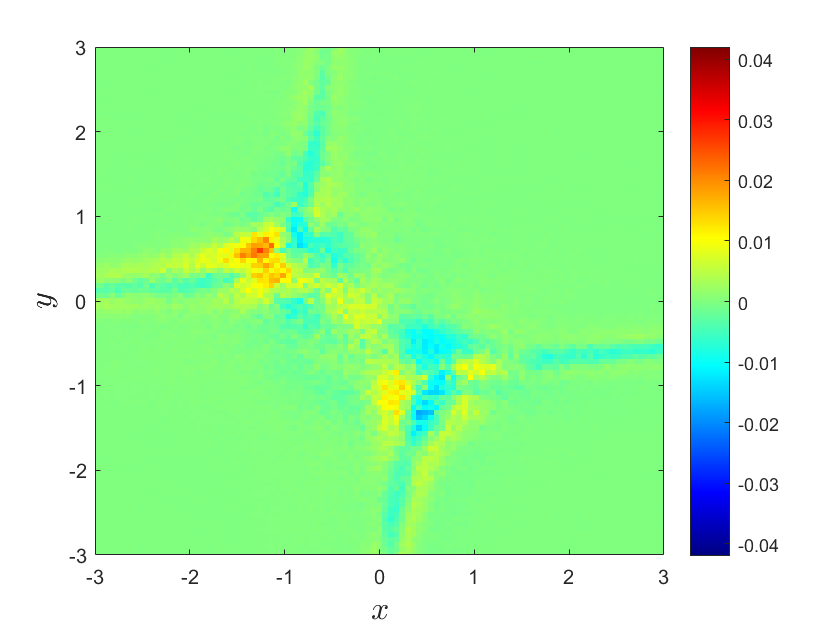}}
\end{minipage}
\hfill
\begin{minipage}[]{0.4 \textwidth}
 \leftline{~~~~~~~\tiny\textbf{Case c}}
\centerline{\includegraphics[width=6cm,height=5cm]{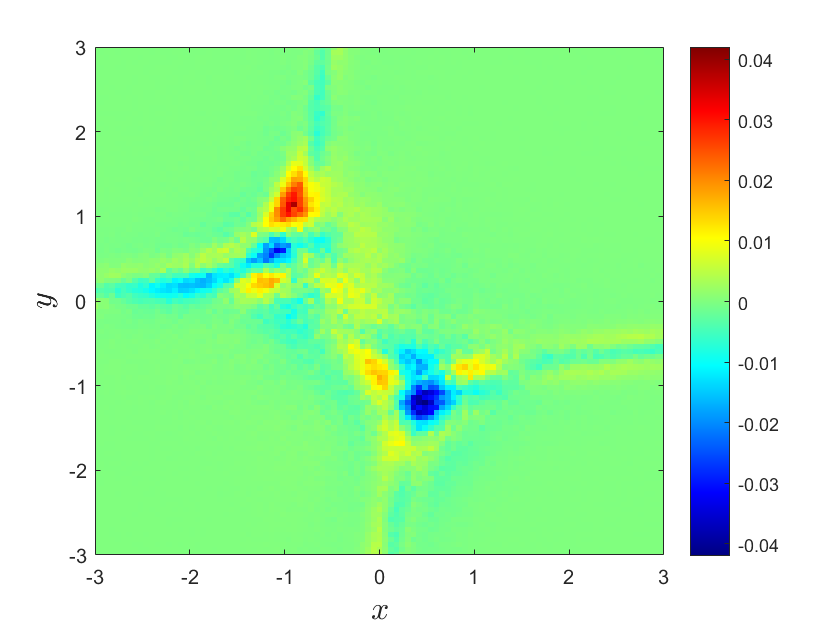}}
\end{minipage}
\caption{ \textbf{Inverse problem II - 2D L\'{e}vy noise.} Top left: Reference PDF density when $t=1$; (a) PDF error for case a; (b) PDF error for case b; (c) PDF error for case c. Here $N=7$ snapshots are used as described in the text.}
\label{f2-inverse-2d-levy-case2-error}
\end{figure}

\begin{figure}[H]
\begin{minipage}[]{0.3 \textwidth}
 \leftline{~~~~~~~\tiny\textbf{(a)}}
\centerline{\includegraphics[width=4cm,height=4cm]{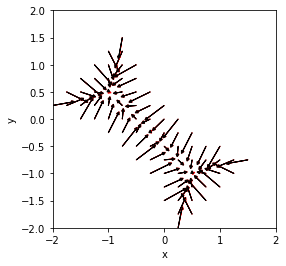}}
\end{minipage}
\hfill
\begin{minipage}[]{0.2 \textwidth}
 \leftline{~~~~~~~\tiny\textbf{(b)}}
\centerline{\includegraphics[width=4cm,height=4cm]{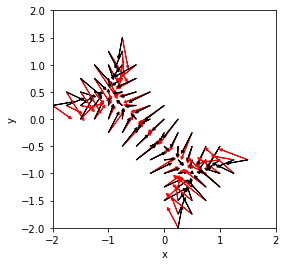}}
\end{minipage}
\hfill
\begin{minipage}[]{0.3 \textwidth}
 \leftline{~~~~~~~\tiny\textbf{(c)}}
\centerline{\includegraphics[width=4cm,height=4cm]{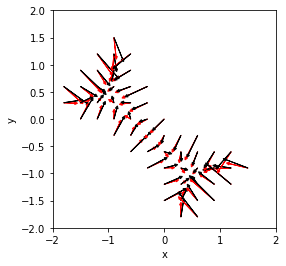}}
\end{minipage}
\caption{\textbf{Inverse problem II - 2D L\'{e}vy noise, case II:}  Learning the drift term using (a) parameters; (b) two NNs;  (c) one NN. Here $N=7$ snapshots are used as described in the text.}
\label{f2-inverse-2d-levy-case2-drift}
\end{figure}

In sub-case (a), we construct the NN for the PDF, then learn the parameter $\lambda_0$, $\lambda_1$, $\lambda_2$, $\lambda_3$, $\varepsilon_x$ and $\varepsilon_y$. 
The results are shown in Figure \ref{f2-inverse-2d-levy-case2-error}(a).
In sub-case (b), we construct the NN for the PDF and two different NNs to approximate the two drift terms, and learn the parameters $\varepsilon_x$ and $\varepsilon_y$. The results are shown in Figure \ref{f2-inverse-2d-levy-case2-error}(b). In sub-case (c), we construct the NN for the PDF and one NN to approximate the potential $\Phi$ of the drift terms, and learn the parameters $\varepsilon_x$ and $\varepsilon_y$. The results are shown in Figure \ref{f2-inverse-2d-levy-case2-error}(c).
The phase diagram of the drift term is shown in Figure~\ref{f2-inverse-2d-levy-case2-drift}. The results of parameter estimation are shown in Table \ref{tab:2d-levy-case2}; 

\begin{table*}[ht]
\scriptsize
\begin{center}
\caption{Inverse problem II -- 2D L\'{e}vy noise: Parameter estimation for $N=5$ and $N=7$ snapshots of available data.}
\begin{tabular}{ c cc cc cc cc cc c cc cc cc cc ccc c}
\hline
&& Parameter           & $\lambda_{0}$  & $\lambda_{1}$  &$ \lambda_{2}$ & $\lambda_{3}$ &$ \varepsilon_x $ & $\varepsilon_y $ \\[1ex]
&& True parameter &$1$  &$1$  &$-0.5$  &$-0.5$  &$1$   &$1$        \\[1ex]
\hline
&& Case (a)     &$0.9938$  &$1.0078$  &$-0.4960$  &$-0.5094$  &$1.0269$   &$0.9891$    \\[1ex]
&$N=5$& Case (b)     &$*$  &$*$  &$*$  &$*$  &$0.3053$ &$0.2628$       \\[1ex]
&& Case (c)     &$*$  &$*$  &$*$  &$*$  &$0.8954$ &$0.7850$       \\[1ex]
\hline
&& Case (a)     &$0.9966$  &$0.9972$  &$-0.5069$  &$-0.4988$  &$1.0249$   &$1.0136$    \\[1ex]
&$N=7$& Case (b)     &$*$  &$*$  &$*$  &$*$  &$0.9029$ &$0.9474$       \\[1ex]
&& Case (c)     &$*$  &$*$  &$*$  &$*$   &$0.9052$ &$0.9154$      \\[1ex]
\hline
\end{tabular}\label{tab:2d-levy-case2}
\end{center}
\end{table*}

\subsection{High-dimensional problems}\label{sec:high-dim}

In this section, we present results for 5D problem with Brownian motion. In Appendix F we present results for similar 3D and 4D cases. 
We consider the following 5D SDE:
\begin{align}  \label{stomodel-5D}
  dX^{(i)}_t&= a_i(X_t^{(i)})dt + \sigma_i dB_t^{(i)},~~~i=1,2,3,...,n
 \end{align}
where the $a_i(x)=x-x^3$ and $n=5$.

\begin{figure}[H]
\begin{minipage}[]{0.45 \textwidth}
 \leftline{~~~~~~~\tiny\textbf{(a1)}}
\centerline{\includegraphics[width=5.8cm]{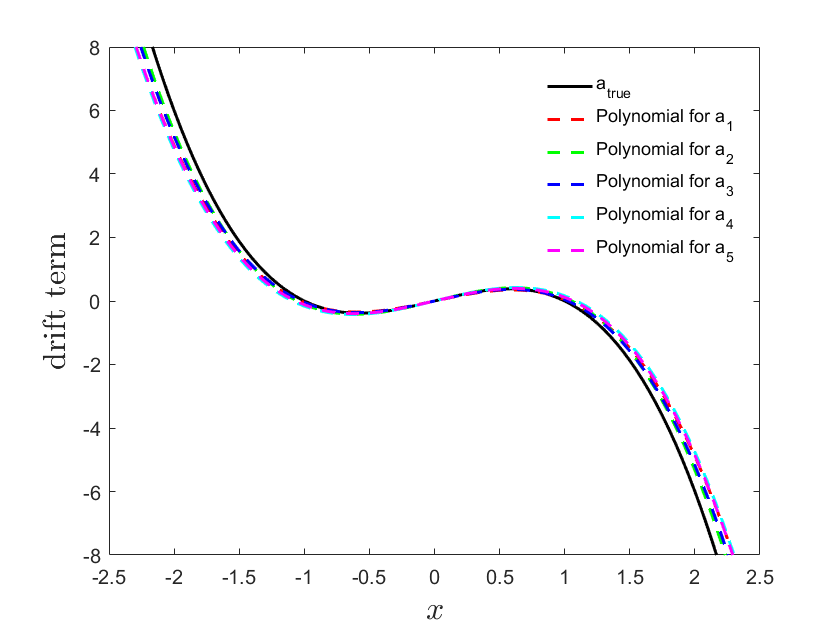}}
\end{minipage}
\hfill
\begin{minipage}[]{0.45 \textwidth}
 \leftline{~~~~~~~\tiny\textbf{(a2)}}
\centerline{\includegraphics[width=5.8cm]{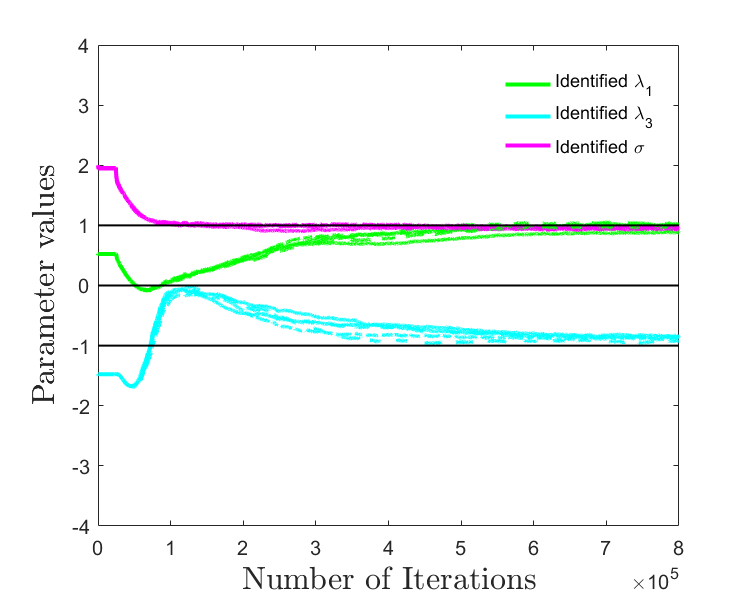}}
\end{minipage}  
\vfill
\begin{minipage}[]{0.45 \textwidth}
 \leftline{~~~~~~~\tiny\textbf{(b1)}}
\centerline{\includegraphics[width=5.8cm]{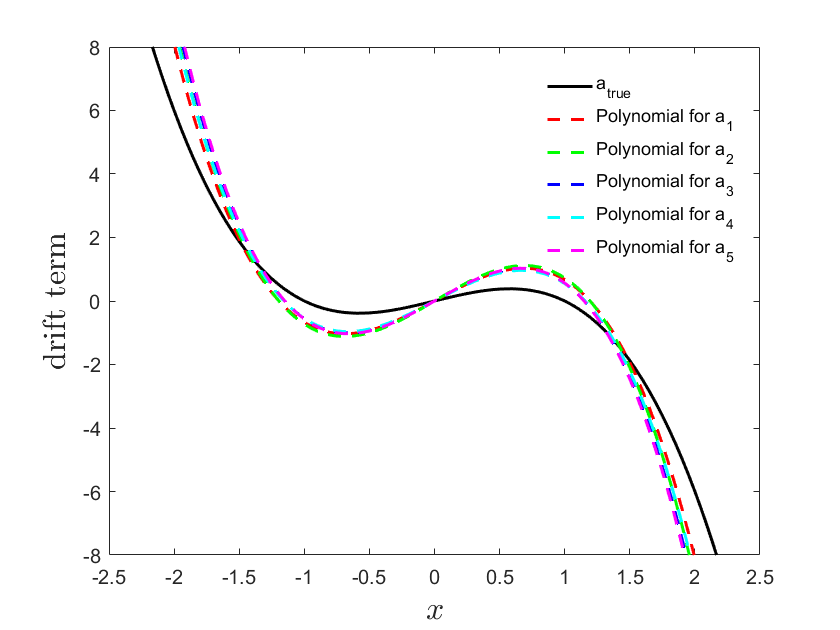}}
\end{minipage}
\hfill
\begin{minipage}[]{0.45 \textwidth}
 \leftline{~~~~~~~\tiny\textbf{(b2)}}
\centerline{\includegraphics[width=5.8cm]{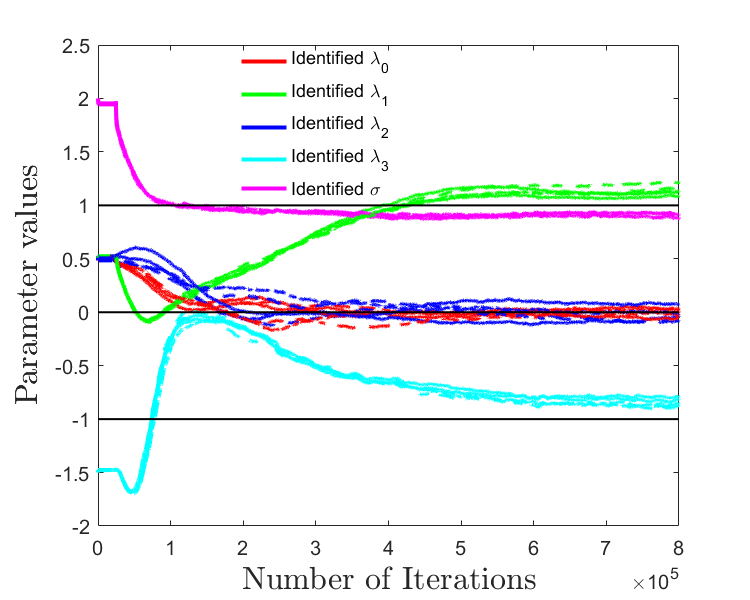}}
\end{minipage}  
\caption{ \textbf{Inverse problem II - 5D Brownian case:} Observation data are available
for $N=7$ snapshots at $t=0.1,0.2,0.3,0.5,0.7,0.9,1$.
The cases (a) and (b) are described in the text.}
\label{f-5d}
\end{figure}

We assume that the observation data are available at $t=0.1,0.2,0.3,0.5,0.7,0.9,1$ i.e., we have $N=7$ snapshots. For each time, we have $100,000$ samples, and we set the minibatch size $b=50,000$ to train the first part in \eqref{eqn:loss_data}. We use the Monte-Carlo method with $200,000$ samples to approximate the integral part of \eqref{eqn:loss_data}  and $10,000$ residual points for the FP equation.

For case (a), we set the drift term 
$a_i(x)=\lambda_{1,i}x+\lambda_{3,i}x^3$, where $i=1,2,3,4,5$. We will learn the parameter $\lambda_{1,i}$,$\lambda_{3,i}$ and $\sigma_i$. The drift terms we learn are shown in Figure \ref{f-5d} (a1) while the estimated parameters are shown in Figure \ref{f-5d} (a2). 
For case (b), we set the drift term 
$a_i(x)=\lambda_{0,i}+\lambda_{1,i}x+\lambda_{2,i}x^2+\lambda_{3,i}x^3$, where $i=1,2,3,4,5$. We will learn the parameter $\lambda_{0,i}$,$\lambda_{1,i}$,$\lambda_{2,i}$,$\lambda_{3,i}$ and $\sigma_i$. The drift terms we learn are shown in Figure \ref{f-5d}(b1) while the estimated parameters are show in Figure \ref{f-5d}(b2).

\section{Summary}

We have developed a new framework based on the physics-informed neural networks (PINNs) and
the Fokker-Planck (FP) equation that enables us to learn the probability density function (PDF) at all times given a few snapshots of the particles. The straightforward approach to accomplishing this would have been to estimate the PDF from the given particle observations and then used this estimated PDF as data in PINNs to infer the solution at all times. We have indeed  implemented this approach but we demonstrated that in most cases and certainly when the data available is small, it is more accurate to include directly the particle observations in the loss function. The key idea connecting the particle observations and the PDF is based on the variational form of the Kullback-Leibler (KL) divergence as shown in Eqn \ref{eqn:pdf_sup}. 
Here, we have considered inverse problems from 1D to 5D, of first type I, where the only unknown is the initial distribution function and we aimed to infer the PDF dynamics at all times, and of type II, where, in addition, the drift terms and noise intensity (Brownian or Levy) are unknown. The drift terms are functions of the position, and hence we have attempted to either model them parametrically assuming a polynomial form or directly using a neural network; for example, for a 2D case we employed two neural networks to approximate the drift terms. Fig. 1 shows a schematic of the PINN that we employ in the current study.

While we did not attempt to investigate the minimum number of snapshots required to be able to solve the above two types of inverse problems, we found that for even a single snapshot and 100 samples we could obtain the initial distribution and infer the PDF dynamics with reasonable accuracy at all times (e.g., for problem I, see Fig. 2). For problem II, both for the Brownian and L\'{e}vy noise,  the noise intensities are more difficult to learn, and hence more snapshots with higher number of samples are required to obtain accurate results. However, even in problem II, we employed a relatively few number of snapshots, e.g., 5 or 7. Overall, parametrizing the drift terms as polynomial functions leads to more accurate results for the drift terms but this requires prior knowledge of the form of the drift terms. With regards to the dimensionality, here we demonstrated that we can easily obtain accurate results in 5D but more work is needed on active learning to 
investigate smart sampling ways in order to achieve similar results in higher dimensions. In a companion work in \cite{YangLiu-GAN-SODE}, we employed generative neural networks (GANs) and the stochastic differential equation governing the particles' trajectories to obtain accurate results in 20D with a relatively small number of samples.

\appendix
\section*{Appendices}
In the following, we present further tests for both the inverse problems I and II driven by either 
Brownian noise or L\'{e}vy noise in 1D and 2D. We also investigate problem II for Brownian noise in 3D and 4D.

\section{Problem I: L\'{e}vy noise}

\begin{figure}[ht]
\begin{minipage}[]{0.2 \textwidth}
 \leftline{~~~~~~~\tiny\textbf{(a1)}}
\centerline{\includegraphics[width=3.5cm,height=3.2cm]{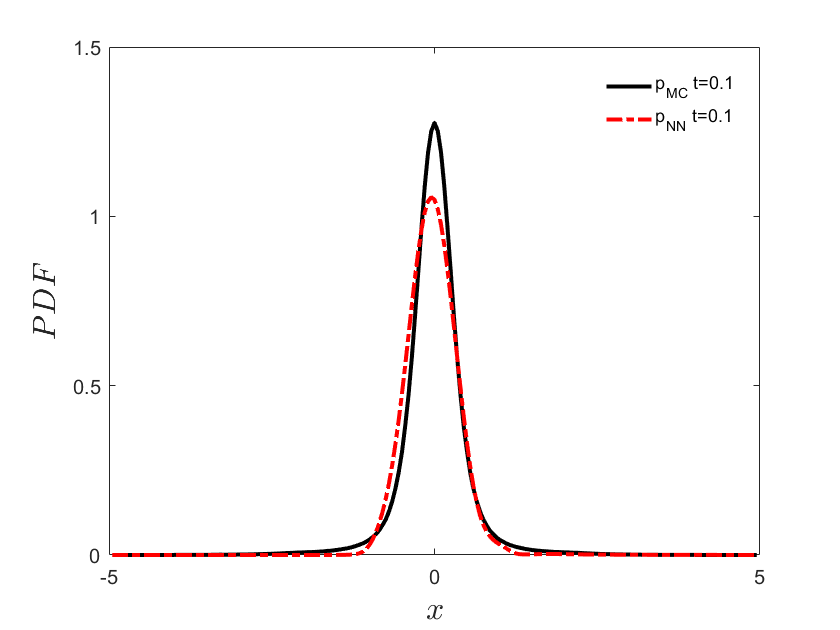}}
\end{minipage}
\hfill
\begin{minipage}[]{0.2 \textwidth}
 \leftline{~~~~~~~\tiny\textbf{(a2)}}
\centerline{\includegraphics[width=3.5cm,height=3.2cm]{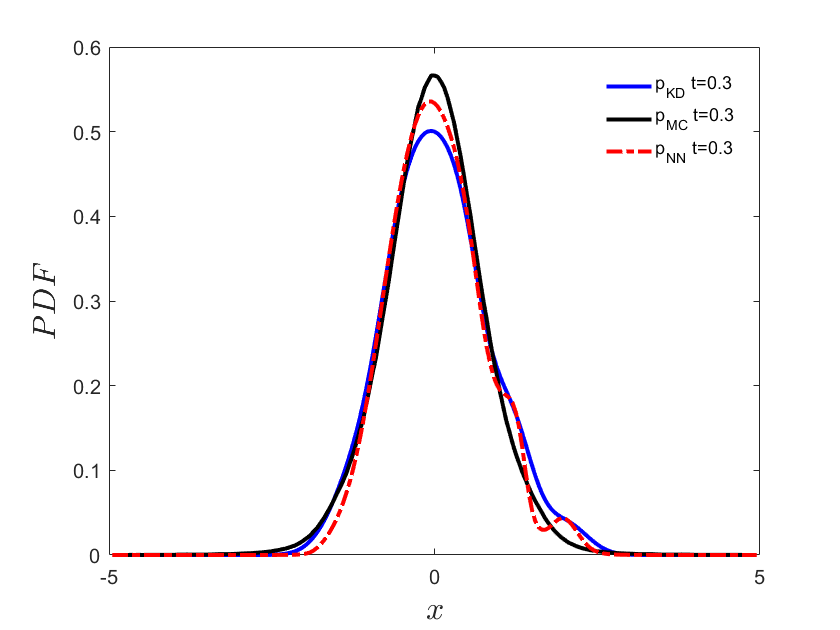}}
\end{minipage}
\hfill
\begin{minipage}[]{0.2 \textwidth}
 \leftline{~~~~~~~\tiny\textbf{(a3)}}
\centerline{\includegraphics[width=3.5cm,height=3.2cm]{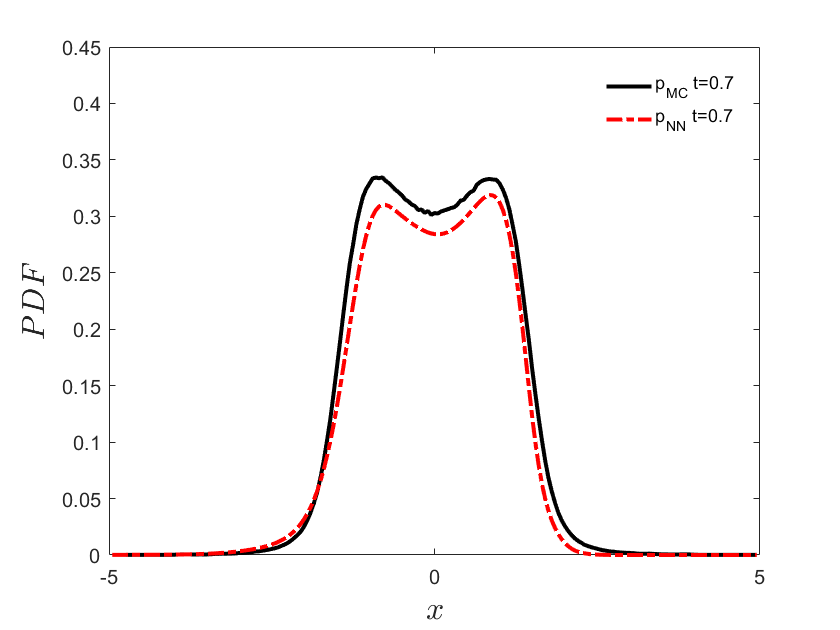}}
\end{minipage}
\hfill
\begin{minipage}[]{0.2 \textwidth}
 \leftline{~~~~~~~\tiny\textbf{(a4)}}
\centerline{\includegraphics[width=3.5cm,height=3.2cm]{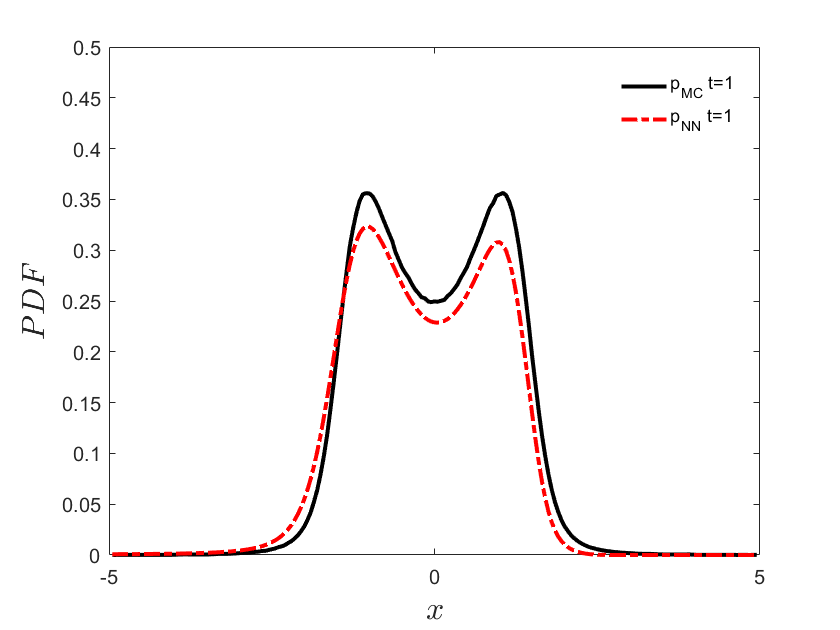}}
\end{minipage}
\hfill
\begin{minipage}[]{0.2 \textwidth}
 \leftline{~~~~~~~\tiny\textbf{(b1)}}
\centerline{\includegraphics[width=3.5cm,height=3.2cm]{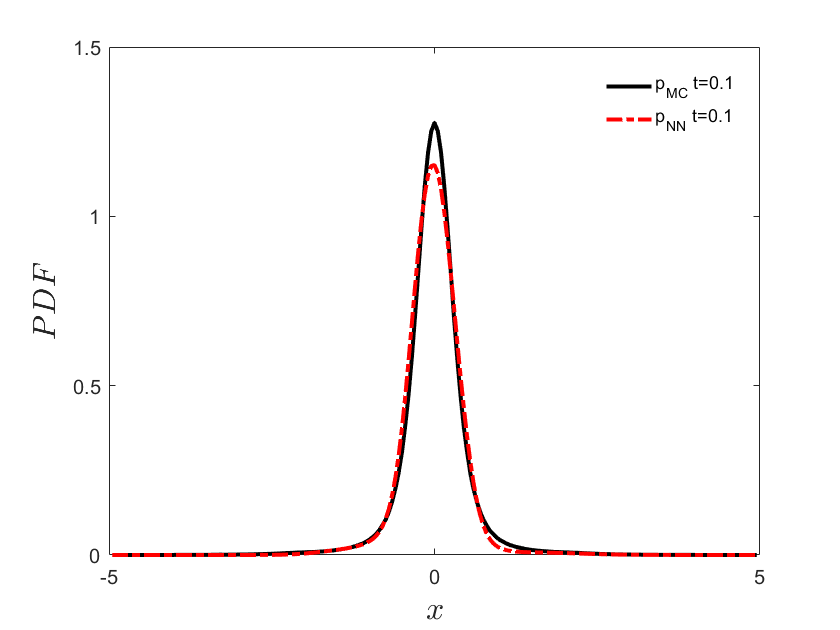}}
\end{minipage}
\hfill
\begin{minipage}[]{0.2 \textwidth}
 \leftline{~~~~~~~\tiny\textbf{(b2)}}
\centerline{\includegraphics[width=3.5cm,height=3.2cm]{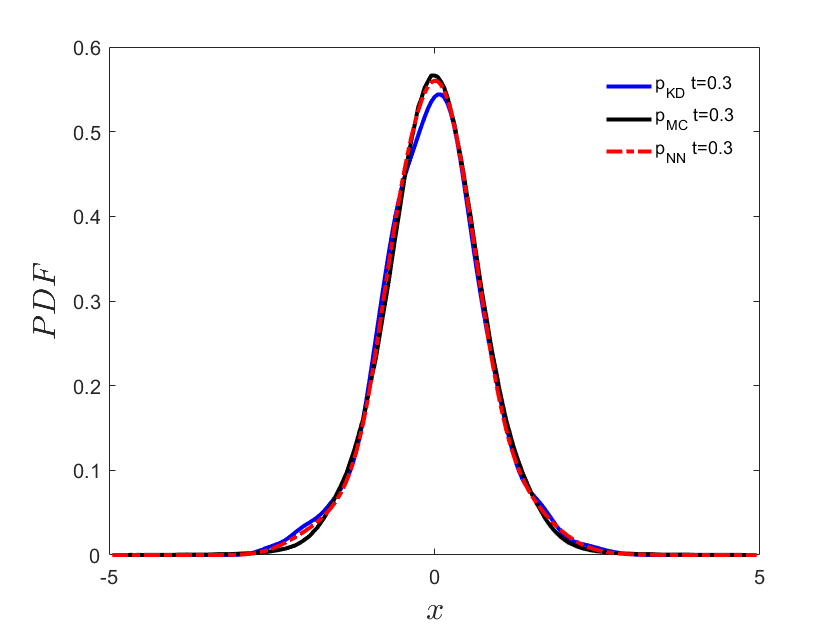}}
\end{minipage}
\hfill
\begin{minipage}[]{0.2 \textwidth}
 \leftline{~~~~~~~\tiny\textbf{(b3)}}
\centerline{\includegraphics[width=3.5cm,height=3.2cm]{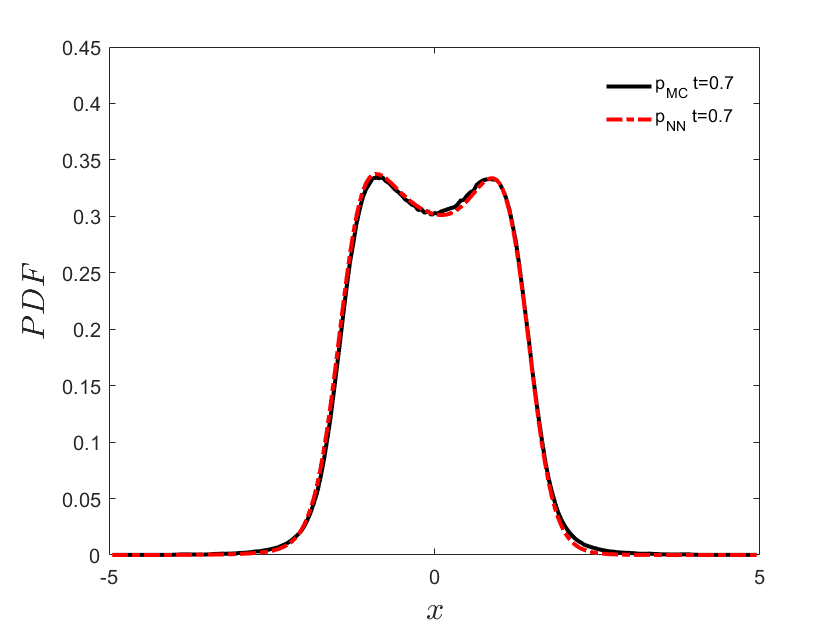}}
\end{minipage}
\hfill
\begin{minipage}[]{0.2 \textwidth}
 \leftline{~~~~~~~\tiny\textbf{(b4)}}
 \centerline{\includegraphics[width=3.5cm,height=3.2cm]{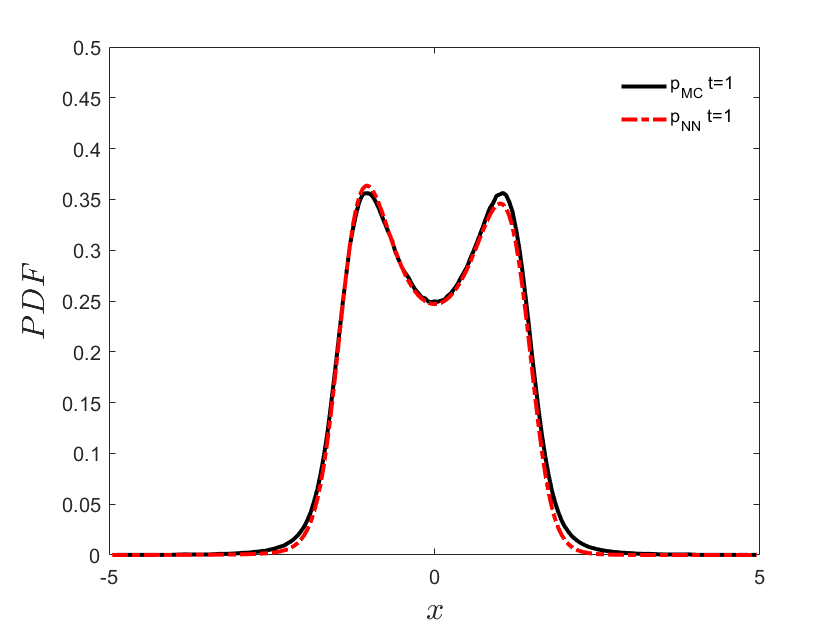}}
\end{minipage}
\begin{minipage}[]{0.2 \textwidth}
 \leftline{~~~~~~~\tiny\textbf{(c1)}}
\centerline{\includegraphics[width=3.5cm,height=3.2cm]{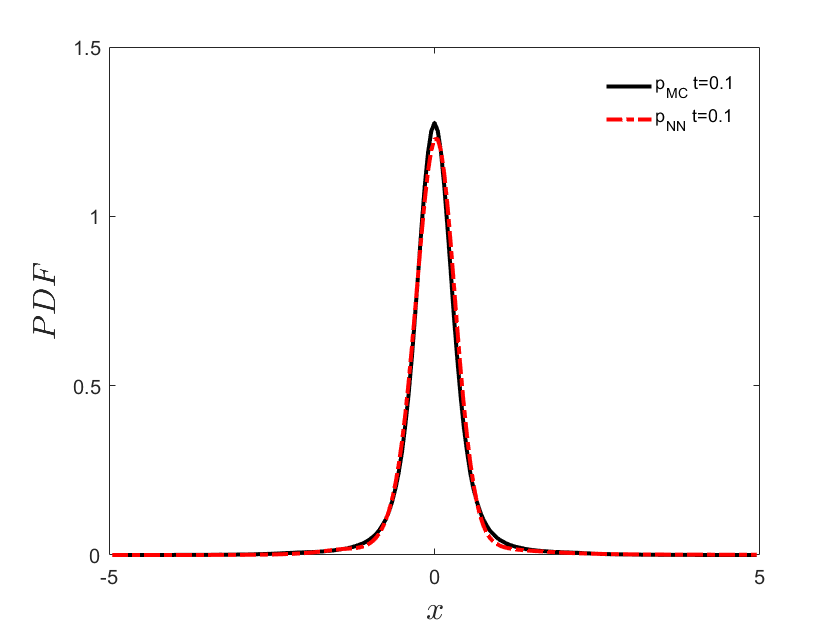}}
\end{minipage}
\hfill
\begin{minipage}[]{0.2 \textwidth}
 \leftline{~~~~~~~\tiny\textbf{(c2)}}
\centerline{\includegraphics[width=3.5cm,height=3.2cm]{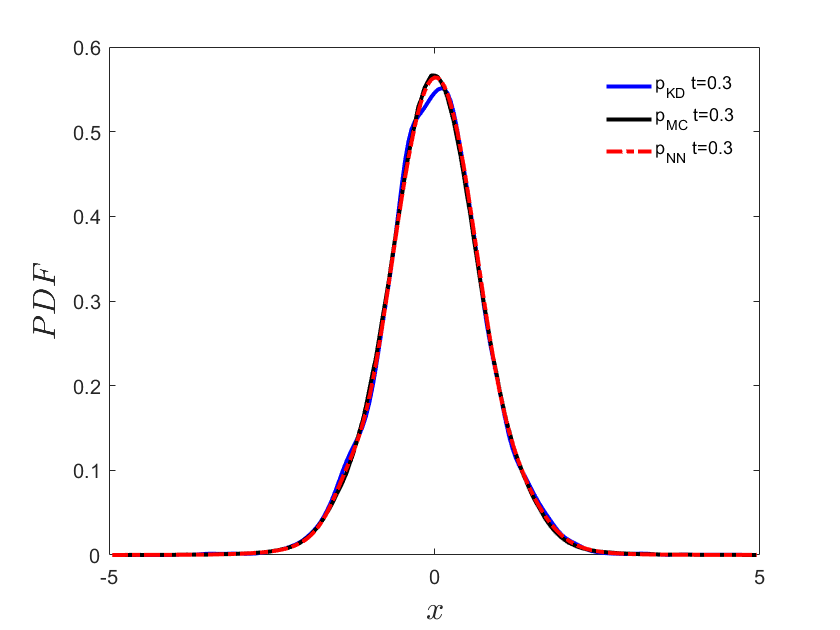}}
\end{minipage}
\hfill
\begin{minipage}[]{0.2 \textwidth}
 \leftline{~~~~~~~\tiny\textbf{(c3)}}
\centerline{\includegraphics[width=3.5cm,height=3.2cm]{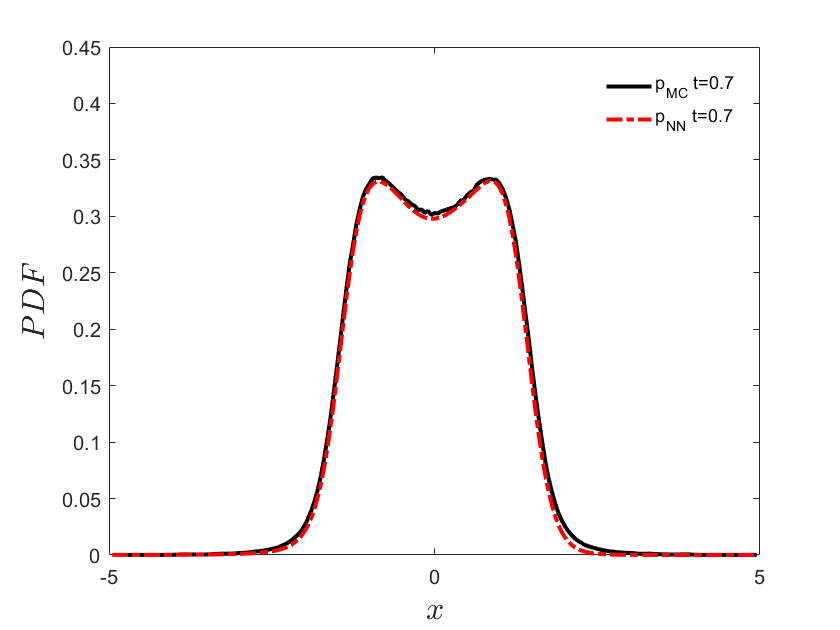}}
\end{minipage}
\hfill
\begin{minipage}[]{0.2 \textwidth}
 \leftline{~~~~~~~\tiny\textbf{(c4)}}
\centerline{\includegraphics[width=3.5cm,height=3.2cm]{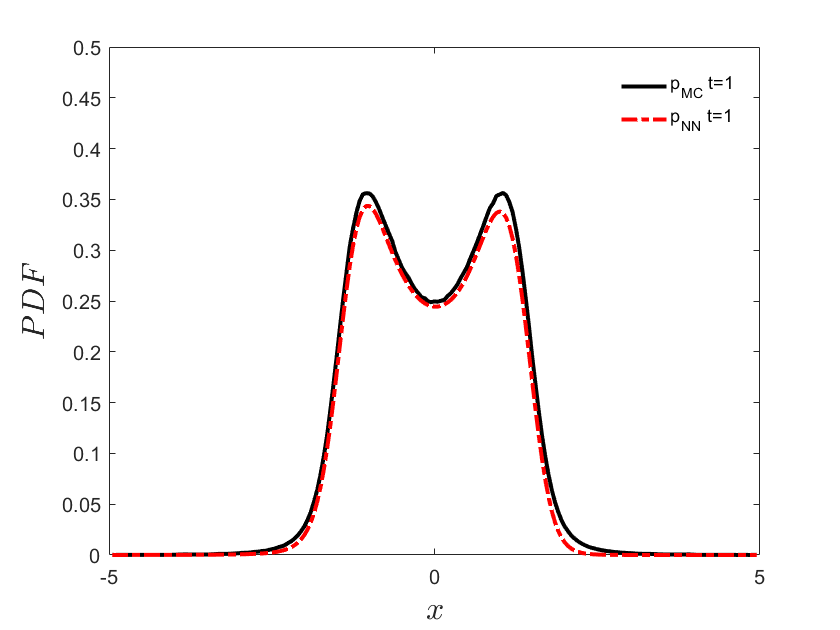}}
\end{minipage}
\caption{\textbf{Appendix A. Inverse problem I - L\'{e}vy noise:} Evolution of PDF for $a(x)=x-x^3$ and $\varepsilon=1$. Observation data of samples of $X_t$ are available at $t=0.3$. (a) 100 samples of $X_t$; (b) 1000 samples of $X_t$; (c) 10000 samples of $X_t$. The red curves are the PINN predictions and the black curves are the reference solutions obtained by MC. Also, shown with blue color at time $t=0.3$ are the PDFs obtained based on the kernel density estimation method.
}
\label{f1-forward-1d-Levy}
\end{figure}

Here, we consider the drift term $a(x)=x-x^3$, $\alpha=1.5$, $\sigma=0$ and $\varepsilon=1$. We use the compound trapezoid formula with $201$ points to approximate the integral part of \eqref{eqn:loss_data} and $800$ points to compute the residual part \eqref{eqn:loss_pde}.
For the inverse problem I, we assume that the observation data is available only for one snapshot at $t=0.3$ with 100 (case a), 1000 (case b) and 10000 (case c) samples.
The results are shown in Figure \ref{f1-forward-1d-Levy}. In the second column, we use the density estimation to obtain the PDF (blue line in this figure). 
Same as in the case with Brownian motion, we also observe convergence with respect to the number of samples, as expected.
%

\begin{figure}[ht]
\begin{minipage}[]{0.3 \textwidth}
\centerline{\includegraphics[width=4cm]{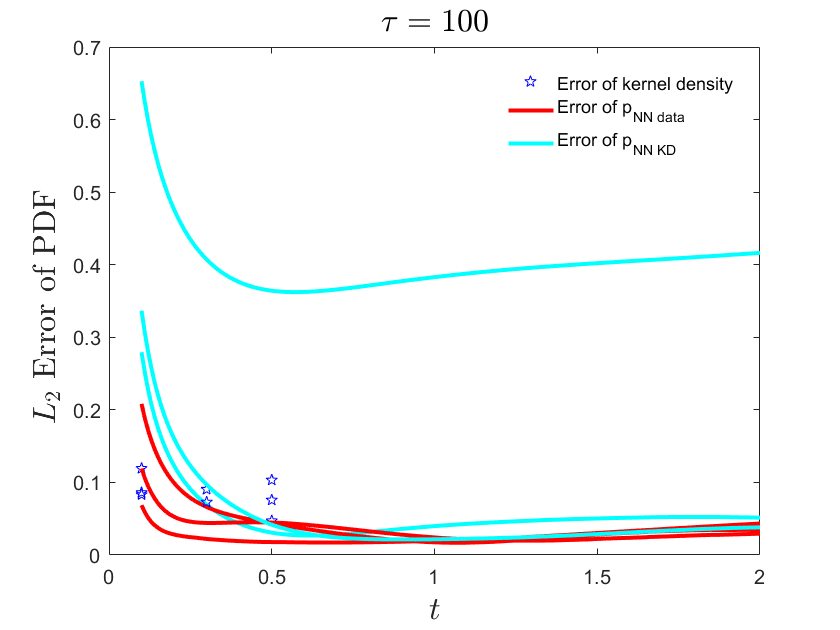}}
\end{minipage}
\hfill
\begin{minipage}[]{0.2 \textwidth}
\centerline{\includegraphics[width=4cm]{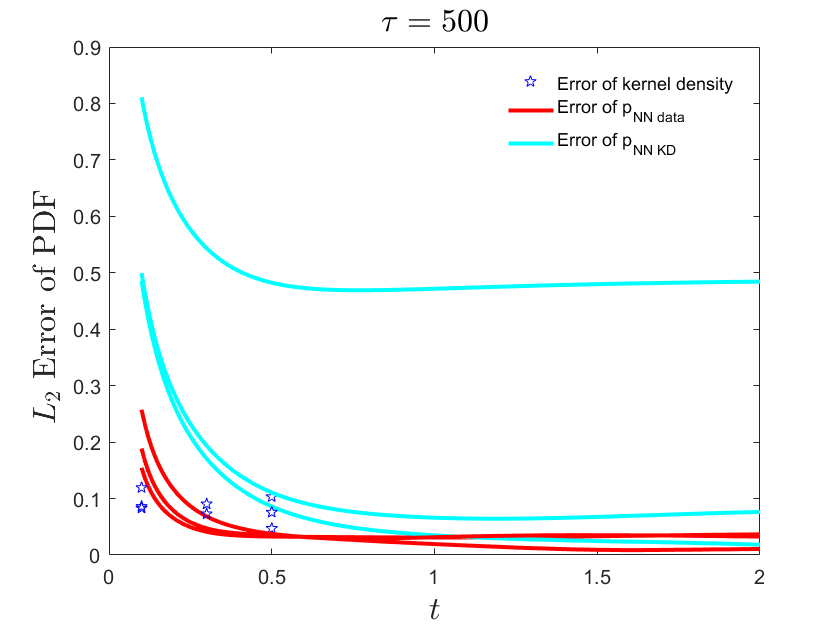}}
\end{minipage}
\hfill
\begin{minipage}[]{0.3 \textwidth}
\centerline{\includegraphics[width=4cm]{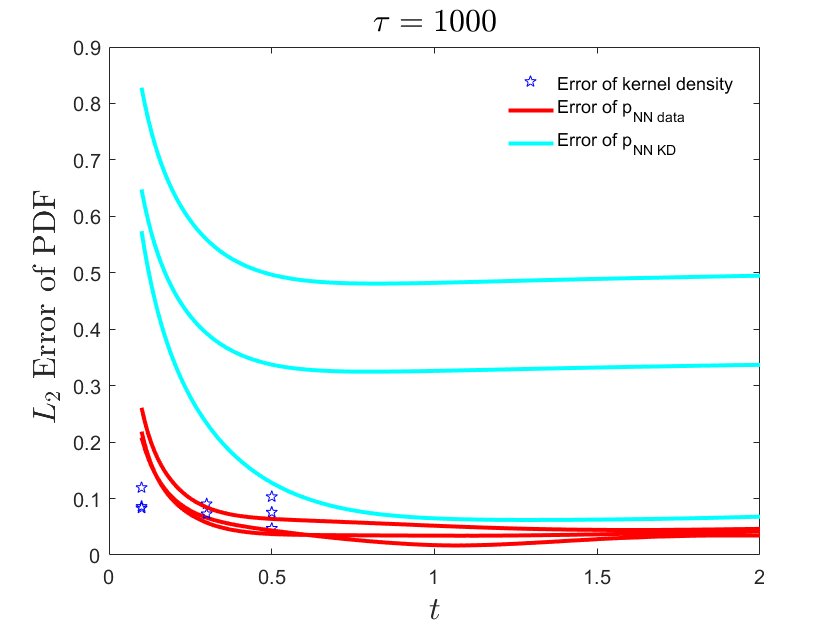}}
\end{minipage}
\begin{minipage}[]{0.3 \textwidth}
\centerline{\includegraphics[width=4cm]{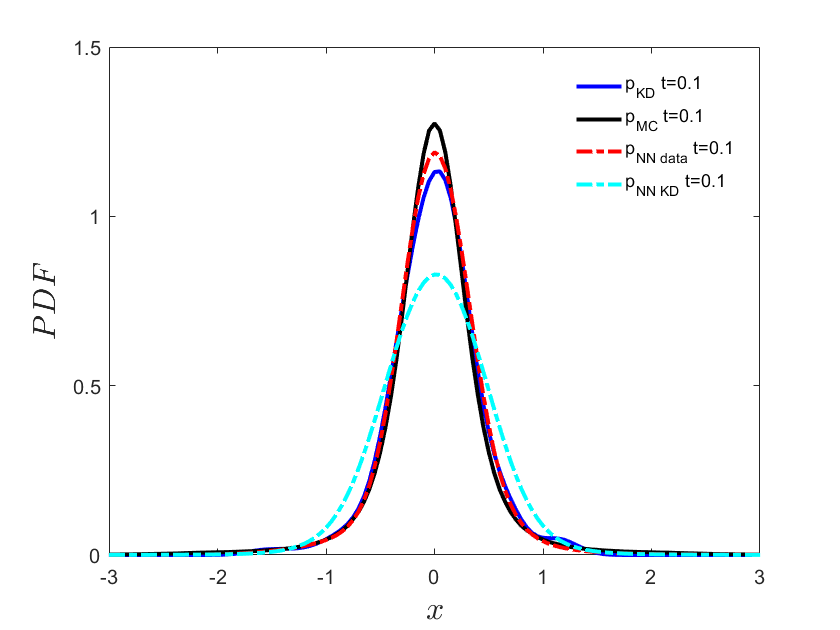}}
\end{minipage}
\hfill
\begin{minipage}[]{0.2 \textwidth}
\centerline{\includegraphics[width=4cm]{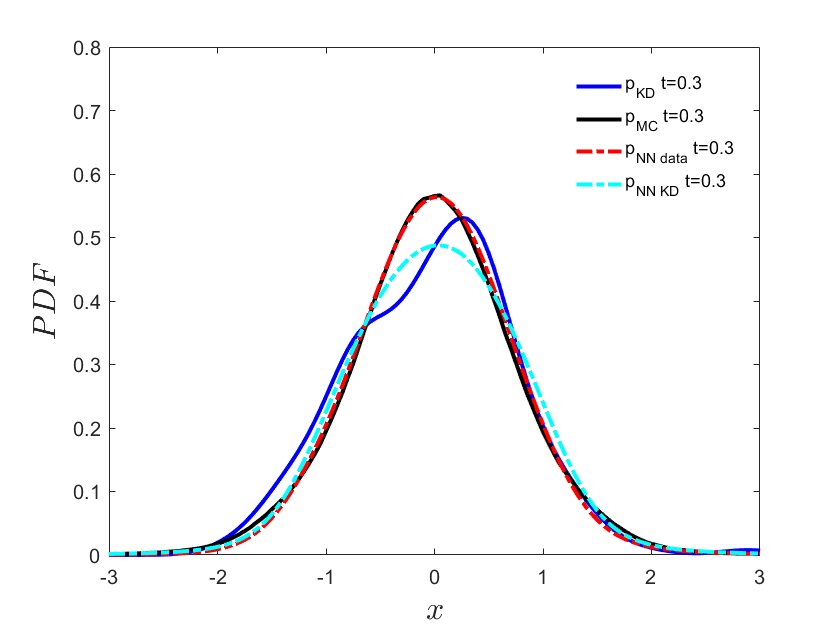}}
\end{minipage}
\hfill
\begin{minipage}[]{0.3 \textwidth}
\centerline{\includegraphics[width=4cm]{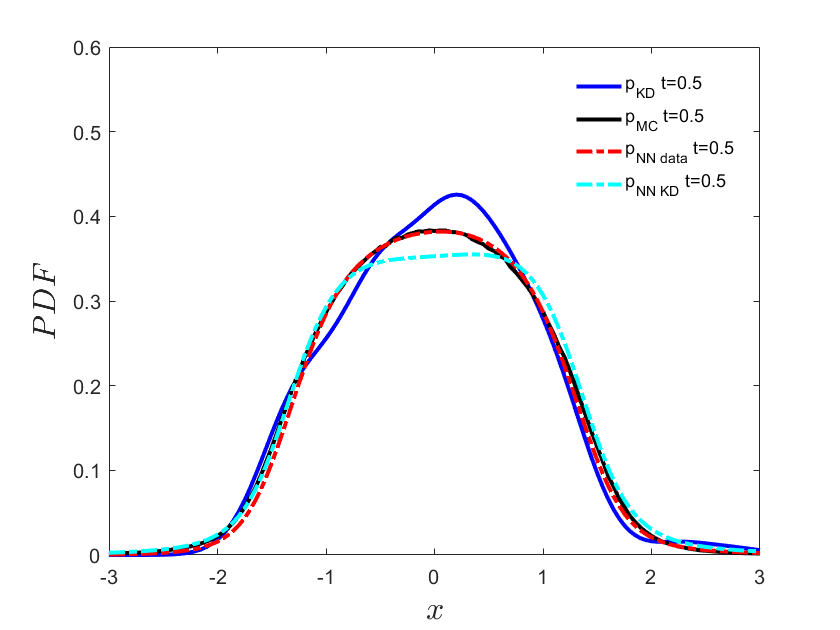}}
\end{minipage}
\begin{minipage}[]{0.3 \textwidth}
\centerline{\includegraphics[width=4cm]{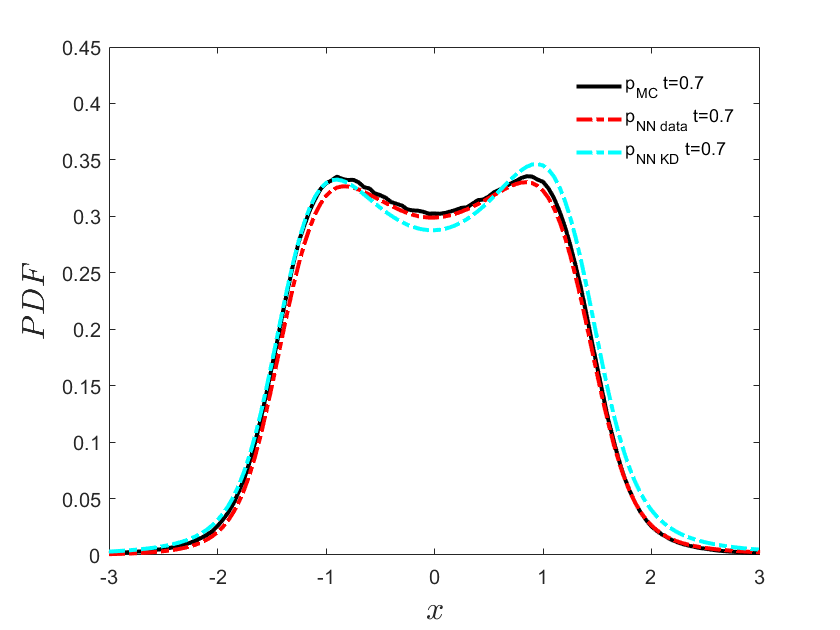}}
\end{minipage}
\hfill
\begin{minipage}[]{0.2 \textwidth}
\centerline{\includegraphics[width=4cm]{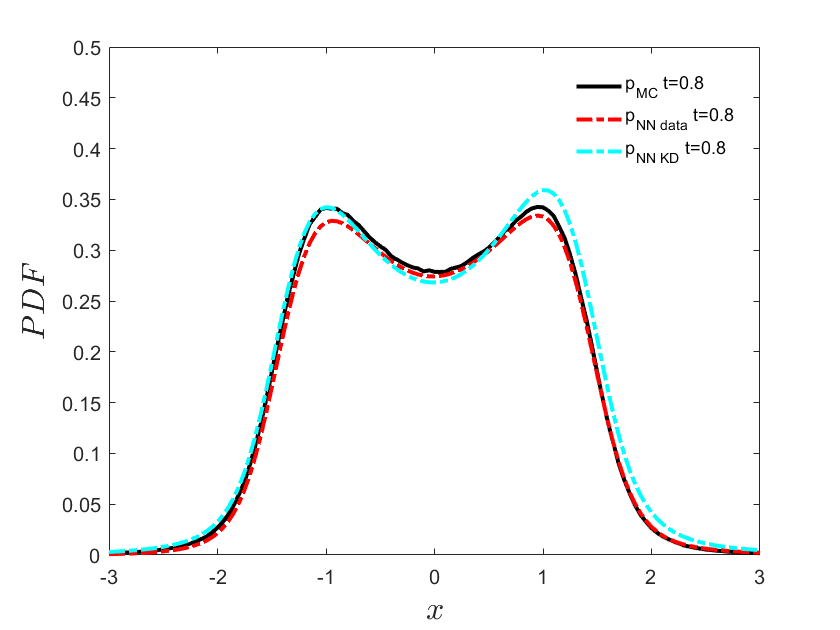}}
\end{minipage}
\hfill
\begin{minipage}[]{0.3 \textwidth}
\centerline{\includegraphics[width=4cm]{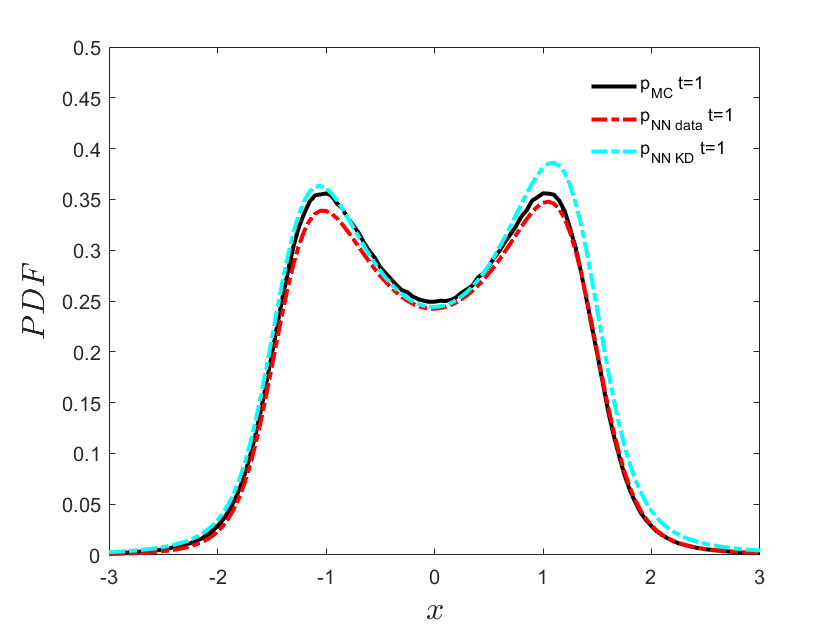}}
\end{minipage}
\caption{\textbf{Appendix A. Inverse problem I - L\'{e}vy noise:} Error as a function of time for different values of the weight (first row) and evolution of PDF for $a(x)=x-x^3$ and $\varepsilon=1$ (second and third rows). 
First row: the different color curves correspond to different runs. The blue stars represent the error of $p_{KD}$.
Second and third rows:
Observation data of 200 samples of $X_t$ are available at 
$t=0.1,0.3,0.5$. The red curves are the PINN predictions while the black curves are the reference solutions obtained by MC. Also, shown with blue color at time $t=0.1, 0.3, 0.5$ are the PDFs obtained based on the kernel density estimation method, and with cyan color the PINN predictions using the estimated PDF as data in the loss function.
}
\label{Levy-KD-NN}
\end{figure}

 We also compute the inverse problem I of L\'{e}vy noise with observations at multiple snapshots, namely, at $t=0.1,0.3,0.5$. For each time, we have $200$ independent samples. 
 We use the kernel density estimation to approximate the density $p_{KD}$ using the observation data. 
 When we have the $p_{KD}$, we use this approximate density as observation data of PDF and use the neural network to train the model to obtain the PDF at all times; we
set $p_{NN~KD}$ as the output. We use $N_p=201$ equidistant points as the collocation points in \eqref{loss_KD}.
For each case we test different values of weights $\tau$ in \eqref{eqn:loss_all}, and for each $\tau$ we run the code for 3 times with different samples, initialization and random seeds.
In the first row of Figure \ref{Levy-KD-NN}, we present the errors for different $\tau$. We just show one of the three cases with $\tau=100$ in the second and third row of Figure \ref{Levy-KD-NN}. We compare these results with the results by the PINN method using the observation data on the particles directly. The comparison is shown in Figure \ref{Levy-KD-NN}; we can see that the results from the latter method are more accurate.
%

\section{Problem II: Brownian noise}
For the inverse problem II, we compute two cases corresponding to different observation data.  Considering the probability shown in Figure \ref{f1-forward-1d-BM}, we can see that for early times, there is only one peak, but later there are two peaks emerging. Hence, we want to verify if small observation data is sufficient in order to capture the two peaks. To this end, we consider two cases of observation data. In the first case, observation data is available at $t=0.1,0.3,0.5$, while in the other case data is available at $t=0.2,0.5,1$. 
We will use the polynomial ($a(x)=\lambda_0+\lambda_1 x+\lambda_2 x^2+\lambda_3 x^3$) or alternatively a neural network to approximate the drift term. For case (a1), the observation data is available at $t=0.1,0.3,0.5$ and we use the polynomial fit to approximate the drift term. For case (a2), the observation data is available at $t=0.1,0.3,0.5$ and we use the neural network to represent the drift term. For case (b1), the observation data is available at $t=0.2,0.5,1$ and we use the polynomial to approximate the drift term. For case (b2), the observation data is available at $t=0.2,0.5,1$ and we use the neural network to approximate the drift term.

\begin{figure}[H]
\begin{minipage}[]{0.3 \textwidth}
 \leftline{~~~~~~~\tiny\textbf{(a)}}
\centerline{\includegraphics[width=4cm]{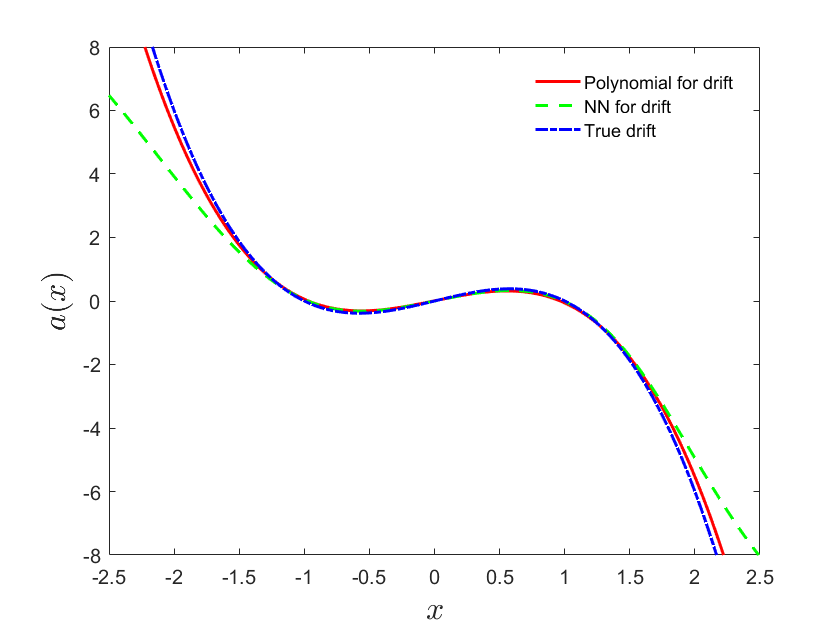}}
\end{minipage}
\hfill
\begin{minipage}[]{0.3 \textwidth}
 \leftline{~~~~~~~\tiny\textbf{(a1)}}
\centerline{\includegraphics[width=4cm]{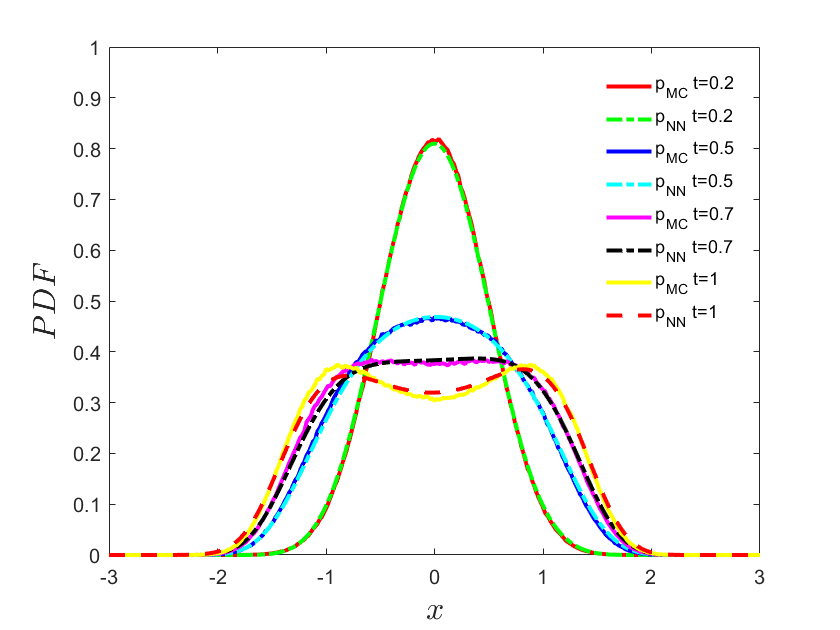}}
\end{minipage}
\hfill
\begin{minipage}[]{0.3 \textwidth}
 \leftline{~~~~~~~\tiny\textbf{(a2)}}
\centerline{\includegraphics[width=4cm]{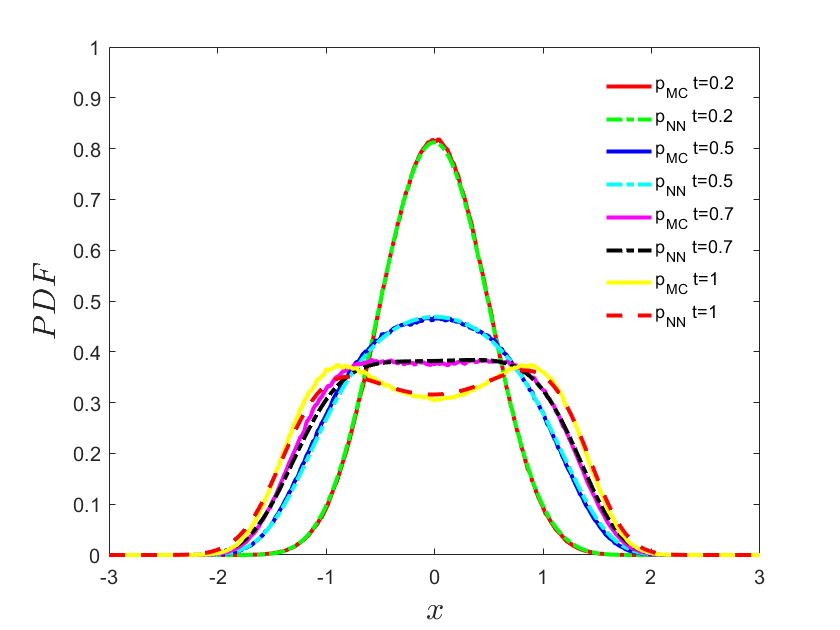}}
\end{minipage}
\vfill
\begin{minipage}[]{0.3 \textwidth}
 \leftline{~~~~~~~\tiny\textbf{(b)}}
\centerline{\includegraphics[width=4cm]{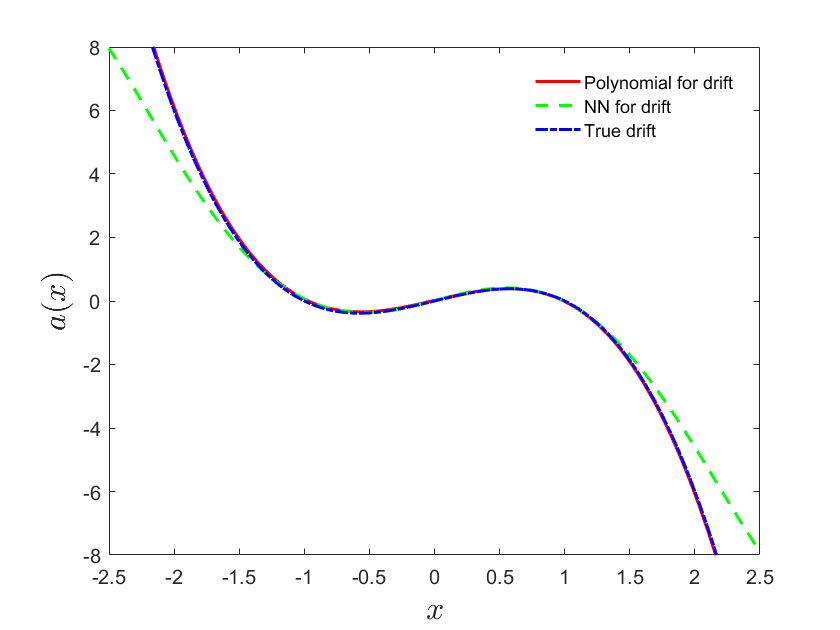}}
\end{minipage}
\hfill
\begin{minipage}[]{0.3 \textwidth}
 \leftline{~~~~~~~\tiny\textbf{(b1)}}
\centerline{\includegraphics[width=4cm]{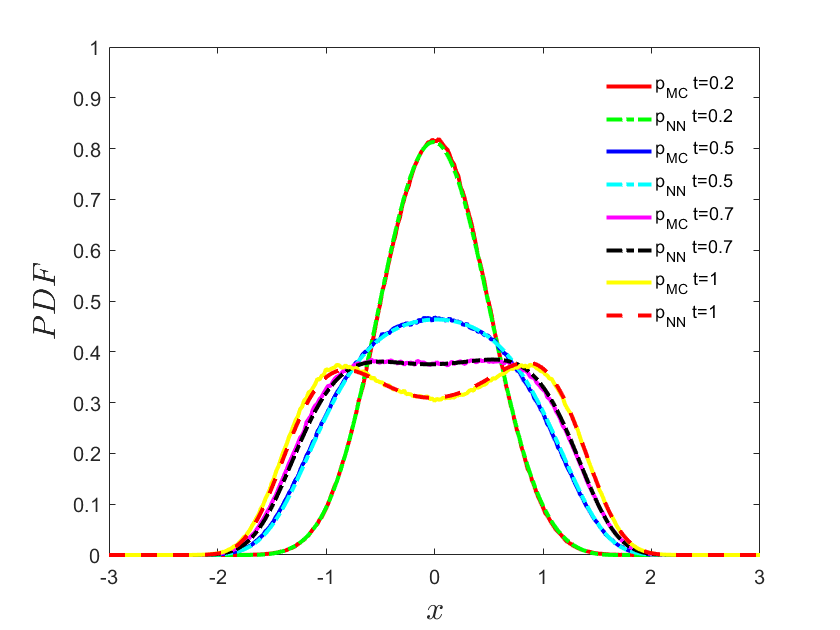}}
\end{minipage}
\hfill
\begin{minipage}[]{0.3 \textwidth}
 \leftline{~~~~~~~\tiny\textbf{(b2)}}
\centerline{\includegraphics[width=4cm]{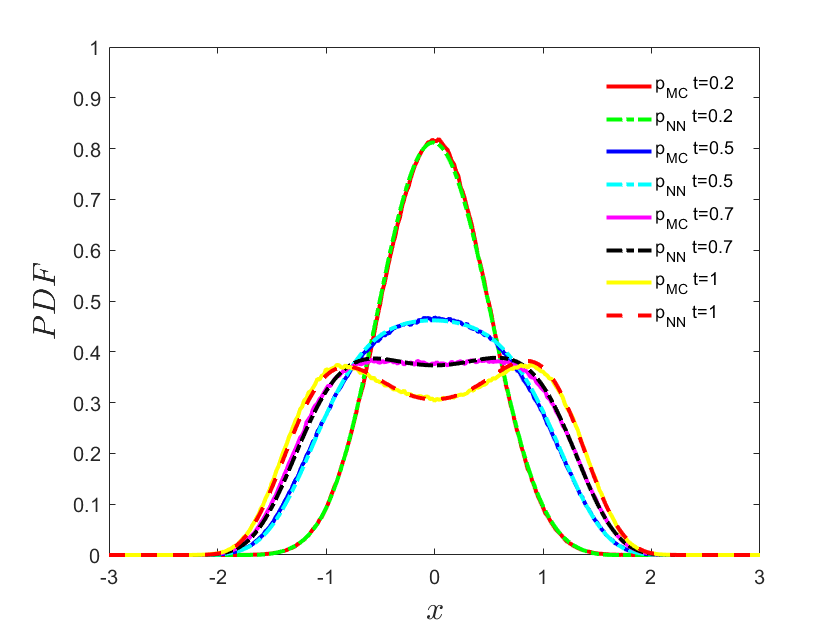}}
\end{minipage}
\caption{\textbf{Appendix B - Inverse problem II: Brownian noise: } Drift term and PDF inference for $a(x)=x-x^3$ and $\sigma=1$ with 10,000 samples of $X_t$ at different times: (a) Observation data at $t=0.1,0.3,0.5$; (b) Observation data at $t=0.2,0.5,1$; (a1) Observation data at $t=0.1,0.3,0.5$ and use of polynomial fit to learn the drift term; (a2) Observation data at $t=0.1,0.3,0.5$ and use of neural network to learn the drift term; (b1) Observation data at $t=0.2,0.5,1$ and use of polynomial fit to learn the drift term; (b2) Observation data at $t=0.2,0.5,1$ and use of neural network to learn the drift term.}
\label{f1-inverse-1d-BM}
\end{figure}

The results are shown in Figure \ref{f1-inverse-1d-BM}. we can see that when the observation data is available at $t=0.1,0.3,0.5$, we can still learn the drift term well, but the error of the drift term is larger than the case when the observation data is available at $t=0.2,0.5,1$. For the PDF, the small observation data can also capture the two peaks of the PDF. The corresponding parameter estimation is shown in Table \ref{tab:1d-bm}. We can see that the error for case (a) is larger than the error of case (b). 
%

\begin{table*}[ht]
\scriptsize
\begin{center}
\caption{Appendix B. Parameter estimation for inverse problem II - Brownian noise.}
\begin{tabular}{ c cc cc cc cc cc c cc cc cc cc ccc c}
\hline
& Parameter           & $\lambda_0$  & $\lambda_1$  &$ \lambda_2$ & $\lambda_3$ &$ \sigma$\\[1ex]
& True parameter &$0$  &$1$  &$0$  &$-1$  &$1$         \\[1ex]
& Case (a1)     &$0.0051$  &$0.8422$  &$-0.0071$  &$-0.8994$  &$1.0347$         \\[1ex]
& Case (a2)     &$*$  &$*$  &$*$  &$*$  &$0.9830$         \\[1ex]
& Case (b1)  &$0.0225$  &$0.9638$  &$-0.0010$  &$-1.0035$  &$1.0138$         \\[1ex]
& Case (b2)  &$*$  &$*$  &$*$  &$*$  &$1.0283$         \\[1ex]
\hline
\end{tabular}\label{tab:1d-bm}
\end{center}
\end{table*}

\section{Problem II: L\'{e}vy noise}
\begin{figure}[ht]
\begin{minipage}[]{0.3 \textwidth}
 \leftline{~~~~~~~\tiny\textbf{(a)}}
\centerline{\includegraphics[width=4cm]{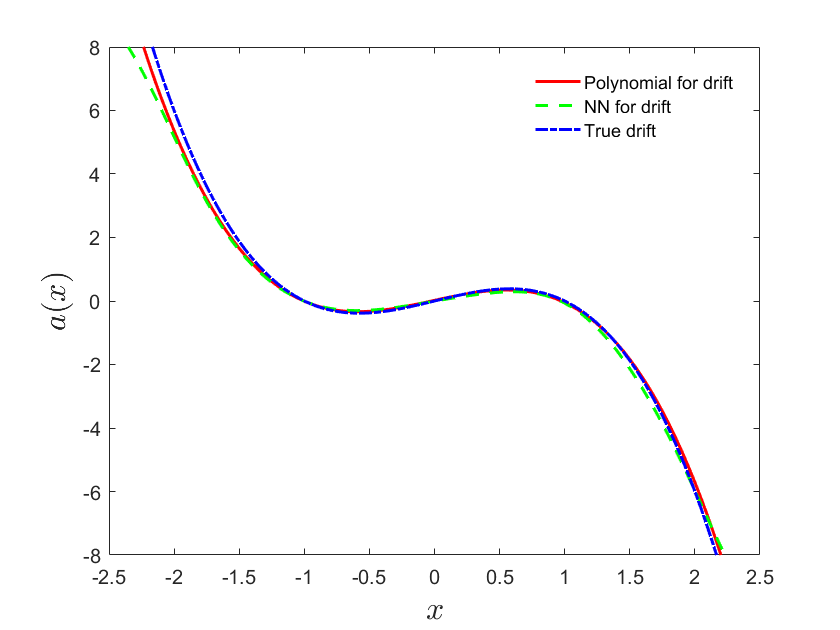}}
\end{minipage}
\hfill
\begin{minipage}[]{0.3 \textwidth}
 \leftline{~~~~~~~\tiny\textbf{(a1)}}
\centerline{\includegraphics[width=4cm]{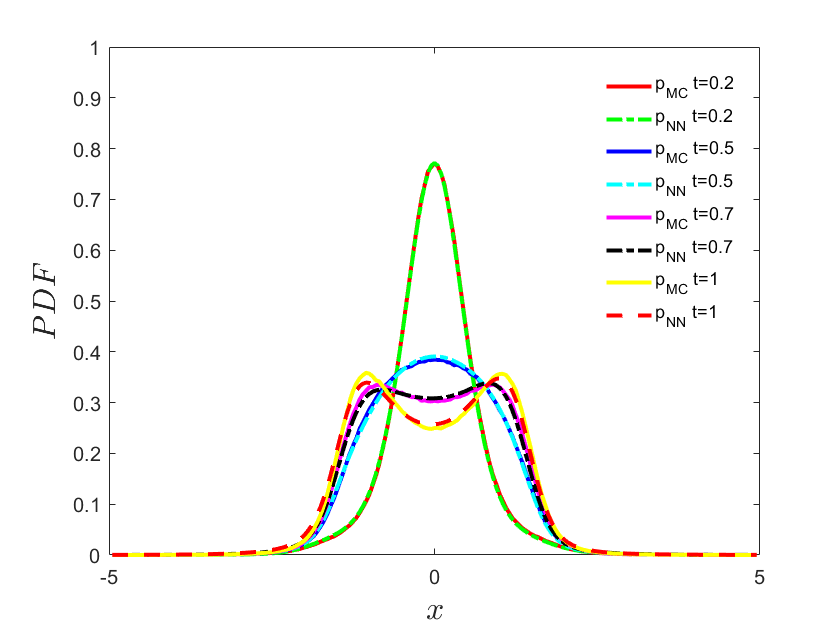}}
\end{minipage}
\hfill
\begin{minipage}[]{0.3 \textwidth}
 \leftline{~~~~~~~\tiny\textbf{(a2)}}
\centerline{\includegraphics[width=4cm]{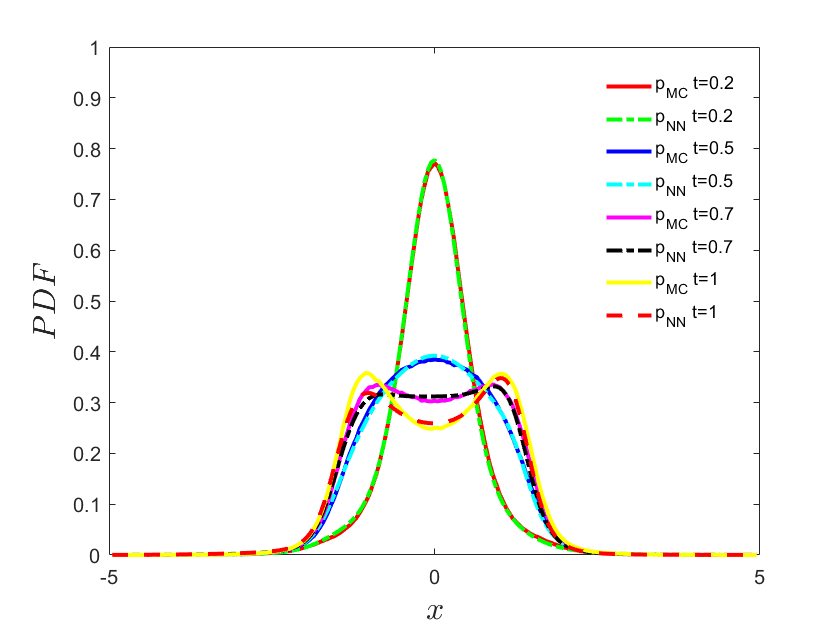}}
\end{minipage}
\vfill
\begin{minipage}[]{0.3 \textwidth}
 \leftline{~~~~~~~\tiny\textbf{(b)}}
\centerline{\includegraphics[width=4cm]{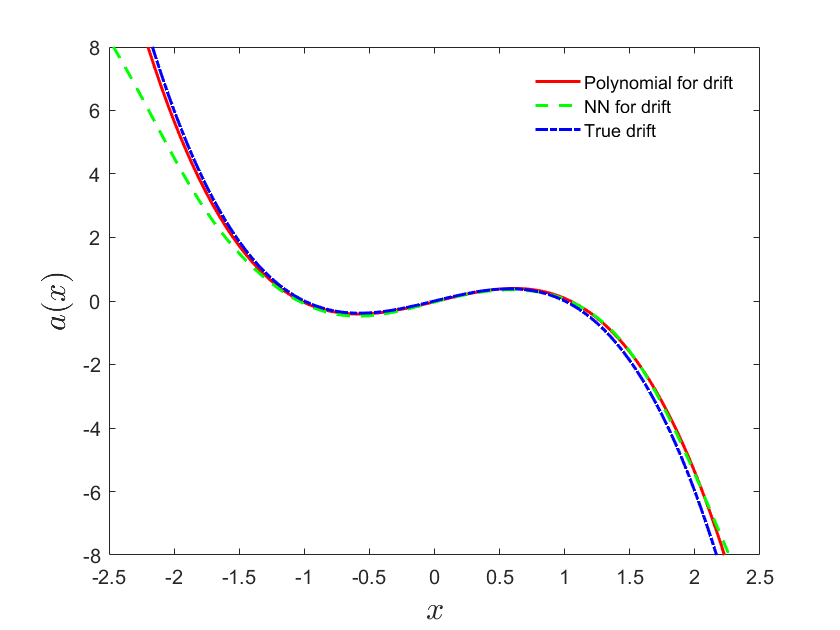}}
\end{minipage}
\hfill
\begin{minipage}[]{0.3 \textwidth}
 \leftline{~~~~~~~\tiny\textbf{(b1)}}
\centerline{\includegraphics[width=4cm]{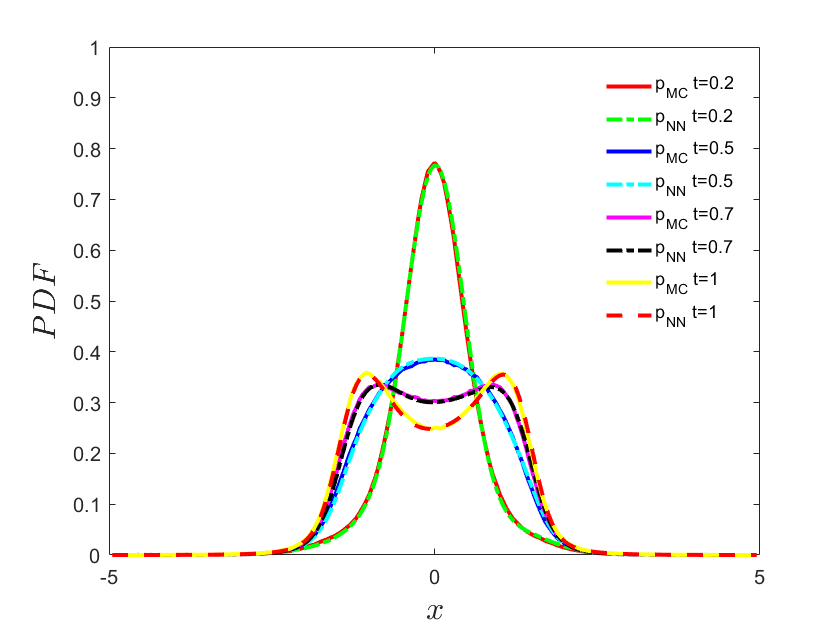}}
\end{minipage}
\hfill
\begin{minipage}[]{0.3 \textwidth}
 \leftline{~~~~~~~\tiny\textbf{(b2)}}
\centerline{\includegraphics[width=4cm]{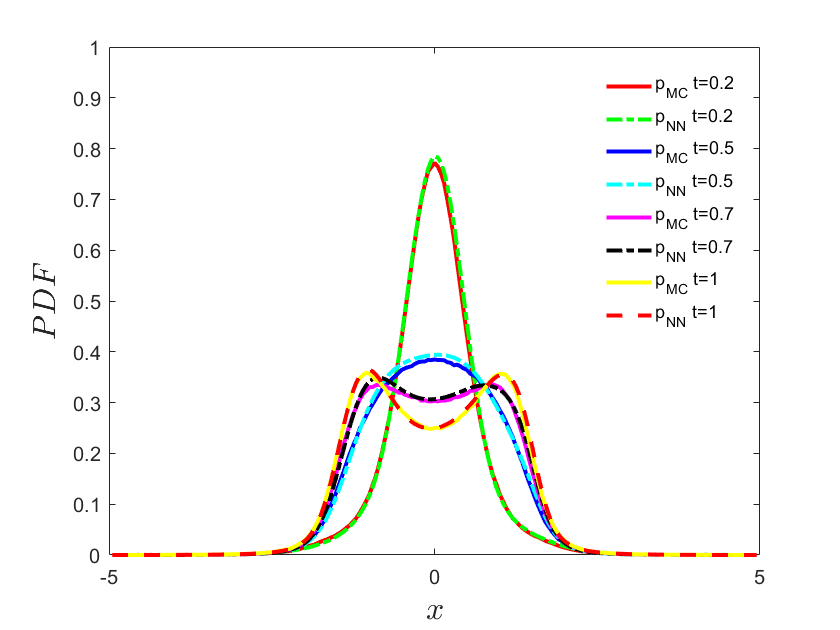}}
\end{minipage}
\caption{\textbf{Appendix C. Inverse problem II - L\'{e}vy noise:} Drift term and PDF inference for $a(x)=x-x^3$ and $\varepsilon=1$ with 10000 samples of $X_t$ at different observation times. (a) Observation data at $t=0.1,0.3,0.5$; (a1) Observation data at $t=0.1,0.3,0.5$ and use of polynomial fit to learn the drift term; (a2) Observation data at $t=0.1,0.3,0.5$ and use of neural network to learn the drift term; (b) Observation data at $t=0.2,0.5,1$; (b1) Observation data at $t=0.2,0.5,1$ and use of polynomial fit to learn the drift term; (b2) Observation data at $t=0.2,0.5,1$ and use of neural network to learn the drift term.}
\label{f1-inverse-1d-Levy}
\end{figure}

For the inverse problem II driven by L\'{e}vy noise, we compute four cases like in the Brownian noise. 
We assume we have $10,000$ available samples at each observation time, and set the minibatch size $b=1,000$ to train the first part in \eqref{eqn:loss_data}. We use the compound trapezoid formula with $201$ points to approximate the integral part and randomly choose $800$ residual points in \eqref{eqn:loss_pde}.
For case (a1), the observation data is available at $t=0.1,~0.3,~0.5$ and we use a polynomial to approximate the drift term. For case (a2), the observation data is available at $t=0.1,~0.3,~0.5$ and we use a neural network to approximate the drift term. For case (b1), the observation data is available at $t=0.2,~0.5,~1$ and we use a polynomial to approximate the drift term. For case (b2), the observation data is available at $t=0.2,~0.5,~1$ and we use a neural network to approximate the drift term. We infer the drift term and the PDF at different times in Figure \ref{f1-inverse-1d-Levy}. The parameter estimation results are shown in Table \ref{tab:1d-levy}.

\begin{table*}[ht]
\scriptsize
\begin{center}
\caption{Appendix C. Parameter estimation for inverse problem II - L\'{e}vy noise.}
\begin{tabular}{ c cc cc cc cc cc c cc cc cc cc ccc c}
\hline
& Parameter           & $\lambda_0$  & $\lambda_1$  &$ \lambda_2$ & $\lambda_3$ &$ \varepsilon$\\[1ex]
& True parameter &$0$  &$1$  &$0$  &$-1$  &$1$         \\[1ex]
& Case (a1)     &$0.0204$  &$0.8969$  &$-0.04876$  &$-0.9166$  &$0.9870$         \\[1ex]
& Case (a2)     &$*$  &$*$  &$*$  &$*$  &$1.0110$         \\[1ex]
& Case (b1)  &$-0.0208$  &$1.0099$  &$0.03828$  &$-0.9434$  &$0.9463$         \\[1ex]
& Case (b2)  &$*$  &$*$  &$*$  & $*$  & $0.9144$         \\[1ex]
\hline
\end{tabular}\label{tab:1d-levy}
\end{center}
\end{table*}

\section{Problem II: 2D Brownian noise}
We consider the 2D example with Brownian noise:
\begin{equation}
d\left( \begin{array}{ccc}
X_t\\
Y_t
\end{array}
\right )=
\left( \begin{array}{c}
a_1(x,y)\\
a_2(x,y)
\end{array}
\right )dt+\left[ \begin{array}{cc}
\sigma_x & 0 \\
0& \sigma_y\\
\end{array}
\right ]  d \left( \begin{array}{c}
B_{1,t}^{\alpha}\\
B_{2,t}^{\alpha}
\end{array}
\right )    
\end{equation}
where $a_1(x,y)=x-x^3$, $a_2(x,y)=y-y^3$, and $\sigma_x=\sigma_y=1$.

we use the compound trapezoid formula with $301\times 301$ points to approximate the integral part of \eqref{eqn:loss_data} and $5000$ points for the residual of the FP equation.
We present four cases to compute the inverse problem. 
For case (a), the observation data is available at $t=0.1,~0.4,~0.7,~1$ and the drift term is approximated by polynomials.
For case (b), the observation data is available at $t=0.1,~0.4,~0.7,~1$ and the drift term is approximated by neural networks.
For case (c), the observation data is
available at $t=0.1,~0.3,~0.5,~0.7,~1$ and the drift term is approximated by polynomials.
For case (d), the observation data is
available at $t=0.1,~0.3,~0.5,~0.7,~1$ and the drift term is approximated by neural networks. For each time snapshot, we have $100,000$ samples, and set the minibatch size as $b=10,000$ for the first part in \eqref{eqn:loss_data}.

\begin{figure}[H]
\begin{minipage}[]{0.45 \textwidth}
 \leftline{~~~~~~~\tiny\textbf{(a)}}
\centerline{\includegraphics[width=5.8cm]{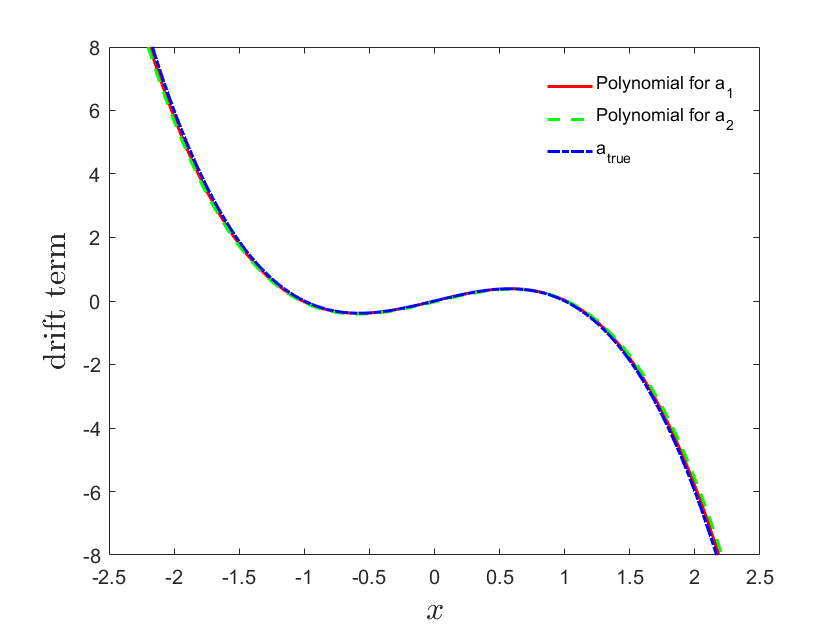}}
\end{minipage}
\hfill
\begin{minipage}[]{0.45 \textwidth}
 \leftline{~~~~~~~\tiny\textbf{(b)}}
\centerline{\includegraphics[width=5.8cm]{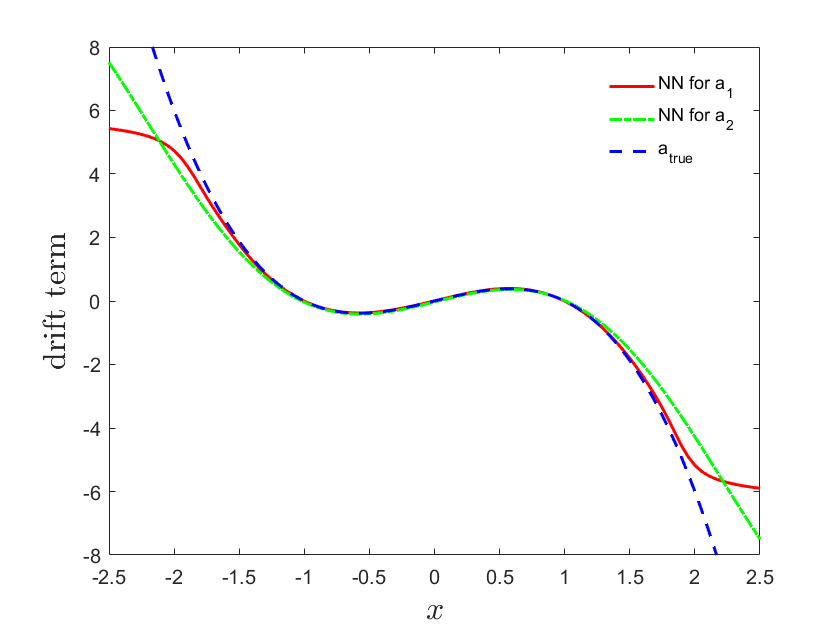}}
\end{minipage}
\vfill
\begin{minipage}[]{0.45 \textwidth}
 \leftline{~~~~~~~\tiny\textbf{(c)}}
\centerline{\includegraphics[width=5.8cm]{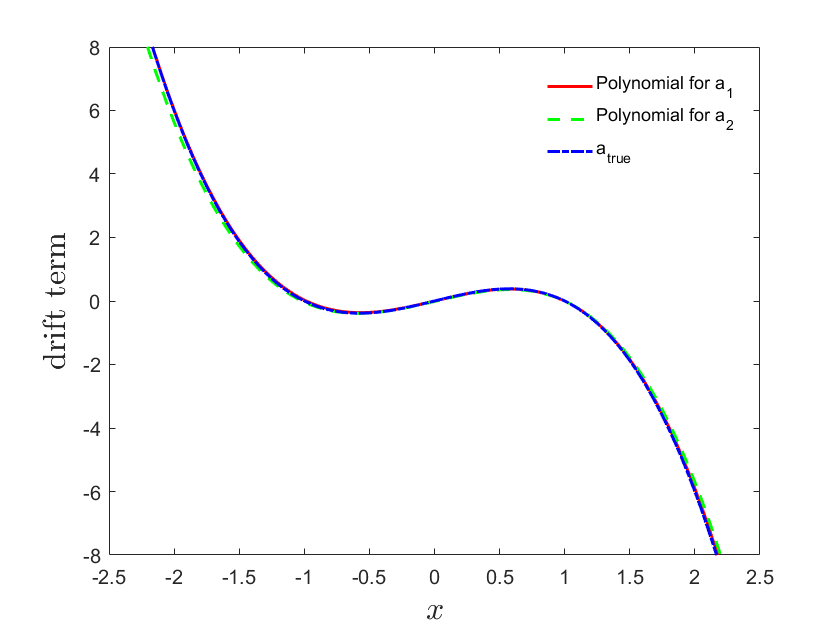}}
\end{minipage}
\hfill
\begin{minipage}[]{0.45 \textwidth}
 \leftline{~~~~~~~\tiny\textbf{(d)}}
\centerline{\includegraphics[width=5.8cm]{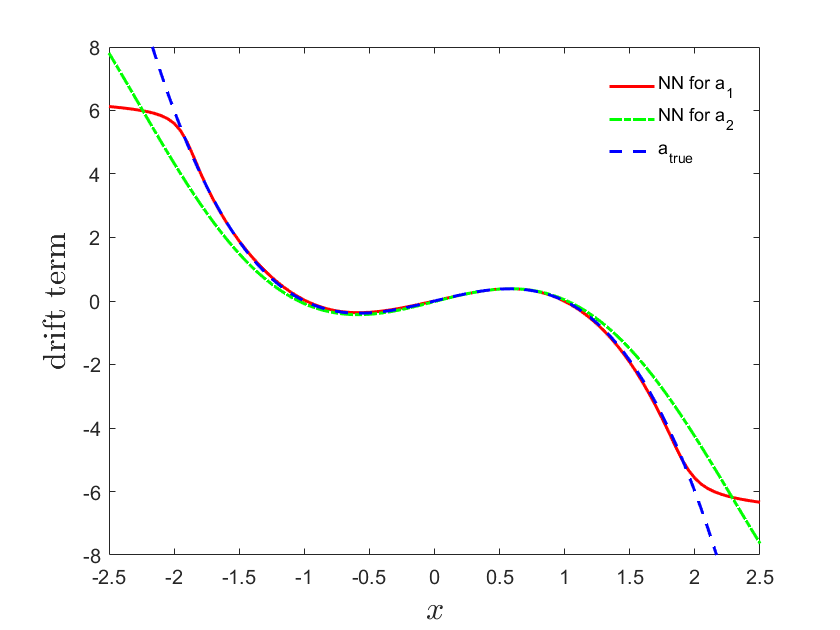}}
\end{minipage}
\caption{\textbf{Appendix D. Inverse problem II - 2D Brownian noise:} Learning the drift term. (a) Observation data at $t=0.1,0.4,0.7,1$ and polynomial fit; (b) observation data at $t=0.1,0.4,0.7,1$ and neural network; (c) observation data at $t=0.1,0.3,0.5,0.7,1$ and polynomial fit; (d) observation data at $t=0.1,0.3,0.5,0.7,1$.}
\label{f1-inverse-2d-BM}
\end{figure}

\begin{figure}[H]
\centerline{\includegraphics[width=6cm,height=5cm]{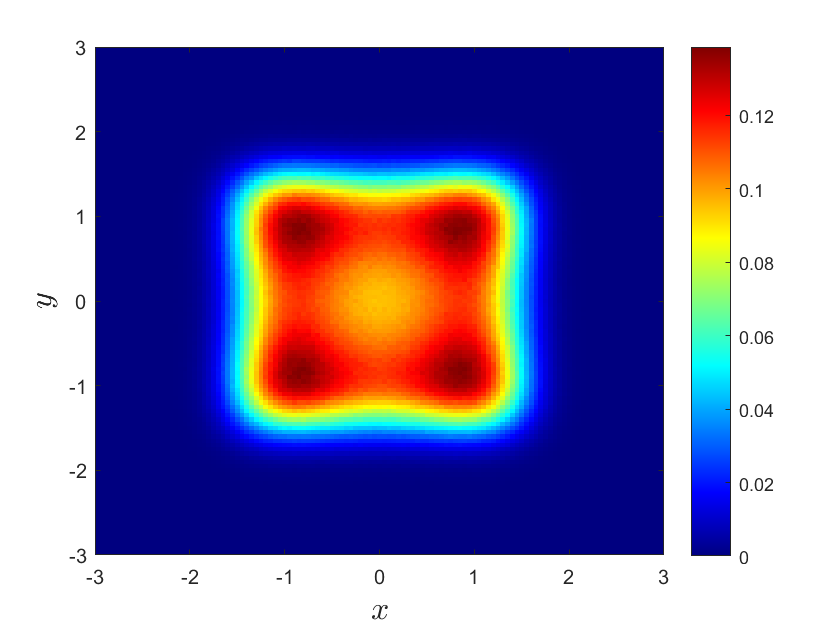}}
\begin{minipage}[]{0.45 \textwidth}
 \leftline{~~~~~~~\tiny\textbf{(a)}}
\centerline{\includegraphics[width=6cm,height=5cm]{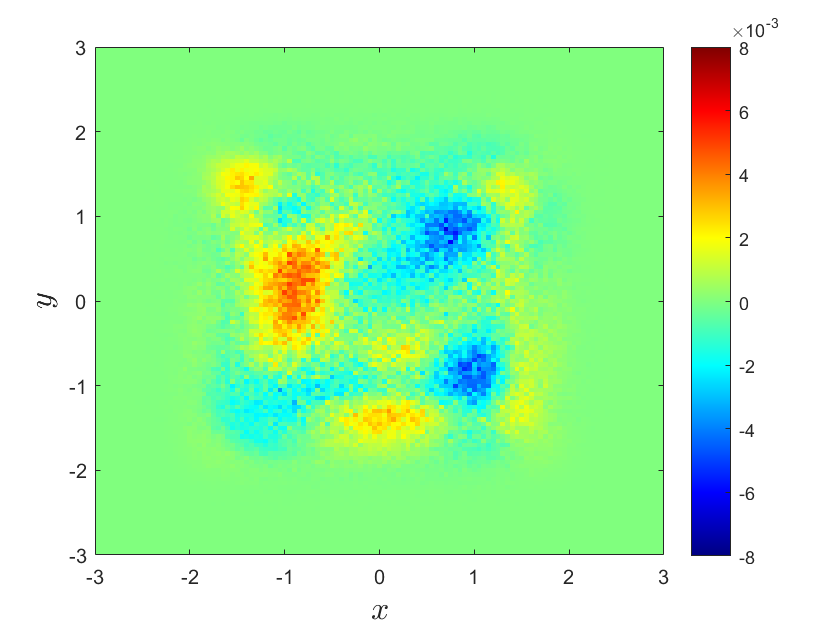}}
\end{minipage}
\hfill
\begin{minipage}[]{0.45 \textwidth}
 \leftline{~~~~~~~\tiny\textbf{(b)}}
\centerline{\includegraphics[width=6cm,height=5cm]{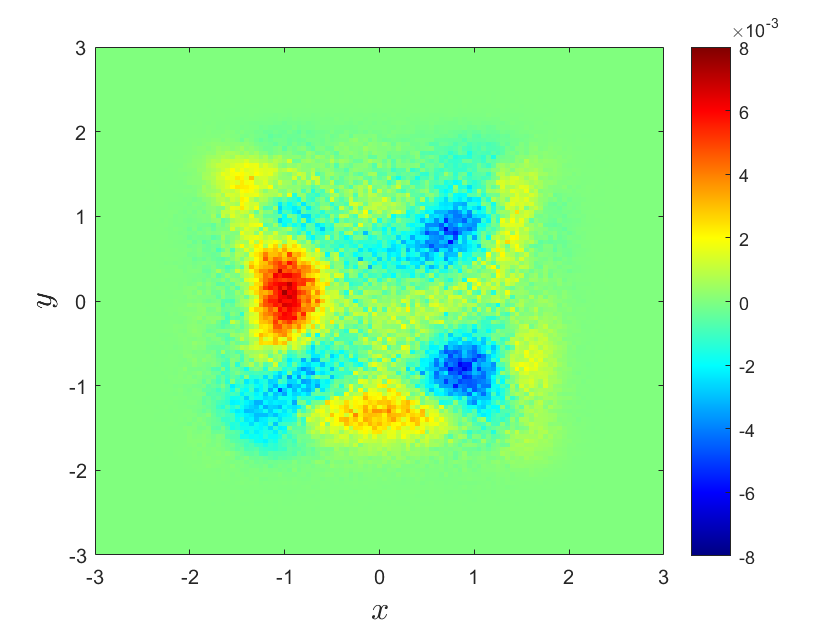}}
\end{minipage}
\begin{minipage}[]{0.45 \textwidth}
 \leftline{~~~~~~~\tiny\textbf{(c)}}
\centerline{\includegraphics[width=6cm,height=5cm]{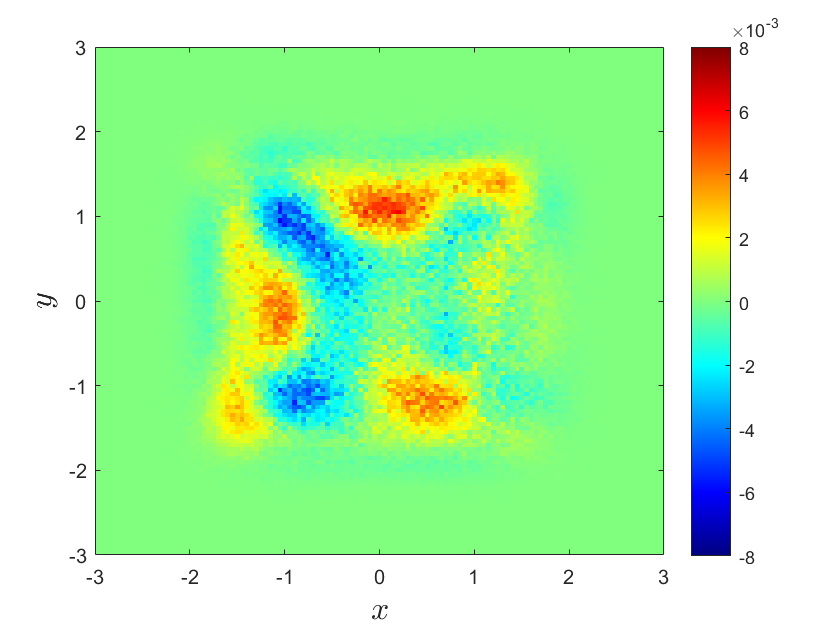}}
\end{minipage}
\hfill
\begin{minipage}[]{0.45 \textwidth}
 \leftline{~~~~~~~\tiny\textbf{(d)}}
\centerline{\includegraphics[width=6cm,height=5cm]{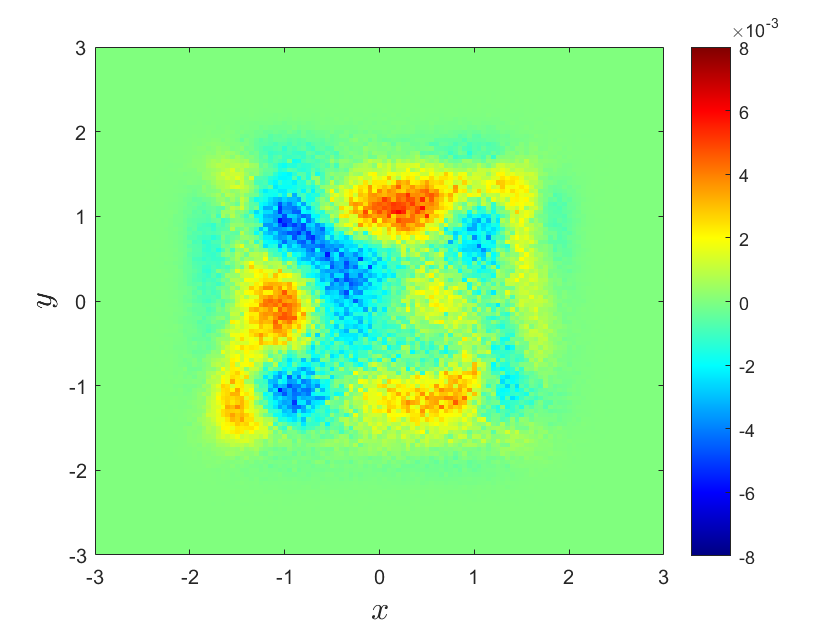}}
\end{minipage}
\caption{\textbf{Appendix D. Inverse problem II - 2D Brownian noise:} Top: PDF solution using the MC method when $t=1$; Case a: PDF for case a; Case b: PDF error for case b; Case c: PDF error for case c; Case d: PDF for case d. }
\label{f2-inverse-2d-BM-error}
\end{figure}

The predictions for the drift terms are shown in Figure \ref{f1-inverse-2d-BM}. We use the MC method to obtain the reference PDF and the results are shown in top of Figure \ref{f2-inverse-2d-BM-error}. Specifically, we compare the error results of PDF at $t=1$ for four cases in Figure \ref{f2-inverse-2d-BM-error}. The parameters we learn are shown in  Table \ref{tab:2d-bm}. We can see that the results of case (a) are better than case (b), while the results of case (c) are better than case (d), especially for the drift term.

\begin{table*}[ht]
\scriptsize
\begin{center}
\caption{Appendix D. Parameter estimation for inverse problem II of 2D Brownian.}
\begin{tabular}{ c cc cc cc cc cc c cc cc cc cc ccc c}
\hline
& Parameter           & $\lambda_{0x} $ /$\lambda_{0y} $ & $\lambda_{1x}$/$\lambda_{1y} $  &$ \lambda_{2x}$ /$\lambda_{2y} $& $\lambda_{3x}$ /$\lambda_{3y} $&$ \sigma_{1}$/$\sigma_{2} $\\[1ex]
& True parameter &$0$  &$1$  &$0$  &$-1$  &$1$      \\[1ex]
& Case (a)     &$0.0022 $  &$0.9966$   &$-0.0049$  &$-0.9755$  &$0.9863$\\[1ex]
&      &$ -0.0264$  &$0.9948$  &$0.0162$ &$-0.9526 $ &$ 0.9731$      \\[1ex]
& Case (b)     &$*$  &$*$  &$*$  &$*$  &$0.9974$ \\[1ex]
&  &$*$  &$*$  &$*$  &$*$  &$0.9557$         \\[1ex]
& Case (c)  &$0.0003$  &$0.9752$  &$0.0140$  &$-0.9898$  &$1.0087$ \\[1ex]
&&$-0.0122$  &$0.9679$  &$-0.0067$  &$-0.9506$  &$0.9845$      \\[1ex]
& Case (d)  &$*$  &$*$  &$*$  &$*$  &$1.0063$  \\[1ex]
&&$*$  &$*$  &$*$  &$*$  &$0.9667$      \\[1ex]
\hline
\end{tabular}\label{tab:2d-bm}
\end{center}
\end{table*}

\section{Problem II: 2D L\'{e}vy noise}
~~~~We consider another example of 2D L\'{e}vy noise:
\begin{equation}
d\left( \begin{array}{ccc}
X_t\\
Y_t
\end{array}
\right )=
\left( \begin{array}{c}
a_1(x,y)\\
a_2(x,y)
\end{array}
\right )dt+\left[ \begin{array}{cc}
\varepsilon_x & 0 \\
0& \varepsilon_y\\
\end{array}
\right ]  d \left( \begin{array}{c}
L_{1,t}^{\alpha}\\
L_{2,t}^{\alpha}
\end{array}
\right )    
\end{equation}
where $a_1(x,y)=x-x^3$, $a_2(x,y)=y-y^3$, and $\varepsilon_x=\varepsilon_y=1$.

\begin{figure}[H]
\begin{minipage}[]{0.45 \textwidth}
 \leftline{~~~~~~~\tiny\textbf{(a)}}
\centerline{\includegraphics[width=5.8cm]{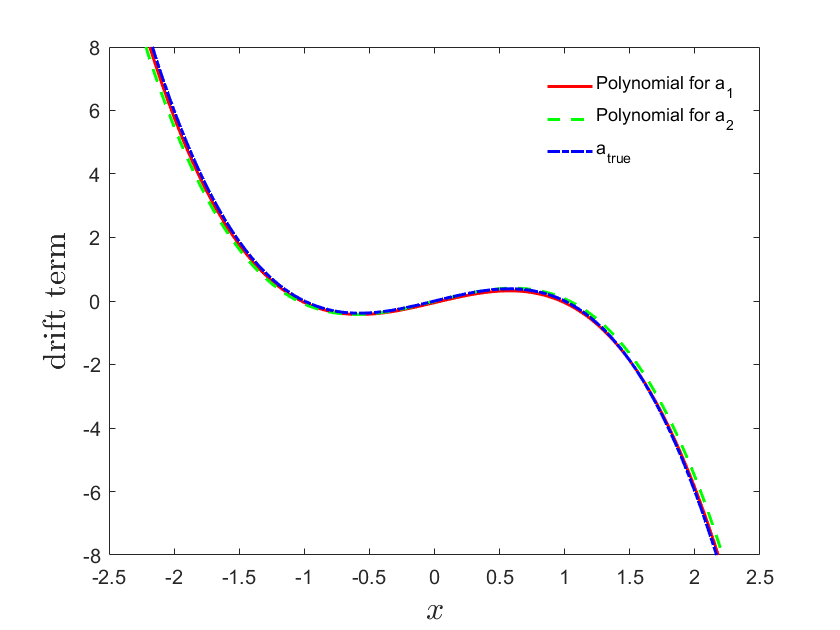}}
\end{minipage}
\hfill
\begin{minipage}[]{0.45 \textwidth}
 \leftline{~~~~~~~\tiny\textbf{(b)}}
\centerline{\includegraphics[width=5.8cm]{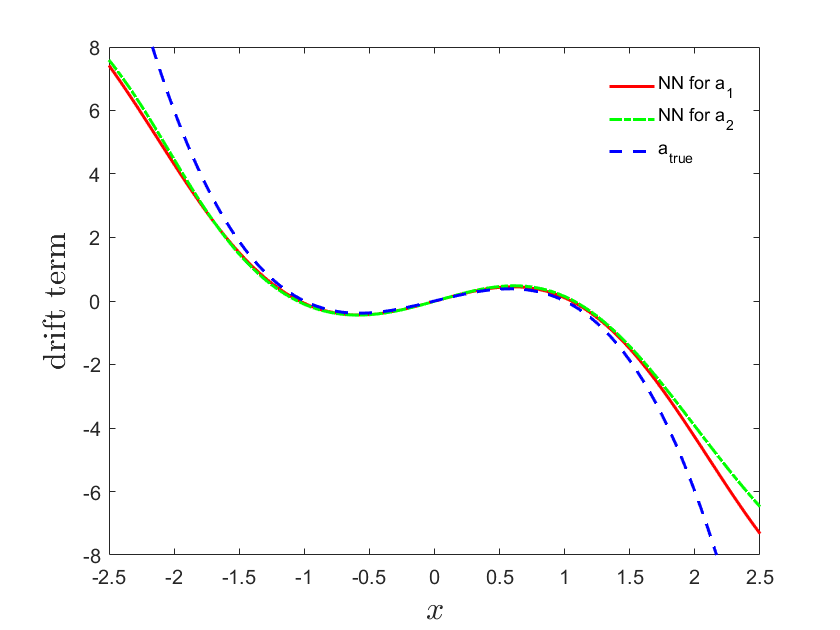}}
\end{minipage}
\hfill
\begin{minipage}[]{0.45 \textwidth}
 \leftline{~~~~~~~\tiny\textbf{(c)}}
\centerline{\includegraphics[width=5.8cm]{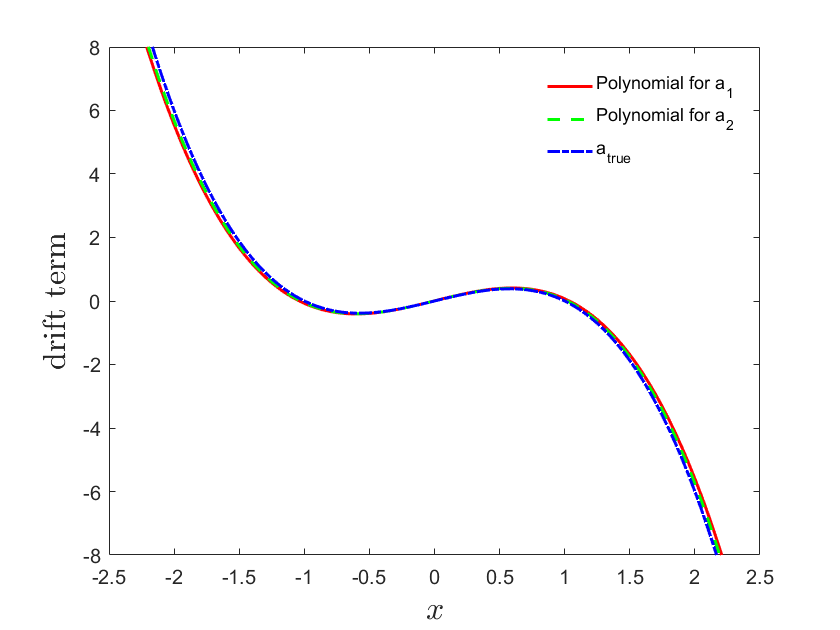}}
\end{minipage}
\hfill
\begin{minipage}[]{0.45 \textwidth}
 \leftline{~~~~~~~\tiny\textbf{(d)}}
\centerline{\includegraphics[width=5.8cm]{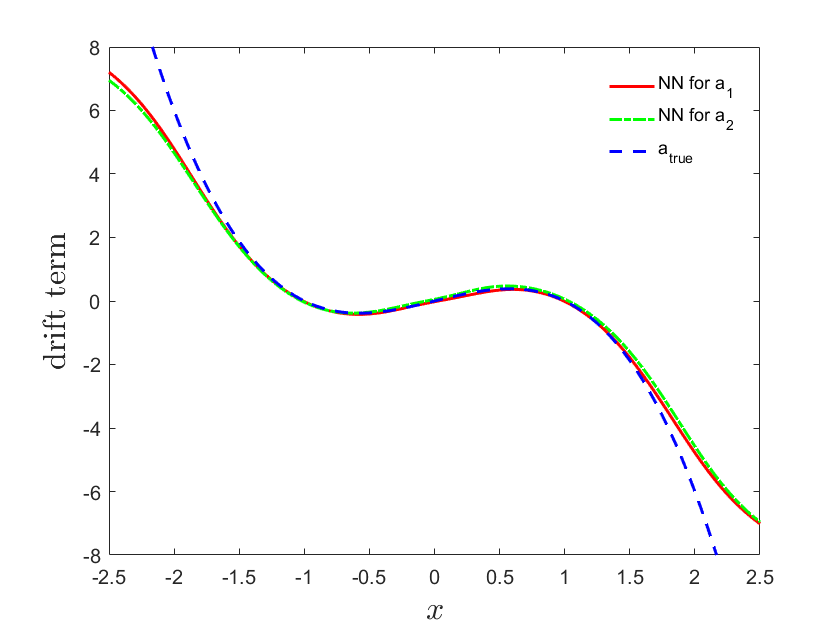}}
\end{minipage}
\caption{\textbf{Appendix E. Inverse problem II - 2D L\'{e}vy noise:} Learning the drift term from
observation data at: (a) $t=0.1,0.4,0.7,1$ and using polynomial fit; (b) $t=0.1,0.4,0.7,1$ and using two neural networks; (c) $t=0.1,0.3,0.5,0.7,1$ and using polynomial fit; (d) $t=0.1,0.3,0.5,0.7,1$ and using two neural networks.}
\label{f1-inverse-2d-Levy}
\end{figure}

\begin{figure}[H]
\centerline{\includegraphics[width=6cm,height=5cm]{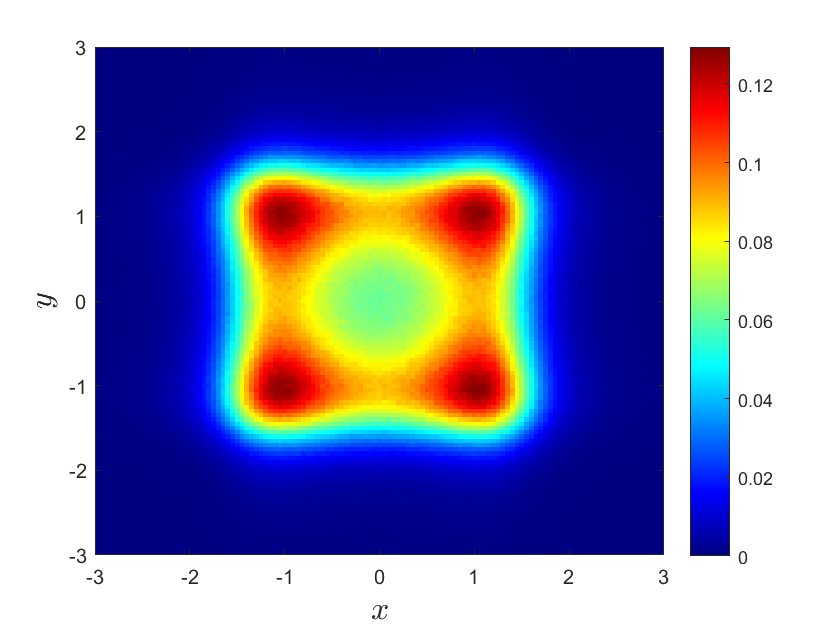}}
\begin{minipage}[]{0.45 \textwidth}
 \leftline{~~~~~~~\tiny\textbf{(a)}}
\centerline{\includegraphics[width=5.8cm]{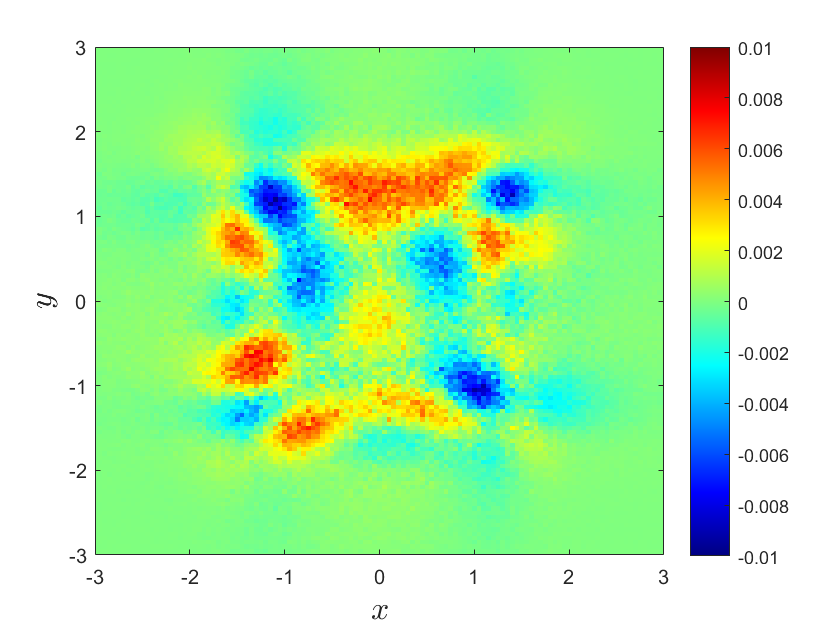}}
\end{minipage}
\hfill
\begin{minipage}[]{0.45 \textwidth}
 \leftline{~~~~~~~\tiny\textbf{(b)}}
\centerline{\includegraphics[width=5.8cm]{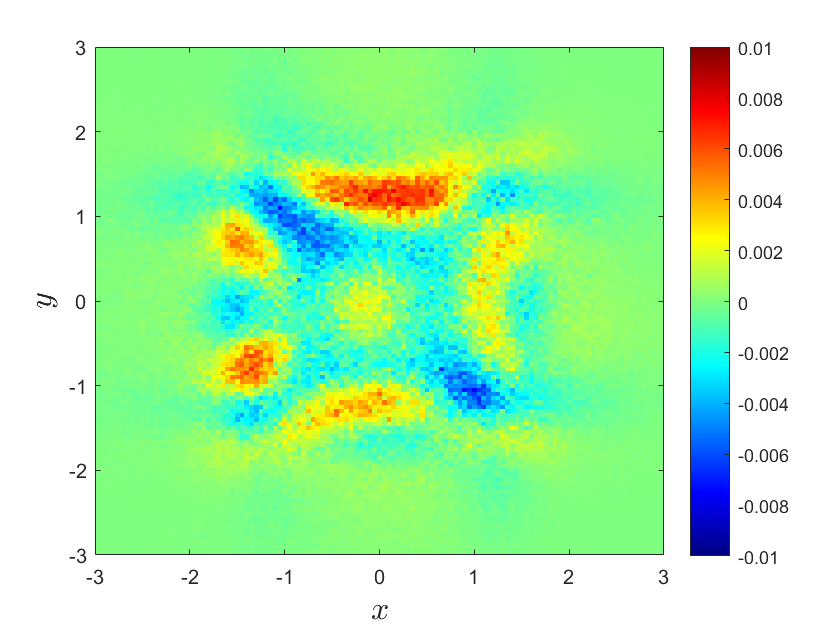}}
\end{minipage}
\hfill
\begin{minipage}[]{0.45 \textwidth}
 \leftline{~~~~~~~\tiny\textbf{(c)}}
\centerline{\includegraphics[width=5.8cm]{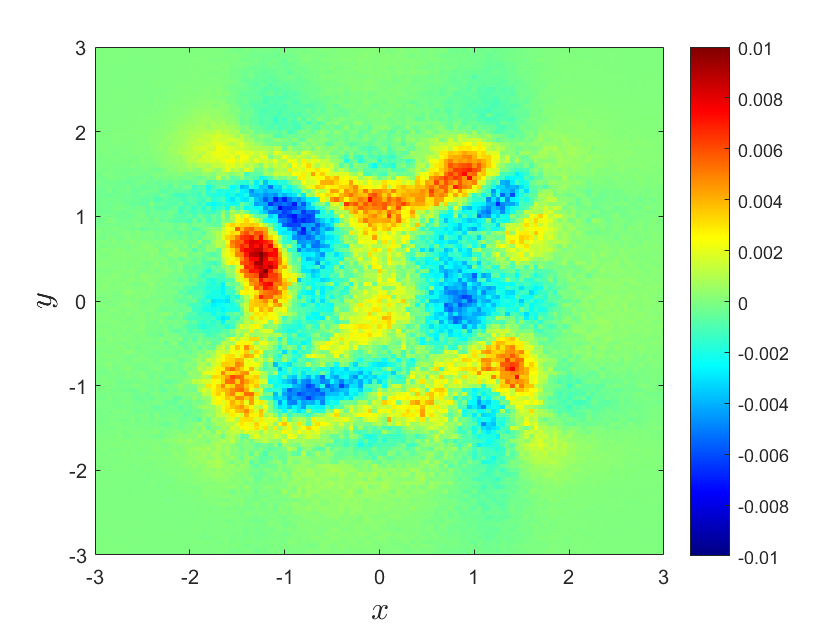}}
\end{minipage}
\hfill
\begin{minipage}[]{0.45 \textwidth}
 \leftline{~~~~~~~\tiny\textbf{(d)}}
\centerline{\includegraphics[width=5.8cm]{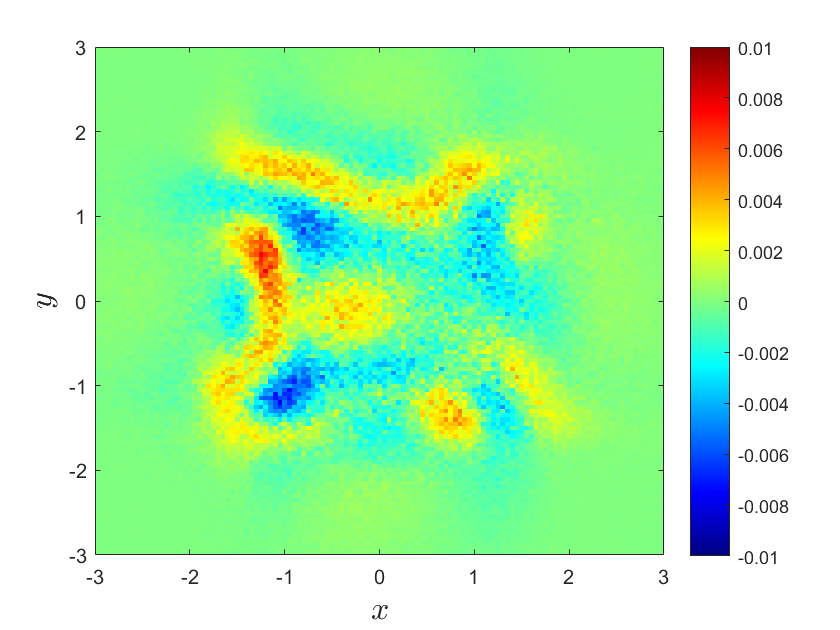}}
\end{minipage}
\caption{\textbf{Appendix E. Inverse problem II - 2D L\'{e}vy noise:} Learning the drift term. Top: Reference PDF when $t=1$. Observation data at: (a) $t=0.1,0.4,0.7,1$ and using polynomial fit; (b) $t=0.1,0.4,0.7,1$ and using a neural network; (c) $t=0.1,0.3,0.5,0.7,1$ and using polynomial fit; (d)  $t=0.1,0.3,0.5,0.7,1$ and using a neural network.}
\label{f2-inverse-2d-Levy-error}
\end{figure}

We use the compound trapezoid formula with $201\times 201$ points to approximate the integral part of \eqref{eqn:loss_data} and $160,000$ residual points for the FP equation.
For each time snapshot, we have $100,000$ samples, and set the minibatch size as $b=5,000$ for the first part in \eqref{eqn:loss_data}. For case (a), the observation data is available at $t=0.1,~0.4,~0.7,~1$ and we use polynomials to approximate the drift term. For case (b), the observation data is available at $t=0.1,~0.4,~0.7,~1$ and we use two  neural networks to approximate the drift term. For case (c), the observation data is available at $t=0.1,~0.3,~0.5,~0.7,~1$  and we use polynomials to approximate the drift term. For case (d), the observation data is 
available at $t=0.1,~0.3,~0.5,~0.7,~1$ and we use two neural networks to approximate the drift term. The results are shown in Figure \ref{f1-inverse-2d-Levy}, \ref{f2-inverse-2d-Levy-error} and Table \ref{tab:2d-levy}. 

\begin{table*}[ht]
\scriptsize
\begin{center}
\caption{Appendix E. Parameter estimation for inverse problem II of 2D L\'{e}vy case II.}
\begin{tabular}{ c cc cc cc cc cc c cc cc cc cc ccc c}
\hline
& Parameter           & $\lambda_{0x}$ /$\lambda_{0y}$ & $\lambda_{1x}$ /$\lambda_{1y}$ &$ \lambda_{2x}$ /$\lambda_{2y}$& $\lambda_{3x}$/$\lambda_{3y}$ &
$ \varepsilon_{x}$/$\varepsilon_{y}$\\[1ex]
& True parameter &$0$  &$1$  &$0$  &$-1$  &$1$   \\[1ex]
& Case (a)     &$-0.0514$  &$0.9699$  &$0.0004$  &$-0.9726$  &$0.9836$   \\[1ex]
& &$-0.0090$  &$1.0300$  &$-0.0033$  &$-0.9462$  &$0.9187$      \\[1ex]
& Case (b)     &$*$  &$*$  &$*$  &$*$  &$0.9353$   \\[1ex]
&  &$*$  &$*$  &$*$  &$*$  &$0.9358$      \\[1ex] 
& Case (c)     &$0.0045$  &$1.0246$  &$-0.0070$  &$-0.9528$  &$0.9291$ \\[1ex]
& &$0.0018$  &$1.0147$  &$-0.0091$  &$-0.9684$   &$0.9614$     
\\[1ex]
& Case (d)     &$*$  &$*$  &$*$  &$*$  &$0.9948$    \\[1ex]
& &$*$  &$*$  &$*$  &$*$   &$0.9601$      \\[1ex]
\hline
\end{tabular}\label{tab:2d-levy}
\end{center}
\end{table*}

\section{Problem II: 3D and 4D Brownian noise}
We consider the 3D and 4D case of \eqref{stomodel-5D}, i.e., $n=3$ and $n = 4$. Similar as in case (b) in Section~\ref{sec:high-dim}, we use a cubic polynomial to parameterize the drift. 

\begin{figure}[H]
\begin{minipage}[]{0.45 \textwidth}
 \leftline{~~~~~~~\tiny\textbf{(a1)}}
\centerline{\includegraphics[width=5.8cm]{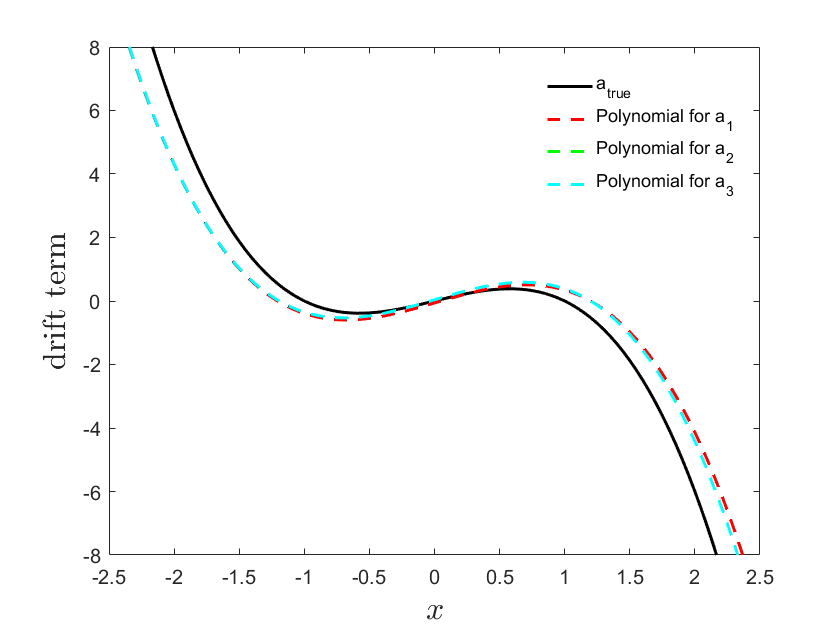}}
\end{minipage}
\hfill
\begin{minipage}[]{0.45 \textwidth}
 \leftline{~~~~~~~\tiny\textbf{(a2)}}
\centerline{\includegraphics[width=5.8cm]{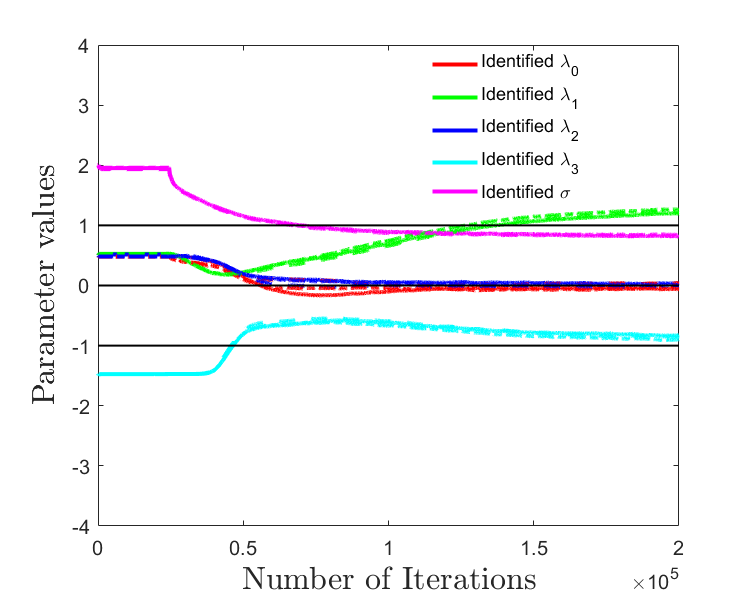}}
\end{minipage}
\begin{minipage}[]{0.45 \textwidth}
 \leftline{~~~~~~~\tiny\textbf{(b1)}}
\centerline{\includegraphics[width=5.8cm]{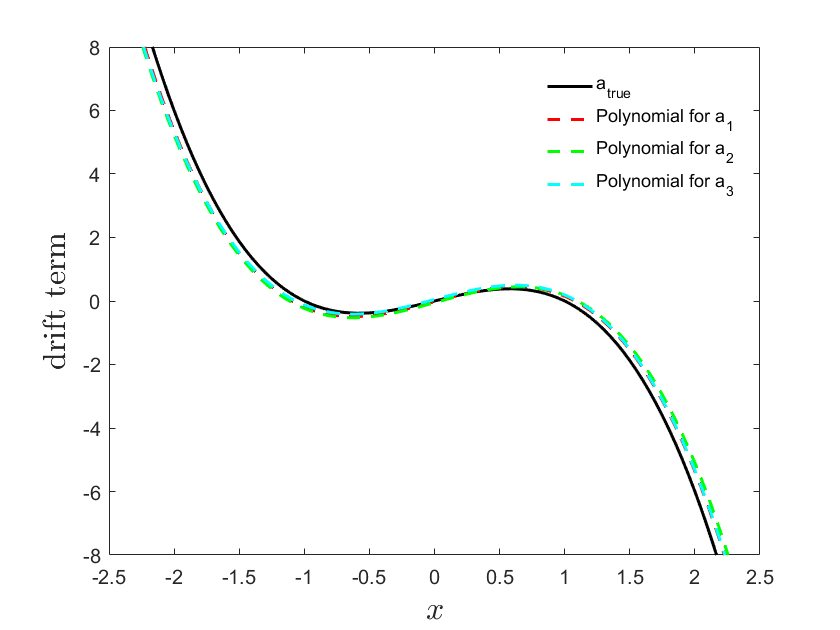}}
\end{minipage}
\hfill
\begin{minipage}[]{0.45 \textwidth}
 \leftline{~~~~~~~\tiny\textbf{(b2)}}
\centerline{\includegraphics[width=5.8cm]{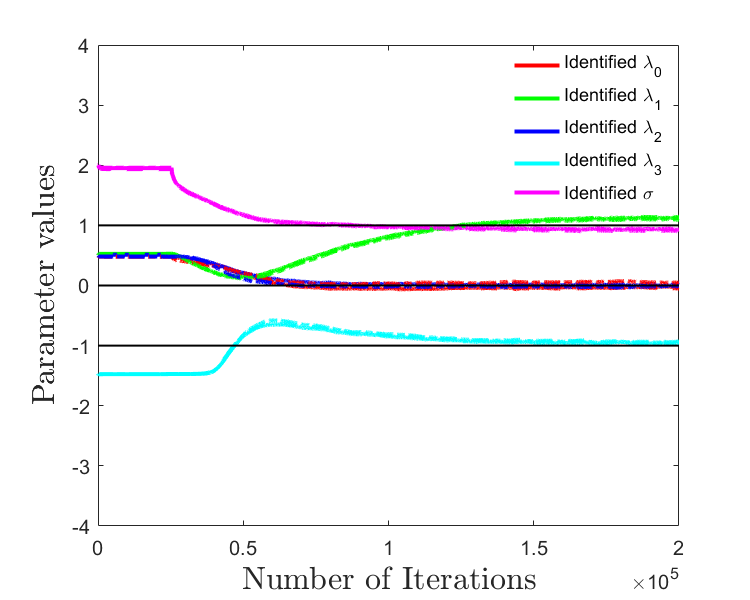}}
\end{minipage}
\caption{ \textbf{Appendix F. Inverse problem II - 3D Brownian noise:} Parameter evolution versus iteration number for two different observation data sets. Here $\lambda_{i}$ denotes the parameters $\lambda_{ix1},\lambda_{ix2},\lambda_{ix3},\lambda_{ix4}$, and $\sigma$ denotes the parameters $\sigma_{x1},\sigma_{x2},\sigma_{x3},,\sigma_{x4}$. Learning the drift term from observation data at (a) $t=0.1,0.3,0.5,0.7,1$; (b) $t=0.1,0.2,0.3,0.5,0.7,0.9,1$.}
\label{f-3d-drift}
\end{figure}

For the 3D case, we consider having observations at $t=0.1,0.3,0.5,0.7,0.9$, i.e., $N=5$, or $t=0.1,0.2,0.3,0.5,0.7,0.9,1$, i.e., $N=7$.  For each time snapshot, we have $100,000$ samples available and set the minibatch size as $10,000$. For the integral part of \eqref{eqn:loss_data}, the compound trapezoid formula with $101\times 101\times 101$ points are used. We use $5,000$ residual points for \eqref{eqn:loss_pde}. The results are shown in Figure \ref{f-3d-drift}. 

\begin{figure}[H]
\begin{minipage}[]{0.45 \textwidth}
 \leftline{~~~~~~~\tiny\textbf{(a1)}}
\centerline{\includegraphics[width=5.8cm]{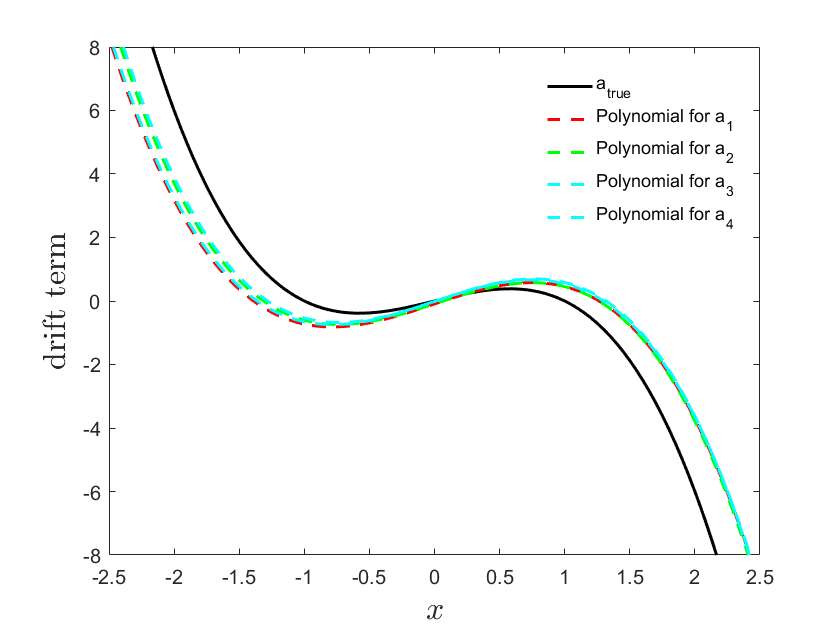}}
\end{minipage}
\hfill
\begin{minipage}[]{0.45 \textwidth}
 \leftline{~~~~~~~\tiny\textbf{(a2)}}
\centerline{\includegraphics[width=5.8cm]{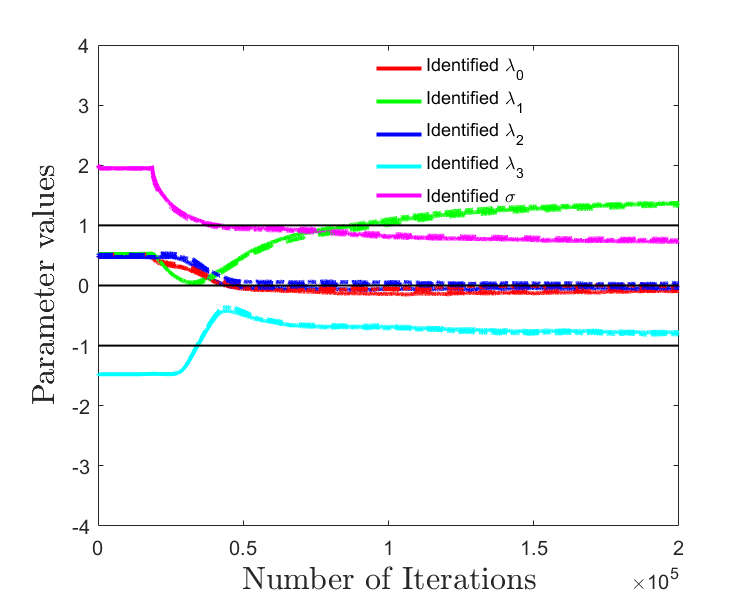}}
\end{minipage}
\hfill
\begin{minipage}[]{0.45 \textwidth}
 \leftline{~~~~~~~\tiny\textbf{(b1)}}
\centerline{\includegraphics[width=5.8cm]{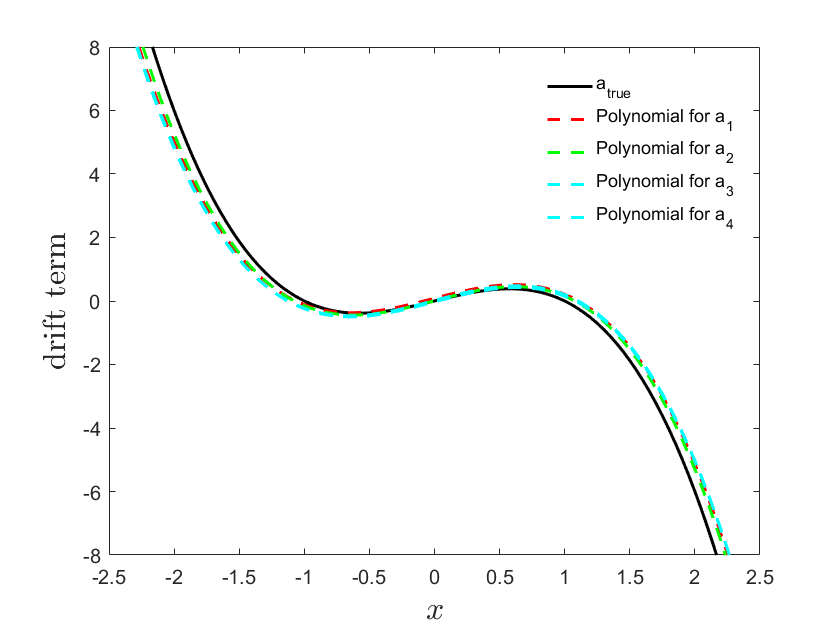}}
\end{minipage}
\hfill
\begin{minipage}[]{0.45 \textwidth}
 \leftline{~~~~~~~\tiny\textbf{(b2)}}
\centerline{\includegraphics[width=5.8cm]{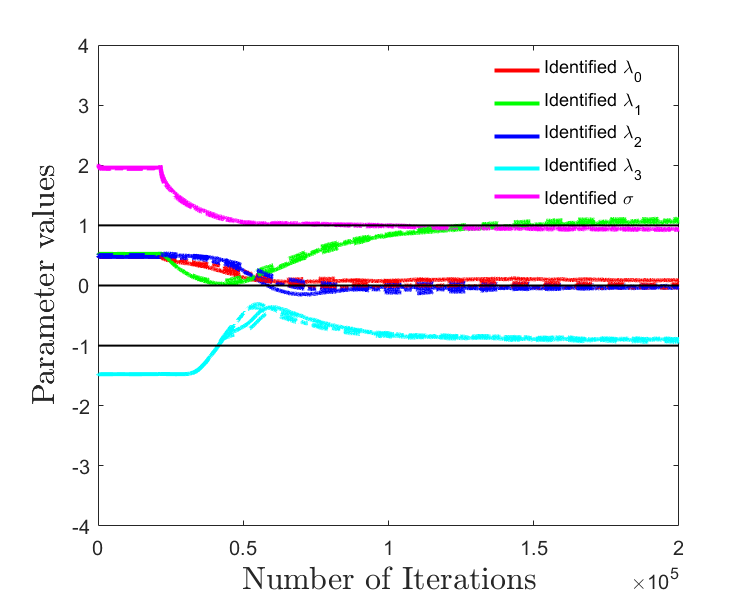}}
\end{minipage}
\caption{ \textbf{Appendix F. Inverse problem II - 4D Brownian noise:} Parameter evolution versus iteration number for two different observation data sets. Here $\lambda_{i}$ denotes the parameters $\lambda_{ix},\lambda_{iy},\lambda_{iz},\lambda_{ip}$, and $\sigma$ denotes the parameters $\sigma_{x},\sigma_{y},\sigma_{z},\sigma_{p}$, where $i=1,2,3,4$. Observation data at (a) $t=0.1,0.3,0.5,0.7,1$; (b) $t=0.1,0.2,0.3,0.5,0.7,0.9,1$.}
\label{f-4d-drift}
\end{figure}

For the 4D case, we consider having observations at $t=0.1,0.3,0.5,0.7,1$, i.e., $N=5$, or $t=0.1,0.2,0.3,0.5,0.7,0.9,1$, i.e., $N=7$. For each time snapshot, we have $10,0000$ samples available and set the minibatch size as $10,000$. Monte Carlo method with $1000,000$ samples are used to approximate the integral part of \eqref{eqn:loss_data}. We use $20,000$ residual points for \eqref{eqn:loss_pde}. The results are shown in Figure \ref{f-4d-drift}.

\bibliographystyle{plain}
\bibliography{references}
\end{document}